\documentclass[11pt,a4paper]{article} 
\pdfoutput=1
\usepackage{jheppub}

\usepackage{amsmath,amsfonts}
\usepackage{latexsym}
\usepackage{amssymb}
\usepackage{hhline}
\usepackage{color}
\usepackage{verbatim}
\usepackage{graphicx}
\usepackage{epstopdf}
\usepackage{textcomp}

\makeatletter
\@addtoreset{equation}{section}
\makeatother

\newcommand{\be}{\begin{equation}}
\newcommand{\ee}{\end{equation}}
\newcommand{\bea}{\begin{eqnarray}}
\newcommand{\eea}{\end{eqnarray}}
\newcommand{\Tr}{\operatorname{Tr}}

\preprint{BI-TP 2014/07\\
\hspace*{12.5cm}DESY 14/039\\
\hspace*{12.5cm}FTUAM 14/12\\
\hspace*{11.cm}IFT-UAM/CSIC-14-025\\
}

\title{\centering Dark Matter versus $h\to \gamma\gamma$ and $h\to \gamma Z$ \\
  with supersymmetric triplets}

\author[1]{Chiara Arina}
\author[2]{, V\'ictor Mart\'in-Lozano}
\author[3,4]{and Germano Nardini}

\affiliation[1]{Institut d'Astrophysique de Paris, 98bis boulevard Arago, 75014 Paris (France)}
\affiliation[2]{Instituto de F\'isica Te\'orica UAM/CSIC and Departamento de F\'isica Te\'orica, Universidad Aut\'onoma de Madrid, 28049 Madrid (Spain)}
\affiliation[3]{Deutsches Elektronen Synchrotron, Notkestrasse 85, D-22603 Hamburg (Germany)}
\affiliation[4]{Fakult\"at f\"ur Physik, Universit\"at Bielefeld, D-33615 Bielefeld (Germany)}

\abstract{The Triplet extension of the MSSM (TMSSM) alleviates the
  little hierarchy problem and provides a significant enhancement of
  the loop-induced diphoton rate of the lightest CP-even Higgs h. In
  this paper we pursue the analysis of the TMSSM Higgs phenomenology
  by computing for the first time the h to Z and gamma
  decay. Interestingly we find that the rates of loop-induced decays
  are correlated and their signal strengths can rise up to 40\% --
  60\% depending on the channel. We furthermore study the dark matter
  phenomenology of the TMSSM. The lightest neutralino is a good dark
  matter candidate in two regions. The first one is related to the
  Higgs and Z resonances and the LSP is mostly Bino. The second one is
  achieved for a mass larger than 90 GeV and the LSP behaves as the
  well-tempered neutralino. An advantage of the triplet contribution
  is that the well-tempered neutralino can be a Bino-Triplino mixture,
  relieving the problem of achieving $M_2\sim M_1$ in unified
  scenarios. The dark matter constraints strongly affect the Higgs
  phenomenology, reducing the potential enhancements of the diphoton
  and of the Z + gamma channels by 20\% at most. In the near future, dark
    matter direct searches and collider experiments will probe most of
    the parameter space where the neutralino is the dark matter
    candidate.
    }


\begin{document}

\maketitle
\flushbottom

%%%%%%%%%%%%%%%%%%%%%%%%%%%%%%%%%%%%%%%%%%%%%%%%%%%%%%%%%%%%%%%%%%%

\section{Introduction} \label{sec:intro}

The discovery of the Higgs boson~\cite{CMS:yva,ATLAS:2013sla} has
closed a long era: its mass is no more a free parameter. Its value
$m_h\simeq 126\,$GeV is in agreement with the mass range predicted in
supersymmetric scenarios~\cite{Espinosa:1992hp}. Nevertheless, the
minimal version of these models, the so-called Minimal Supersymmetric
Standard Model (MSSM), turns out to be ailing by the LHC
discovery. The value $m_h\simeq 126\,$GeV is indeed well above the one
that the MSSM {\it naturally} predicts, and heavy third generation
squarks and large stop mixing are required to reproduce the measured
mass~\cite{Arbey:2011ab,Carena:2011aa,Kang:2012bv,Fan:2014txa}. The MSSM
electroweak sector therefore needs an unpleasant amount of fine tuning
and a little hierarchy problem plagues the model.

In non-minimal supersymmetric scenarios this problem can be
alleviated.  They can indeed involve new contributions (absent in the
MSSM) that rise the tree-level prediction of the Higgs mass. For this
reason smaller radiative corrections and less tuning in the
electroweak sector are required. The drawback of this important
achievement is (partial) loss of predictivity since extra free
parameters have been introduced. A compromise between naturalness and
predictivity is thus to consider scenarios extending the MSSM as
little as possible.

If one does not enlarge the gauge symmetry group of the Standard Model
(SM), the only extension boosting the tree-level Higgs mass is to
couple new chiral superfields to the Higgs sector of the
superpotential. To this aim only singlets and $SU(2)_L$ triplets with
hypercharges $Y=0,\pm 1$ are allowed by gauge
invariance~\cite{Espinosa:1991wt}. Whereas the former
option has been deeply studied, see
e.g.~\cite{Ellwanger:2009dp} and references therein, the latter
is less known and has received special attention only after ATLAS and
CMS initially measured
sizeable deviations in the diphoton Higgs
rate~\cite{Aad:2012tfa,Chatrchyan:2012ufa}. Indeed, the triplet
superfield involves extra charginos that can largely enhance the
diphoton
channel~\cite{antonio1,Kang:2013wm,Basak:2013eba} without requiring peculiar features such as large
deviations in the main Higgs decay rates, huge stop masses, ultra
  light charginos or very heavy Higgsinos as it occurs in other
scenarios~\cite{Carena:2011aa,Batell:2013bka,Casas:2013pta,Belanger:2014roa}. In
particular, such an enhancement can be achieved in both decoupling and
non-decoupling regime (i.e.~with large and small CP-odd Higgs mass
$m_A$) while resembling the dominant SM Higgs
couplings~\cite{antonio2}.

Although the observed Higgs signal
strengths~\cite{CMS:yva,ATLAS:2013sla} might appear SM-like because of
an accidental compensation between production and decay rates that per
se differ from the SM predictions, it is still worth to analyze
scenarios where each Higgs decay but the loop-induced ones, and
  Higgs production is SM-like. In this simplified approach, indeed, it is easier to highlight the
origin of a potential deviation (in loop-induced channels) that lies
on the top of the global suppression/enhancement present in all
channels. Such a deviation is somehow expected since loop-induced
processes are particularly sensitive to new physics that can not
  perturb the dominant Higgs channels. This method has been applied
in ref.~\cite{antonio1} to show that charginos can provide up to
45\% diphoton enhancement in the $Y=0$ Triplet extension of the
MSSM (TMSSM).

The same approach is applied in the present paper. We extend the
analysis of ref.~\cite{antonio1} to a broader parameter space and we
find that a slightly larger enhancement of about 60\% can be achieved
via chargino contributions.  More interestingly, we show that this
departure from the SM prediction is tightly correlated to the
deviation in the $h\to Z\gamma$ channel.  In any case, the
$\Gamma(h\to Z\gamma)$ rate can never be larger than about 1.4 times
its SM value~\footnote{For studies on the $Z\gamma$ channel in other
  non-minimal supersymmetric frameworks see
  i.e.~refs.~\cite{Cao:2013ur,Belanger:2014roa}.}.

These upper bounds are obtained without imposing any Dark Matter (DM)
constraint on the TMSSM field content. Nevertheless they are
compatible with the DM observables if the Higgs phenomenology is
somehow disentangled from the DM puzzle. This is achieved for instance
by invoking gravitinos, axions and axinos as DM
candidates~\cite{Covi:1999ty,Baer:2008yd,Steffen:2008qp,Feng:2010gw},
or by postulating cosmological scenarios with non-standard DM
production~\cite{Gelmini:2010zh}. On the contrary, if the DM candidate
is required to be the Lightest Supersymmetric Particle (LSP) of the
TMSSM within the traditional cosmological assumptions, the above
bounds should be revisited. To this aim we pursue the analysis of the
Higgs phenomenology for the case having the lightest neutralino as DM
particle. In order to capture the most stringent features related to
the $h\to Z\gamma$ and $h\to \gamma\gamma$ enhancements, we require
the relic density to rely only on the chargino, neutralino and SM
  fields. In other words, besides analyzing the DM annihilation
  via  Higgs and $Z$ boson resonances, we study a kind of
well-tempered neutralino in the TMSSM.

By definition the well-tempered neutralino in the MSSM is a tuned mixture of gaugino and Higgsino that achieves the
correct relic density away from resonances and coannihilations with other supersymmetric
particles~\cite{ArkaniHamed:2006mb}. The successful parameter space
consists of either the Bino and Higgsinos, or the Bino and Wino having
almost degenerate mass terms. Other issues however jeopardize these
two scenarios: the former is strongly constrained by limits on the DM
Spin-Independent (SI) elastic scattering, and the latter seems
unnatural since supersymmetry breaking mechanisms unlikely lead to
degenerate Bino and Wino soft masses.

Introducing the TMSSM fermionic triplet, hereafter dubbed Triplino,
provides new features to the DM phenomenology. In fact, the
  Triplino can play the role of the Wino component, making the tuning
between Bino, Wino and Higgsinos masses unnecessary and opening up a
new viable DM parameter space for the well-tempered
neutralino. Moreover, the 
Triplino mass parameter is a superpotential term that in
principle can be produced by supersymmetry breaking sources different
from those generating the gaugino masses~\footnote{For instance, one
  can produce the gaugino masses via gauge mediation and the mass
  parameters of Higgsinos and Triplinos via the Giudice-Masiero
  mechanism. Notice that the TMSSM does not seems to be in tension
  with gauge mediation due to the Higgs mass $m_h\approx 126$. Indeed,
  no large trilinear parameters are required to naturally achieve the
  observed Higgs mass~\cite{antonio1}.}.

Interestingly, we find that in the TMSSM the DM constraints strongly
impact the loop-induced Higgs processes. Independently on the regions where the LSP achieves the observed relic
density 
the $h \to \gamma \gamma$ and $h\to Z\gamma$ enhancements cannot be
larger than 20\%. This mostly occurs because larger enhancements need
light Higgsino and Triplino mass parameters, which tend to
push the SI elastic scattering off nuclei of the lightest neutralino
above the LUX exclusion limit~\cite{Akerib:2013tjd}.

The rest of the paper is organized as follows. In
section~\ref{sec:model} we review the basic features of the TMSSM and
its most natural parameter space. We also present some improvements in
the determination of the lightest Higgs mass $m_h$.
Section~\ref{sec:higgssign} describes the Higgs signatures in the
TMSSM, with emphasis to the Higgs invisible width and loop-induced
decay channels. In particular, the first calculation of the
$\Gamma(h\to Z\gamma)$ width in the TMSSM is presented
here. Section~\ref{sec:num} is dedicated to set up the method of our
numerical analysis, as well as the parameter
choice. Section~\ref{sec:resh} studies in detail the signal strengths
of loop-induced Higgs decays and their correlation. We then move to
discuss the DM phenomenology and its impact on the Higgs
signatures in section~\ref{sec:dm}. Section~\ref{sec:Concl} is finally
devoted to summarize our findings.

%%%%%%%%%%%%%%%%%%%%%%%%%%%%%%%%%%%%%%%%%%%
\section{The TMSSM model}\label{sec:model}
\subsection{Generic Features}
\label{sec:features}
In the TMSSM the matter content of the MSSM is extended by a $Y=0$
$SU(2)_L$-triplet superfield
\be
\Sigma=\left(
\begin{array}{cc}
  \xi^0/\sqrt{2} & \xi_2^+\\
  \xi_1^-&  -\xi^0/\sqrt{2}
\end{array}
\right)
\label{Sigma}~.
\ee
In comparison with the MSSM, the TMSSM superpotential and
soft-breaking Lagrangian contain respectively two and three extra
renormalizable
terms~\cite{Espinosa:1991wt,DiChiara:2008rg}:
\bea
\label{super}
W_{\rm TMSSM}&=&W_{\rm MSSM}+\lambda H_1 \cdot \Sigma H_2+\frac{1}{2} \mu_\Sigma \Tr\Sigma^2~,\\
\mathcal L_{\rm TMSSM_{\rm SB}} &=&\mathcal L_{\rm MSSM_{\rm SB}} +
m_4^2 \Tr(\Sigma^\dag\Sigma) + \left[B_\Sigma \Tr(\Sigma^2) +
  \lambda A_\lambda H_1\cdot\Sigma H_2 + \mathrm{h.c.}\right]~,
\eea
where $A \cdot B \equiv \epsilon_{ij} A^iB^j$ with
$\epsilon_{21}=-\epsilon_{12}=1$ and $\epsilon_{22}=\epsilon_{11}=0$.
For sake of simplicity we assume no sources of CP violation and
consequently all parameters are taken as real.

In general the neutral scalar component $\xi^0$ acquires a VEV
$\langle \xi^0\rangle$. Electroweak precision observables impose
$\langle \xi^0\rangle\lesssim 4\,$GeV at 95\% CL~\cite{PDG,
  antonio2} which, unless of a tuning on the parameters, corresponds
to the hierarchy
\be \displaystyle |A_\lambda|,\,|\mu| \,,|\mu_\Sigma| \lesssim
10^{-2}
\frac{m^2_\Sigma+\lambda^2 v^2/2}{\lambda v}~,\label{approximation}
\ee
with $m_\Sigma^2\equiv m_4^2+\mu_\Sigma^2+B_\Sigma\mu_\Sigma$.  For
$A_\lambda,\,\mu$ and $\mu_\Sigma$ at the electroweak scale, such a
hierarchy requires $m_\Sigma\gtrsim
2\,$TeV~\cite{antonio2}~\footnote{For discussions on the naturalness of
  such a hierarchical scenario see refs.~\cite{antonio1,antonio2}.}.
As a consequence, the mixing between the MSSM Higgs sector and the
scalar triplet is rather small and it can be safely neglected for
$m_\Sigma\gtrsim 5\,$TeV~\cite{antonio1}. These values of $m_\Sigma$
as well as the hierarchy in eq.~\eqref{approximation} will be assumed
in the following. This in particular allows to take $\langle
\xi^0\rangle\approx 0$.

As the $\Sigma$ scalar components decouple from the Higgs fields
$H_{1}$ and $H_{2}$, which interact with the down and up right-handed
quarks respectively, the Higgs sector at the electroweak scale looks
like the one of the MSSM with some $\mathcal O(\lambda^2v^2)$ shifts
in the tree-level mass spectrum. By imposing the minimization
conditions for the electroweak symmetry breaking, it turns
out~\cite{antonio2}
\begin{eqnarray}
  m_3^2&=&m_A^2 \sin\beta\cos\beta ~,\\
  m_Z^2&=&\frac{m_2^2-m_1^2}{\cos 2\beta}-m_A^2+\lambda^2 v^2/2~,\label{min_MZ}\\
  m_A^2&=&m_1^2+m_2^2+2|\mu|^2+\lambda^2 v^2/2~,\label{min_MA}\\
  m_H^\pm&=&m_A^2+m_W^2+\lambda^2 v^2/2~,
\end{eqnarray}
where $\tan\beta=v_2/v_1$, $v=\sqrt{v_1^2+v_2^2}=174\,$GeV, $m_Z$ and
$m_W$ are the $Z$ and $W$ vector boson masses, and $m_1^2$, $m_2^2$ and
$m_3^2$ are the usual MSSM soft parameters of the Higgs fields
$H_{1,2}$ whose neutral components are decomposed as
$H_i^0=v_i+(h_i+i\chi_i)/\sqrt{2}$. Moreover, the CP-even squared mass
matrix in the basis $(h_2,h_1)$ is given by
\be
  \mathcal M^2_{h,H}=\left(
\begin{array}{cc}
  m_A^2 \cos^2\beta+m_Z^2\sin^2\beta &(\lambda^2
  v^2-m_A^2-m_Z^2)\sin\beta\cos\beta \\
  (\lambda^2 v^2-m_A^2-m_Z^2)\sin\beta\cos\beta&m_A^2  
  \sin^2\beta+m_Z^2\cos^2\beta 
\end{array}
\right)~.
\label{scalarmass}
\ee
The contributions $\mathcal O(\lambda^2v^2)$ lift the lightest
eigenvalue $m_h^2$ and the little hierarchy problem can be then
alleviated with respect to the MSSM. This can be easily seen in the
limit $m_A\to\infty$ where
\begin{equation}
  \label{tree-level}
m^2_{h,{tree}}=m_Z^2\cos^2 2\beta+\frac{\lambda^2}{2} v^2 \sin^2 2\beta~.
\end{equation}
The $\mathcal O(\lambda^2v^2)$ term can provide a sizeable boost to
$m_h$. In particular, no large radiative corrections are required to
catch $m_h\simeq 126\,$GeV for large $\lambda$ and small $\tan\beta$~\footnote{Nevertheless, large values of $\lambda$ generate a Landau
  pole and the TMSSM may require an ultraviolet completion to maintain
  perturbativity up to the unification scale.}.

On the other hand, some rather large radiative corrections to the
Higgs sector are unavoidable due to the lack of experimental evidence
of stops and gluinos. Within the specific assumptions the experimental
analyses are based on~\cite{ATLAS:2013oea}, stop and gluino bounds in
the presence of any lightest neutralino mass are quite stringent,
namely $m_{\widetilde t}\gtrsim 650\,$GeV and $M_3\gtrsim
1.4$~TeV~\cite{SUS-013-011, SUS-013-024, atlas} (for loopholes see
e.g.~refs.~\cite{Franceschini:2012za, Fuks???}). Their radiative
corrections to the Higgs sector are then sizeable and need to be
stabilized at the expense of a certain amount of fine tuning in the model (for details see
e.g.~ref.~\cite{Papucci:2011wy}).

A further important source of tuning comes from the triplet if
$m_\Sigma$ is large. We require this to be subdominant to the gluino
and stop ones in order to alleviate the little hierarchy problem as
much as possible. Notice that this condition does not prevent from
$m_\Sigma>m_{\widetilde t}$ since triplets have less degrees of freedom
and (typically) smaller coupling to $H_{u,d}$ than stops. In this
respect, the parameter choice $m_\Sigma\gtrsim 5\,$TeV, $m_{\widetilde t}\gtrsim 650\,$GeV and $\lambda\lesssim 1$ is
allowed~\cite{antonio1,antonio2}.

In order to simplify our analysis, we will restrict the parameter
space to a subset where all the above issues are taken into account.
We will focus on the parameter region
\bea 
m_\Sigma=&& \mathrm{5\,TeV}~,\quad A_t=A_b=0~,\quad M_3=
\mathrm{1.4\,TeV}~,\quad m_A=\mathrm{1.5\,TeV}\,,\label{set1}\\
&&\lambda\lesssim 1~,\quad \tan\beta \sim\mathcal O(1) ~,\quad
\widetilde{m}\gtrsim \mathrm{750\,GeV}~,\label{set2}
\eea
with $\widetilde m= m_U=m_D=m_Q$. This choice indeed {\it (i)}
alleviates the little hierarchy problem as it boosts $m_h$ with
subdominant $\Sigma $ radiative corrections. Moreover, as far as
$\mu_\Sigma$ and $\mu$ are not too large, it {\it (ii)} naturally
satisfies the hierarchy \eqref{approximation} and {\it (iii)} allows
to neglect the mixing between the $\Sigma$ scalars and the low energy
sector. All sleptons are considered heavy enough not to interact with
the relevant SUSY spectrum; numerically they have been taken to be 3
TeV. The precise parameter space we consider is defined in
section~\ref{sec:num}, together with all observational constraints
used in this analysis.

\subsection{The Higgs Mass}
\label{sec:Higgs}

Nowadays the LHC measurement of the Higgs mass is very accurate. The
most recent analyses present 2-$\sigma$ uncertainties of about 1\% on
the central value $m_h\simeq
125.6\,$GeV~\cite{ATLAS:2013mma,CMS:yva}. Such accuracy goes much
further than the typical precision that beyond-the-SM theoretical
papers achieve. These works indeed are more aimed to capture the
qualitative features of new frameworks than to accurately evaluate
their predictions.

In this spirit, seminal works on the TMSSM have analyzed the Higgs
sector at tree-level
approximation~\cite{Espinosa:1991wt,Espinosa:1992hp,
  DiChiara:2008rg}. Dominant one-loop corrections coming from stops
and scalar triplets, as well as one-loop contributions from heavy
Higgsinos and Triplinos, have been included only
recently~\cite{antonio1,antonio2,Bandyopadhyay:2013lca}. Despite
these efforts, the theoretical uncertainties on the TMSSM Higgs mass
spectrum is far from being comparable with the experimental one.

A pragmatic approach to this problem is to absorb the (potentially
large) theoretical error on $m_h$ into an effective uncertainty on the
high energy parameters, especially on the $m_U$, $m_Q$ and $m_\Sigma$
soft-breaking terms (and on the trilinear parameters if they are
allowed to be large). It is however problematic to quantify the latter
uncertainty and how it propagates to the physical observables. For
instance, big effects can arise in the DM relic density in the
neutralino-stop coannihilation region, or in the SI cross-section when
stop mediation dominates the interaction. On the other hand, less
dramatic effects arise when the parameters absorbing the Higgs
theoretical uncertainty provide sub-leading corrections to the
observables.  In order to reduce these uncertainties, here we improve
the recent TMSSM Higgs mass
calculations~\cite{antonio1,antonio2,Bandyopadhyay:2013lca} and consider loop
effects in the whole mass spectrum.

For this purpose we use the \texttt{SARAH-3.3.0}
program~\cite{Staub:2009bi,Staub:2010jh} to obtain the full two-loop
Renormalization Group Equations (RGEs). The code, which works in the
$\overline{\rm DR}$ renormalization scheme, also provides the full
one-loop ElectroWeak-Symmetry Breaking (EWSB) conditions and full
one-loop spectrum to which we include some $\mathcal O(h_t^2g_3^2)$
and $\mathcal O(h_t^4)$ two-loop contributions.

\begin{figure}[t]
\begin{minipage}[t]{0.49\textwidth}
\centering
\includegraphics[width=1.\columnwidth,trim=0mm 0mm 0mm 0mm, clip]{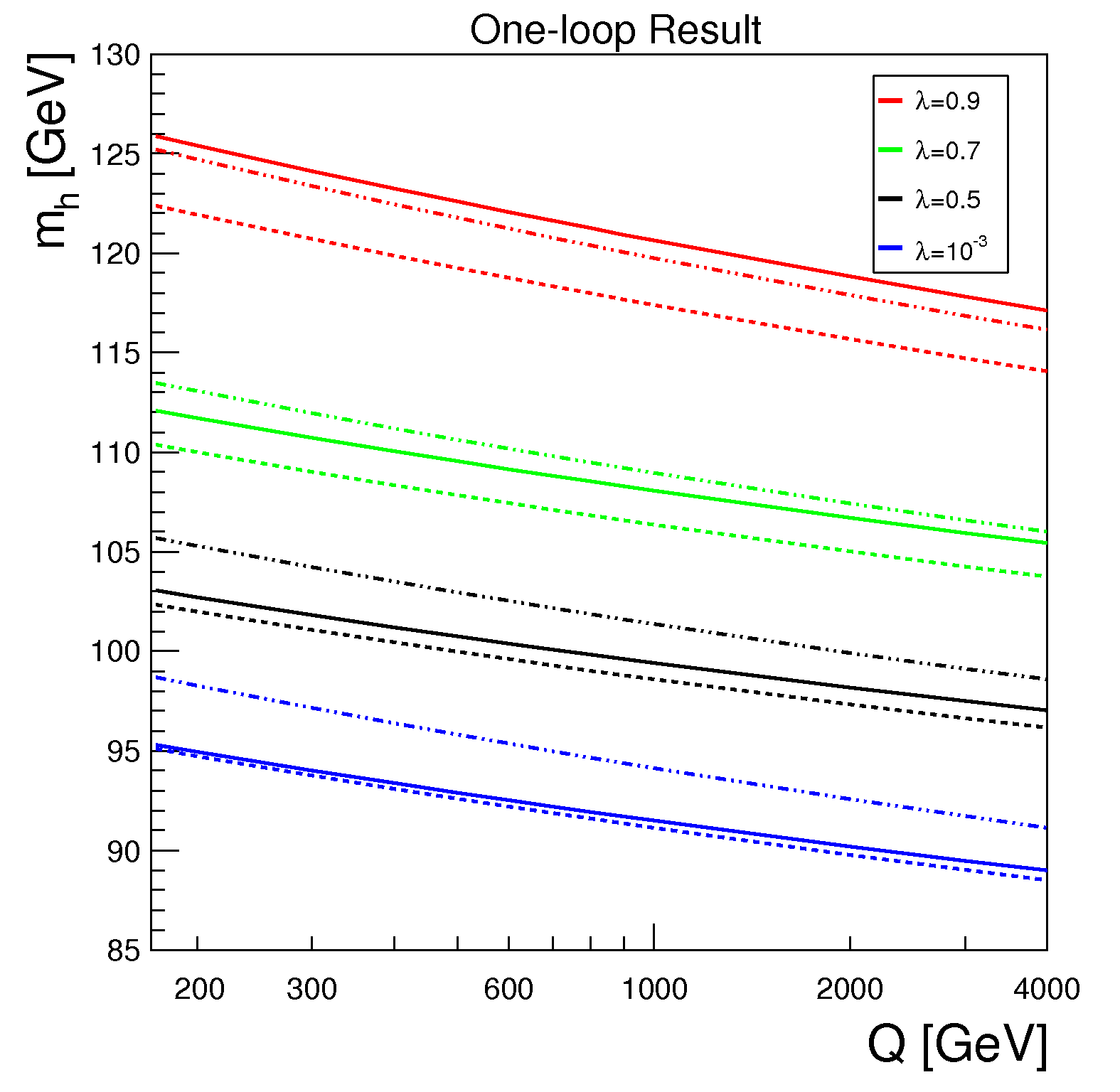}
\end{minipage}
\hspace*{0.2cm}
\begin{minipage}[t]{0.49\textwidth}
\includegraphics[width=1.\columnwidth,trim=0mm 0mm 0mm 0mm, clip]{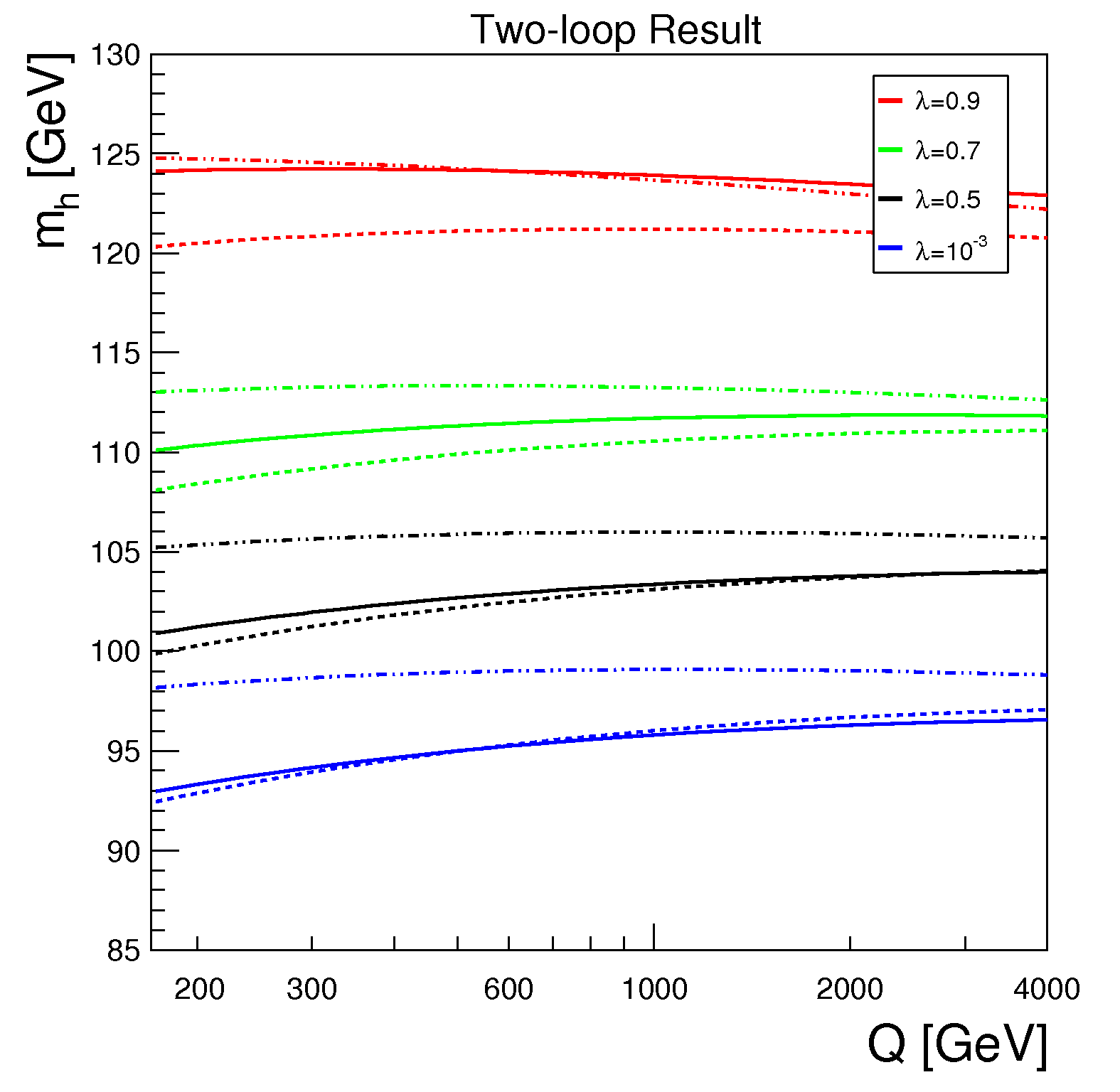}
\end{minipage}
\caption{{\it Left:} The Higgs mass $m_h$ as a function of the SUSY
  renormalization scale $Q$ in the one-loop approximation. {\it
    Right:} Same as left in the two-loop approximation. Same color
  identifies same input value of $\lambda$, as labelled. A subset of
  parameters is fixed at: $\mu_\Sigma= M_1=150$ GeV, $m_A=1.5\,$TeV,
  $m_\Sigma=5\,$TeV, $A_t=A_b=0$ and $M_3=1.4\,$TeV. Solid lines
  (dashed lines) [dotted-dotted-dashed lines] are evaluated for
  $\widetilde m=700\,$GeV and $\mu= M_2=300\,$GeV ($\widetilde
  m=700\,$GeV and $\mu=M_2=1\,$TeV) [$\widetilde m=\mu=M_2=1\,$TeV].}
\label{fig:scaleDepen}
\end{figure}

The RGEs are solved numerically by the
\texttt{SPheno-3.2.4}~\cite{Porod:2003um,Porod:2011nf} code.  The
solution fulfills the above EWSB conditions at the electroweak scale
$m_Z$, as well as some experimental constraints (e.g.~the quark mass
spectrum; for details see refs.~\cite{Porod:2003um,Porod:2011nf}).  It
is univocally determined once we choose the values of the residual
free parameters of the theory~\footnote{The quantities $M_1^2$,
  $M_2^2$ and $A_\lambda$ are fixed as functions of the other
  parameters through the EWSB equations with $B_\Sigma=0$.}. These
inputs are given (and we will quote them) at the SUSY renormalization
scale, $Q$.

Once the RGEs are solved, all running parameters and couplings at the
scale $Q$ are known. These are used to determine the pole mass
spectrum. In this way, we determine the pole mass $m_h$ at full
one-loop plus $\mathcal O(h_t^2g_3^2)+\mathcal O(h_t^4)$ two-loop
order on top of the two-loop RGE resummation~\footnote{We include the
  $\mathcal O(h_t^2 g_3^2)$ and $\mathcal O(h_t^4)$ two-loop effects
  since we expect $\mathcal O(h_t^2 \lambda^2)$ corrections to be
  subdominant in the regime $\lambda\lesssim 1$ and small $\tan\beta$
  due to the color factors and $h_t^2=m_t^2/\sin^2\beta\lesssim
  \lambda^2$. These $\mathcal O(h_t^2 g_3^2)$ and $\mathcal O(h_t^4)$
  corrections match with those of the MSSM and are therefore easy
    to implement in~\texttt{SPheno} (for details see
  ref.~\cite{Allanach:2004rh} and references therein).}.

The renormalization scale dependence $m_h(Q)$ highlights the
improvement in the Higgs mass calculation and it is presented in
figure~\ref{fig:scaleDepen} for several values of $\lambda$ and the
parameter setting in eq.~\eqref{set1} with $\mu_\Sigma= M_1=
150\,$GeV. In the figure solid (dotted) [dotted-dotted-dashed] lines
are plotted for $\widetilde m=700\,$GeV and $\mu= M_2= 300\,$GeV
($\widetilde m=700\,$GeV and $\mu= M_2= 1\,$TeV) [$\widetilde m=\mu=
M_2= 1\,$TeV]. The scale dependence is strongly reduced by going from
one-loop (left panel) to two-loop (right panel) order. The addition of
the $\mathcal O(h_t^2g_3^2)+\mathcal O(h_t^4)$ contributions is then
crucial to improve the result, as it is well known in the
MSSM~\footnote{Notice that the lines with very small $\lambda$
  reproduce the MSSM result except of modifications due to the 
  extra $SU(2)_L$-charged content provided by the triplet.}
(cf.~curves at $\lambda=10^{-3}$), whereas the undetermined $\mathcal
O(\lambda^2 h_t^2)$ corrections seem to be subdominant even at
$\lambda\approx 1$.

Figure~\ref{fig:scaleDepen} also guides in the choice of $Q$. The
$\mathcal O(h_t^2g_3^2)+\mathcal O(h_t^4)$ corrections are minimized
at $Q$ nearby the electroweak scale, and $m_h(Q\approx m_t)$ is then
expected to be quite stable under further radiative
corrections. Although the exact number slightly depends on the
parameter choice, for concreteness we fix $Q=m_t$ in the rest of
the analysis.
 
A last comment concerns the chargino and neutralino parameters. As
shown in the figure, if (part of) the chargino spectrum is heavy, relevant negative corrections to $m_h$ can
arise~\cite{Bandyopadhyay:2013lca}. For instance, depending on the
value of $\lambda$, $m_h$ is lowered by about 1--\!\!:\,4 GeV by
moving $\mu= M_2$ from 300\,GeV to 1\,TeV when $\widetilde m=700\,$GeV
and $\mu_\Sigma=M_1=150\,$GeV (c.f.~dotted and solid curves of
figure~\ref{fig:scaleDepen}). Of course, this decrement can be
compensated by modifying either $(\lambda,\tan\beta)$ and/or by
increasing $\widetilde m$, as the dotted-dotted-dashed lines
highlight.

%%%%%%%%%%%%%%%%%%%%%%%%%%%%%%%%%%%%%%%%%%%
\section{Higgs signatures}\label{sec:higgssign}

Since our aim is to explore the qualitative capabilities of the TMSSM,
in particular those related to DM features, we do not look for
interplay of Higgs production and decay widths to overcome the LHC
bounds. We instead try to work well within the ballpark allowed by
data, that is, we attempt to reproduce a SM-like Higgs sector.

The first step in this direction is to fix the tree-level Higgs
couplings to SM fields.  They are SM-like if, on the top of our
assumption $m_\Sigma\gg m_h$, it occurs either {\it (i)} $m_A$ is much
larger that $m_h$ or {\it {(ii)}} $\tan\beta$ and $\lambda$ have
values within the so-called alignment region~\cite{antonio2}. Here we
focus on the first possibility. In this case results are independent
of the specific choice of $m_A$ and we can thus fix $m_A=1.5$\,TeV
without lack of generality.

The second step is to check the radiative corrections to the Higgs
couplings coming from non-SM particles. For our parameter choice,
given in eqs.~\eqref{set1} and~\eqref{set2}, loop corrections to
tree-level interactions are negligible.  They may instead be
responsible of important deviations from the SM in loop-induced
processes. For gluon fusion, which is the main Higgs production
mechanism at LHC, no relevant deviation arises in our analysis since
squarks are assumed rather heavy and $\tan\beta$ is small. Therefore,
the total Higgs production is SM-like. On the contrary, charginos may
be light and eventually the $\Gamma(h\to\gamma\gamma)$ and
$\Gamma(h\to Z\gamma)$ widths may depart from their SM
values. However, these two processes are not yet well measured due to
lack of statistics and of indirect impact on other processes: in
practice $\Gamma(h\to\gamma\gamma)$ and $\Gamma(h\to Z\gamma)$ are so
small that they play no role in the branching ratios of other Higgs
decays. For this reason we do not force them to be SM-like, as we aim
to do with the dominant Higgs channels.

Finally, one has to guarantee that no new relevant Higgs decay process
is open. This typically occurs when the mass of the lightest
neutralino is sufficiently small to allow for the
$h\to\widetilde\chi_1^0\widetilde\chi_1^0$ channel. In such a case,
any signal strength $R_{XY}\equiv{\rm BR}(h\to XY)/{\rm BR_{SM}}(h\to
XY)$ calculated by disregarding the invisible width, should be
corrected by the factor $1
-{\rm BR}(h\to\widetilde\chi^0\widetilde\chi^0)$~\footnote{This
  definition of $R_{XY}$ is based on the fact that the Higgs
  production is SM-like for the setting in eqs.~\eqref{set1}
  and~\eqref{set2}.}. As the branching ratio
${\rm BR}(h\to\widetilde\chi^0\widetilde\chi^0)$ is bounded by ATLAS
and CMS analyses~\cite{Aad:2014iia,CMS:2013yda}, it is worth
to estimate it.

\subsection{The $h\to\widetilde\chi_1^0 \widetilde\chi_1^0$ channel}
The Higgs decay channel into a pair of lightest neutralinos is open
for $m_{\widetilde\chi_1^0} < m_h/2$. Its width is given by
\begin{equation}
\Gamma(h\to\widetilde\chi^0_1\widetilde\chi^0_1)=\frac{G_F
  m_W^2}{2\sqrt{2}\pi}\,m_h\left(1-\frac{4m_{\widetilde\chi^0_1}^2}{m_h^2}\right)^{3/2}g_{h\chi^0_1\chi^0_1}^2~,
\label{invisib}
\end{equation}
where 
\begin{equation}\label{eq:ghh}
  g_{h\chi^0_1\chi^0_1}=(N_{12}-\frac{g_1}{g_2} N_{11})(\sin\beta N_{14}-\cos\beta N_{13})+
  \frac{\lambda}{g_2}N_{15}(N_{14}\sin\beta+N_{13}\cos\beta)\,.
\end{equation}
Here the quantities $N_{1i}$ are the components of the lightest
(unitary) eigenvector of the neutralino mass matrix ${\mathcal
  M}_{\widetilde\chi^0}$ which is determined at one-loop after the
RGEs flow achieved via \texttt{SPheno} and \texttt{SARAH} as explained
in section~\ref{sec:Higgs}. The quantity $m_{\widetilde\chi^0_1}$ is
the pole mass of the lightest eigenstate of ${\mathcal
  M}_{\widetilde\chi^0}$. At tree level ${\mathcal
  M}_{\widetilde\chi^0}$ reduces to
\begin{equation} \label{eq:mnmass}
{\mathcal M}^{tree}_{\widetilde\chi^0} = \left( 
\begin{array}{ccccc}
M_1 &0 &-\frac{1}{2} g_1 v_1  &\frac{1}{2} g_1 v_2  &0\\ 
0 &M_2 &\frac{1}{2} g_2 v_1  &-\frac{1}{2} g_2 v_2  &0\\ 
-\frac{1}{2} g_1 v_1  &\frac{1}{2} g_2 v_1  &0 &- \mu  &-\frac{1}{2} v_2 \lambda \\ 
\frac{1}{2} g_1 v_1  &-\frac{1}{2} g_2 v_2  &- \mu  &0 &-\frac{1}{2} v_1 \lambda \\ 
0 &0 &-\frac{1}{2} v_2 \lambda  &-\frac{1}{2} v_1 \lambda  & \mu_\Sigma \end{array} 
\right) ~.
\end{equation} 

Notice that due to the LEP chargino mass constraint
$m_{\widetilde\chi_1^+}\gtrsim 100\,$GeV, a lightest neutralino with
mass $m_{\widetilde\chi_1^0} < m_h/2$ must be predominantly Bino.  The
coupling $g_{h\chi^0_1\chi^0_1}$ is then dominated by the Higgsino and
Bino mixings, namely $N_{11}N_{13}$ and $N_{11}N_{14}$.  Consequently,
for a given set of parameters, the experimental constraint on
BR$(h\to\widetilde\chi^0_1\widetilde\chi^0_1)$ turns out to be a lower
bound on $\mu$.

When $m_{\widetilde\chi_1^0}$ is even smaller, namely lighter than
$m_Z/2$, also the LEP bound $\Gamma(Z\to \widetilde\chi_{1}^{0}
\widetilde\chi_{1}^{0})\lesssim 2\,$MeV ~\cite{ALEPH:2005ab} has to be
taken into account. As the constraint on BR$(h\to
\widetilde\chi_1^0\widetilde\chi_1^0)$, it imposes a lower bound on
$\mu$ once the other parameters are fixed. We quantify it by the
expression
\begin{equation}
\Gamma(Z\to {\widetilde\chi}_{1}^{0}{\widetilde\chi}_{1}^{0})=
\frac{1}{12\pi}\frac{G_F}{\sqrt{2}}m_Z^3
\left(
1-\frac{4 m_{\widetilde\chi_{1}^{0}}^2}{m_Z^2}
\right)^{3/2}
\left(|N_{13}|^2-|N_{14}|^2\right)^2 ~.
\label{eq:Zlr}
\end{equation}

Moreover, when Higgsinos are extremely heavy and the lightest
(Bino-like) neutralino is below the threshold of about 20 GeV, the
further channel $h\to \widetilde\chi_2^0\widetilde\chi_1^0$ may be
kinematically open without any dangerous enhancement to the invisible
width of the Higgs or $Z$ bosons.  However, being $\mu$ very large,
the chargino $\widetilde\chi_1^\pm$ and the neutralino
$\widetilde\chi_2^0$ are almost degenerate. The LHC analysis on three
leptons plus missing energy~\cite{SUS-013-011} excludes the parameter
region of this scenario where $\widetilde\chi_2^0$ mostly decays into
$Z^{(*)}\widetilde\chi_1^0$. In the remaining region where the channel
$\widetilde\chi_2^0\to h^{(*)}\widetilde\chi_1^0$ competes, it is
instead unclear what the experimental limits are. Determining them
would require a specific analysis that goes beyond the scope of this
study and we then conservatively focus on the region with BR$(h\to
\widetilde\chi^0\widetilde\chi^0)={\rm BR}(h\to
\widetilde\chi_1^0\widetilde\chi_1^0)$.

\subsection{The $h\to\gamma\gamma$ channel} 
\label{sec:hgg}

Since the Higgs production is SM-like, the diphoton signal strength
$R_{\gamma\gamma}$ depends only on BR$(h\to\gamma\gamma)$. For our
setting in eqs.~\eqref{set1} and~\eqref{set2} only charginos can
induce deviations from the SM prediction of
$\Gamma(h\to\gamma\gamma)$. Their contributions to $R_{\gamma\gamma}$
have been already calculated by means of the low-energy
approximation~\cite{antonio1,antonio2} or in the $M_2$
decoupling limit~\cite{DiChiara:2008rg}, starting from the
tree-level chargino mass matrix
\be
\mathcal
M_{\widetilde\chi^\pm}^{tree}=
\left(
\begin{array}{ccc} M_2& g _{2}v\sin\beta& 0\\ g _{2}v\cos\beta&\mu& -\lambda v\sin\beta\\0&\lambda v\cos\beta& \mu_\Sigma\end{array}
\right)~.
\label{charginos}
\ee
It has been observed that maximal diphoton enhancement occurs when all
chargino mass parameters are light (compatibly with the chargino mass
bound) and moreover, in the regime of very small $\tan\beta$ and large
$\lambda$ (linked one to each other by the Higgs mass constraint),
when Triplino and Higgsino mass parameters are
degenerate~\cite{antonio2}.

In the present analysis we improve the previous estimate by including
loop-corrections in $\mathcal M_{\widetilde\chi^\pm}$. In many cases
these radiative contributions increase the lightest chargino mass by
about 10\% with respect to its tree-level value. They can hence be
important when one cuts the allowed parameter space due to the LEP
bound $m_{\widetilde\chi^\pm_i}\gtrsim 100\,$GeV.

When only charginos provide new (sizeable) contributions to the
diphoton channel and the Higgs production is SM-like,
$R_{\gamma\gamma}$ is given by
\begin{eqnarray}
\label{Rgg}
& & R_{\gamma\gamma}=\left| 1+\frac{
    {A}^{\gamma\gamma}_{\widetilde\chi^\pm_{1,2,3}}}{A^{\gamma\gamma}_W+A^{\gamma\gamma}_t}
\right|^2\,,\\
&   & {A}^{\gamma\gamma}_{\widetilde\chi^\pm_{1,2,3}}=\sum_{i=1}^3 \frac{2 M_W}{\sqrt{2}\,m_{\widetilde\chi^\pm_i}}
(g_{h\widetilde\chi^+_i\widetilde\chi^-_i}^L + g_{h\widetilde\chi^+_i\widetilde\chi^-_i}^R)
A_{1/2}(\tau_{\widetilde\chi^\pm_i})\,,
\end{eqnarray}
where $A_{1/2}$ is the spin-1/2 scalar function (see
e.g.~ref.~\cite{Djouadi:2005gj} for its explicit expression) with
argument $\tau_{\widetilde\chi^\pm_i}=m_h^2/4
m_{\widetilde\chi^\pm_i}^2$, and
$g_{h\widetilde\chi^+_i\widetilde\chi^-_i}$ is the lightest Higgs
effective coupling to charginos. The quantities $A^{\gamma\gamma}_W$
and $A^{\gamma\gamma}_t$ are the $W$-boson and top-quark contributions
whose values are respectively -8.3 and 1.9 for $m_h\simeq 126\,$GeV.

In the procedure we apply, which corresponds to the one
\texttt{SPheno} and \texttt{SARAH} employ, $R_{\gamma\gamma}$ is
calculated by plugging the  chargino pole masses into
$A_{1/2}(\tau_{\widetilde\chi^\pm_i})$. Moreover, the couplings
$g_{h\widetilde\chi^+_i\widetilde\chi^-_i}^{R}$ and
$g_{h\widetilde\chi^+_i\widetilde\chi^-_i}^{L}$ are the particular
case $i=j$ of the expressions
\begin{comment}%
\be
\label{g_hchacha}
g_{h\widetilde\chi^+_i\widetilde\chi^-_j} = ~~g_{h\widetilde\chi^+_i\widetilde\chi^-_j}^L ~P_L ~+~ g_{h\widetilde\chi^+_i\widetilde\chi^-_j}^R ~P_R
\ee
%
with
%
\end{comment}
\bea
g_{h\widetilde\chi^+_i\widetilde\chi^-_j}^L &=& \frac{1}{\sqrt{2}} \left[\left(
U_{j1}V_{i2}-\frac{\lambda}{g_2} U_{j2}V_{i3}\right)  \sin\beta+ \left(U_{j2}V_{i1}+\frac{\lambda}{g_2} U_{j3}V_{i2}\right)\cos\beta \right]~, \label{g_hchachaL}
\\
g_{h\widetilde\chi^+_i\widetilde\chi^-_j}^R &=& \frac{1}{\sqrt{2}} \left[\left(
U_{i1}V_{j2}-\frac{\lambda}{g_2} U_{i2}V_{j3}\right)  \sin\beta+ \left(U_{i2}V_{j1}+\frac{\lambda}{g_2} U_{i3}V_{j2}\right)\cos\beta \right]~, \label{g_hchachaR}
\eea
where $U$ and $V$ are the unitary matrices diagonalizing the one-loop
chargino mass matrix $\mathcal M_{\widetilde\chi^\pm}$ such that $U
\mathcal
M_{\widetilde\chi^\pm}V^T=$\,diag$(m_{\widetilde\chi^\pm_1},m_{\widetilde\chi^\pm_2},m_{\widetilde\chi^\pm_3})$.

\subsection{The $h\to Z\gamma$ channel} 
\label{sec:hZg}

LHC constraints on $R_{Z\gamma}$ are still very
weak~\cite{ATLAS:2013rma,Chatrchyan:2013vaa}. Nevertheless, the $h\to
Z\gamma$ channel, likewise the $h\to \gamma\gamma$ decay, is worth to
analyze since it is particularly sensitive to new colorless
electrically-charged particles which do not change the Higgs
production. At the best of our knowledge, in the TMSSM the
$R_{Z\gamma}$ signal strength has never been calculated.

Similarly to the case of $R_{\gamma\gamma}$, for our
setting~\eqref{set1} and~\eqref{set2} only charginos can move
$\Gamma(h\to Z\gamma)$ from its SM value. This leads to
\be
R_{Z\gamma}=\left| 1+\frac{
    {A}^{Z\gamma}_{\widetilde\chi^\pm_{1,2,3}}}{A^{Z\gamma}_W+A^{Z\gamma}_t}
\right|^2~.
\label{eq:Rzg}
\ee
The contributions $A^{Z\gamma}_W$ and $A^{Z\gamma}_t$ have been first
obtained in refs.~\cite{Cahn:1978nz,Bergstrom:1985hp}. They can be
expressed in term of Passarino-Veltman three-point
functions and turn out to be $A^{Z\gamma}_W=-12$ and $A^{Z\gamma}_t=0.6$ for
$m_h\simeq 126\,$GeV~\cite{Djouadi:1996yq}.

In the TMSSM the chargino contribution comes from triangular loops
where all three chargino mass-eigenstates run in and can be flipped
from one to another at the vertices (both
clockwise and anti-clockwise helicity directions must be taken into
account). No flipping however occurs at the vertex involving the
photon. For this reason only up to two chargino mass-eigenstates run
inside a given loop and each diagram involves a loop integration
that is formally similar to those arising in the MSSM
calculation (where only two charginos exist). Consequently the study of the $\Gamma(h\to
Z\gamma)$ is a straightforward generalization to three charginos of the MSSM expression given in ref.~\cite{Djouadi:1996yq}.

In the view of the above considerations, we can generalize the
procedure of ref.~\cite{Djouadi:1996yq} and we obtain
\be
\label{AZg}
A^{Z\gamma}_{\widetilde\chi^\pm_{1,2,3}}= 
\sum_{j,k=1}^3 
\frac{g_2\, m_{\widetilde\chi^\pm_j}}{g_1\, m_Z}  
~f\!\left(m_{\widetilde\chi^\pm_j},m_{\widetilde\chi^\pm_k},m_{\widetilde\chi^\pm_k}\right)
(g_{h\widetilde\chi^+_j \widetilde\chi^-_i}^L+g_{h\widetilde\chi^+_j \widetilde\chi^-_i}^R) ( g_{Z\widetilde\chi^+_j \widetilde\chi^-_i}^L +  g_{Z\widetilde\chi^+_j \widetilde\chi^-_i}^R)
~,
\ee
in which: $f$ is a linear combination of Passarino-Veltman functions
defined in ref.~\cite{Djouadi:1996yq}; $m_{\widetilde\chi^\pm_j}$ are
pole masses; $g_{h\widetilde\chi^+_j \widetilde\chi^-_i}^{L}$ and
$g_{h\widetilde\chi^+_j \widetilde\chi^-_i}^{R}$ are provided in
eqs.~\eqref{g_hchachaL} and~\eqref{g_hchachaR};
$g_{Z\widetilde\chi^+_j \widetilde\chi^-_i}^L$ and
$g_{Z\widetilde\chi^+_j \widetilde\chi^-_i}^R$ are given by
\bea
g_{Z\chi^{+}_i\chi^{-}_j}^{R} &=& - \left(V_{i1} V_{j1}^*+ \frac{1}{2} V_{i2} 
V_{j2}^* + V_{i3} V_{j3}^*- \delta_{ij} s_W^2 \right)~,  \\ 
g_{Z\chi^{+}_i\chi^{-}_j}^{L} &=& - \left( U_{i1} U_{j1}^* + \frac{1}{2} U_{i2} 
U_{j2}^* + U_{i3} U_{j3}^* - \delta_{ij} s_W^2 \right)~. 
\eea

%%%%%%%%%%%%%%%%%%%%%%%%%%%%%%%%%%%%%%%%%%%
\section{Numerical analysis setup}\label{sec:num}

The TMSSM involves several free parameters. Some of them have to be
fixed for practical purposes but play no role in our analysis. This is
the case for the whole slepton sector whose masses are assumed above
the TeV scale not to interfere with the chargino and neutralino
phenomenology we analyze. Other parameters have a minor impact, and
their choice given in eq.~\eqref{set1} is motivated in
section~\ref{sec:features}.  Some have to satisfy the EWSB conditions
(as explained in section~\ref{sec:Higgs}), and finally only the
followings are still undetermined:
\begin{table}[t]
  \caption{Nested Sampling (NS) parameters and their prior ranges. The priors are flat over the indicated range.\label{tab:priors}}
\begin{center}
\begin{tabular}{|c|c|}
\hline
 NS parameters & Prior range\\
\hline
 $ \log_{10}(M_1/{\rm GeV}),\log_{10}(\mu_\Sigma/{\rm GeV})$  & $1 \to 3$ \\
 $ \log_{10}(\mu/{\rm GeV}), \log_{10}(M_2/{\rm GeV})$ & $2 \to 3$  \\ 
 $\widetilde{m}/\rm TeV$ & $0.63 \to  2$ \\
 $\log_{10}(\tan\beta)$ & $0 \to 1$\\
 $\lambda$ &  $0.5 \to  1.2$\\
\hline
\end{tabular}
\end{center}
\end{table}
\begin{equation}
  \{ \theta_i \}  = \{M_1, M_2, \mu, \mu_\Sigma, \widetilde{m}, \tan\beta, \lambda\} \,.
\label{eq:param}
\end{equation}
To accomplish an efficient sampling on these seven parameters, we
adopt an approach based on Bayes' theorem
\begin{equation}
p(\theta_i | d) \propto \mathcal{L}(d|\theta_i) \pi(\theta_i)\,,
\end{equation}
where $d$ are the data under consideration, $\mathcal{L}(d|\theta_i)$
is the likelihood function and
$p(\theta_i | d)$ is the posterior Probability Distribution Function
(PDF). The function $\pi(\theta_i)$ is the prior PDF, it is
independent of data and describes our belief on the values of the
theoretical parameter, before the confrontation with experimental
results.

All priors $\pi(\theta_i)$ used in the analysis and their ranges of
variation are summarized in table~\ref{tab:priors}.  A flat prior is
assumed for the stop parameter $\widetilde m$, with an upper bound at
2 TeV in order not to introduce a large electroweak fine-tuning. For
gaugino, Higgsino and Triplino masses, we instead consider logarithm
priors and values below the TeV scale. Such a choice is aimed to
improve the statistics in the parameter space where charginos tend to
be close to their LEP mass bound, may enhance $R_{\gamma\gamma}$ and
$R_{Z\gamma}$, and may open the window of the lightest neutralino DM
particle. For the same purpose, and in order not to barely reproduce a
MSSM-like phenomenology, we also impose $\tan\beta$ smaller than 10. A
similar reasoning applies for the chosen range for $\lambda$.  Within
such values we expect to fully cover the TMSSM parameter region
where the little hierarchy problem is alleviated (with respect to the
MSSM) and perturbation theory does not break down before the GUT
scale~\cite{antonio1}. On the other hand, we neither exclude a priori
scenarios with Landau poles at the PeV scale because these may be
avoided in ultraviolet completions of the TMSSM.

The likelihood function is the conditional probability of the data
given the theoretical parameters.  The data $d$ used in
$\mathcal{L}(d|\theta_i) $, which are the observables and constraints
summarized in table~\ref{tab:co}, are as follows.

{\it \underline{Collider data}:} We require the lightest CP-even Higgs
mass $m_h$ to be compatible with the ATLAS and CMS
measurements~\cite{ATLAS:2013sla,CMS:yva}, which we (indicatively)
combine by a statistical mean.  Its uncertainty is dominated by the
theoretical error, which is estimated to be around
3\,GeV~\cite{Allanach:2004rh}. We also assume chargino and stop masses
that fulfill the bounds
$m_{\widetilde{\chi}^\pm_1}>101\,$ GeV~\cite{PDG} and $m_{\widetilde
  t_1}>650\,$ GeV~\cite{SUS-013-011}. Finally, we require the invisible
decay width of the $Z$ boson to be smaller than
2\,MeV~\cite{ALEPH:2005ab}.

{\it \underline{DM data}:} We impose the lightest neutralino relic
abundance to match $\Omega_{\rm DM} h^2$ measured by
Plank~\cite{Ade:2013zuv}, as we are interested only in
single-component DM.  Notice that the experimental error on this observable has become incredibly smaller 
than the theoretical one, hence we consider an additional 20\% of theoretical uncertainty~\cite{Boudjema:2011ig}. Furthermore, we enforce the neutralino SI
cross-section off nuclei, $\sigma_n^{SI}$, to be compatible with the
LUX direct detection exclusion bound~\cite{Akerib:2013tjd} at 90\%
CL. For the theoretical prediction of the SI cross-section mediated
by the Higgs boson, we do not introduce uncertainties related to the
strange quark content of the nucleon: we fix the ratio of nucleon mass
and strange quark mass to be $f_s = 0.053$ MeV, accordingly to
ref.~\cite{Junnarkar:2013ac} (for effects due to different choices of
$f_s$ and similar quantities such as $\sigma_{\pi n}$ see
e.g. refs.~\cite{Koch:1982pu,Gasser:1990ap,Bottino:1999ei,Pavan:2001wz,deAustri:2013saa}).

For either the relic density and the Higgs mass we use a Gaussian
likelihood function whose peak corresponds to the measured central
value and whose width reproduces the standard deviation of the
measurement (explicit quantities are quoted in table~\ref{tab:co}).
For the $\sigma_n^{SI}$ constraint we instead implement a Heaviside
likelihood function.  The DM constraints are implemented in the
likelihood function $\mathcal{L}_{\rm DM}(d|\theta_i)$, the collider
constraints are implemented in the likelihood function
$\mathcal{L}_{\rm Coll}(d|\theta_i)$ and the full likelihood is simply
the product of every individual likelihood associated to an
experimental result. Finally, the above stop and chargino mass limits as well as the constraint on the
  $Z$-boson invisible width are
absorbed into the prior PDFs: each parameter point generating a TMSSM
mass spectrum that violates these bounds is discarded.

\begin{table}[t!]
\caption{Summary of the observables and constraints used in this analysis.  \label{tab:co}}
\begin{center}
\begin{tabular}{| c | c | c | c| }
  \hline
Type &  Observable & Measurement/Limit  & Ref.\\
  \hline
{\it \underline{Collider data}} &  $m_h$  & $ 125.85 \pm 0.4$ GeV (exp) $ \pm 3 $ GeV (theo) & ~\cite{ATLAS:2013sla,CMS:yva}\\
 & $\Gamma(Z\to\widetilde{\chi}_1^0\widetilde{\chi}_1^0)$  & $ <2 $ MeV  & ~\cite{PDG} \\
 & $m_{\widetilde{t}_1}$ & $ > 650$ GeV (LHC 90\% CL) & ~\cite{SUS-013-011} \\
 & $m_{\widetilde{\chi}_1^+}$  &  $>  101$ GeV (LEP 95\% CL) & ~\cite{PDG}\\[1mm]
{\it \underline{ DM data}} &  $\Omega_{\rm DM} h^2$  & $ 0.1186 \pm 0.0031 $  (exp) $ \pm 20\%$  (theo) & ~\cite{Ade:2013zuv}  \\
&   $ \sigma_{\rm Xe}^{SI}  $  &  LUX (90\% CL)  & ~\cite{Akerib:2013tjd}\\
  \hline
\end{tabular}
\end{center}
\end{table}

For some given values of the theoretical inputs $\theta_i$ the
collider and DM observables are computed by means of some public
codes. We briefly summarize the programming procedure.  We employ
\texttt{SARAH-3.3.0} and \texttt{SPheno-3.2.4} to calculate the TMSSM mass
spectrum (where radiative corrections are taken into account as
described in section~\ref{sec:Higgs}). Also
$\Gamma(Z\to\widetilde{\chi}_1^0\widetilde{\chi}_1^0)$, BR$(h\to\widetilde{\chi}_1^0\widetilde{\chi}_1^0)$,
$R_{\gamma\gamma}$ and $R_{Z\gamma}$ are determined by dint of
\texttt{SPheno-3.2.4} along the lines of section~\ref{sec:higgssign} (for
$R_{Z\gamma}$ we also use the Passarino-Veltman functions that are
implemented in the \texttt{CPsuperH2.3} libraries~\cite{Lee:2012wa}).
Afterward, the \texttt{SPheno-3.2.4} output is elaborated by
\texttt{micrOMEGAs\_2.4.5}~\cite{Belanger:2013oya}. In this way we compute
the DM observables listed in table~\ref{tab:co}.

To explore the parameter space we link \texttt{SPheno} and
\texttt{micrOMEGAs\_2.4.5} to the nested sampling algorithm
\texttt{MultiNest\_v3.2}~\cite{Feroz:2008xx} (with specifications of 4000
live points and tolerance parameter set to 0.5). This algorithm
produces the posterior samples from distributions with a large number
of parameters and with multi-modal likelihoods more efficiently than
Markov Chain Monte Carlo. At practical level we run two samples: for
analyzing the Higgs phenomenology we use only $\mathcal{L}_{\rm
  Coll}(d|\theta_i)$ (sample 1), whereas
when exploring the DM constraints as well, we use the full likelihood
(sample 2). \texttt{MultiNest\_v3.2} might however populate with an
insufficient number of points regions where the likelihood is
flat. This is relevant for the $R_{\gamma\gamma}$ and $R_{Z\gamma}$
observables, as we do not impose constraints on their values in the
likelihood function. To address this issue we run two additional
samples with $\mathcal{L}(d|\theta_i)_3 = \mathcal{L}_{\rm
  Coll}(d|\theta_i)\times\mathcal{L}_{\gamma\gamma}(d|\theta_i)$ and
$\mathcal{L}(d|\theta_i)_4 = \mathcal{L}_{\rm
  Coll}(d|\theta_i)\times\mathcal{L}_{\rm DM}(d|\theta_i)
\times\mathcal{L}_{\gamma\gamma}(d|\theta_i)$ (for the case without and
with the DM constraints respectively, sample 3 and sample 4). These
two likelihood functions include a {\it fake} information associated
to an extra Gaussian likelihood function
$\mathcal{L}_{\gamma\gamma}(d|\theta_i)$ with $R_{\gamma\gamma}=1.6\pm
0.2$ to ensure a efficient exploration of region with large 
$h\to \gamma\gamma$ and $h\to \gamma Z$ signal strengths~\footnote{We
  check that the upper bounds on $R_{\gamma\gamma}$ and $R_{Z\gamma}$
  we will obtain do not change by requiring large $R_{Z\gamma}$
  instead of high $R_{\gamma\gamma}$.}. We do not provide a
statistical analysis of the samples but show the result for points
drawn randomly from the posterior PDF, which are provided in the
\texttt{$\ast$post\_equal\_weight.dat} file constructed by
\texttt{MultiNest\_v3.2}, hence we can safely combine the samples
originating from different run with different likelihood functions.
Before discussing our findings, let us mention some experimental
bounds that we do not enforce in the sampling phase.

Different bounds on BR$(h\to\widetilde{\chi}_1^0\widetilde{\chi}_1^0)$
exist in the
literature~\cite{Aad:2014iia,CMS:2013yda,Espinosa:2012vu,Belanger:2013kya,Giardino:2013bma,
  Ellis:2013lra}. Imposing any of them would make
our results of difficult interpretation if a different bound should be
considered. We thus prefer not imposing any cut on
BR$(h\to\widetilde{\chi}_1^0\widetilde{\chi}_1^0)$ and just presenting
its value in the results we present.

We are also aware of the bounds on the chargino and neutralino
  masses based on simplified models: in scenarios with
  $m_{\widetilde{\chi}^0_1} \lesssim 100$\,GeV and
  $m_{\widetilde{\chi}_1^\pm}\simeq m_{\widetilde{\chi}_2^0}$, data
  analyses impose $m_{\widetilde{\chi}_1^\pm}\gtrsim 350$\,GeV at 95\%
  CL if ${\widetilde{\chi}_2^0}$ decays 100\% into $Z$ boson, or
  $m_{\widetilde{\chi}_1^\pm}\gtrsim 170$\,GeV if
  ${\widetilde{\chi}_2^0}$ decays 100\% into $h$
  boson~\cite{Aad:2014nua,Khachatryan:2014qwa}. However, due to 
  its not straightfoward intepretation in the generic TMSSM parameter space, we do not
  impose such constraints. Instead, in the post processing phase of
  the samples we verify that these bounds do not apply in the
  interesting ballpark of our analysis. Indeed, in particular when we
  achieve a relevant diphoton enhancement, the fields
  $\widetilde{\chi}^0_2$ and $\widetilde{\chi}^\pm_2$ are mixed states
  and consequently {\it (i)} their masses are not degenerate and {\it
    (ii)} the neutralino decay channels into $Z$ and $h$ bosons can
  compete especially due the Triplino component and its potentially
  sizeable coupling $\lambda$.

Finally, we also consider the $B_s \to \mu^+ \mu^-$ and $B \to X_s
\gamma$ observables and the neutralino Spin-Dependent (SD)
cross-section off protons and neutrons. We use \texttt{SPheno-3.2.4} and
\texttt{micrOMEGAs\_2.4.5} respectively to calculate them.  As expected in
scenarios with low $\tan \beta$, these $B$-meson signatures are in
full agreement with experiments~\cite{Aaij:2012nna,Amhis:2012bh}. {We
  will compare the results for SD cross-section with COUPP and
  XENON100 limits~\cite{Behnke:2012ys,Aprile:2013doa} on proton and
  neutron respectively and comment them in section~\ref{sec:dm}.

%%%%%%%%%%%%%%%%%%%%%%%%%%%%%%%%%%%%%%%%%%%

\section{$R_{Z\gamma}$ and $R_{\gamma\gamma}$ without DM contraints}\label{sec:resh}

\begin{figure}[t]
\begin{minipage}[t]{0.49\textwidth}
\centering
\includegraphics[width=1.\columnwidth,trim=0mm 0mm 0mm 0mm, clip]{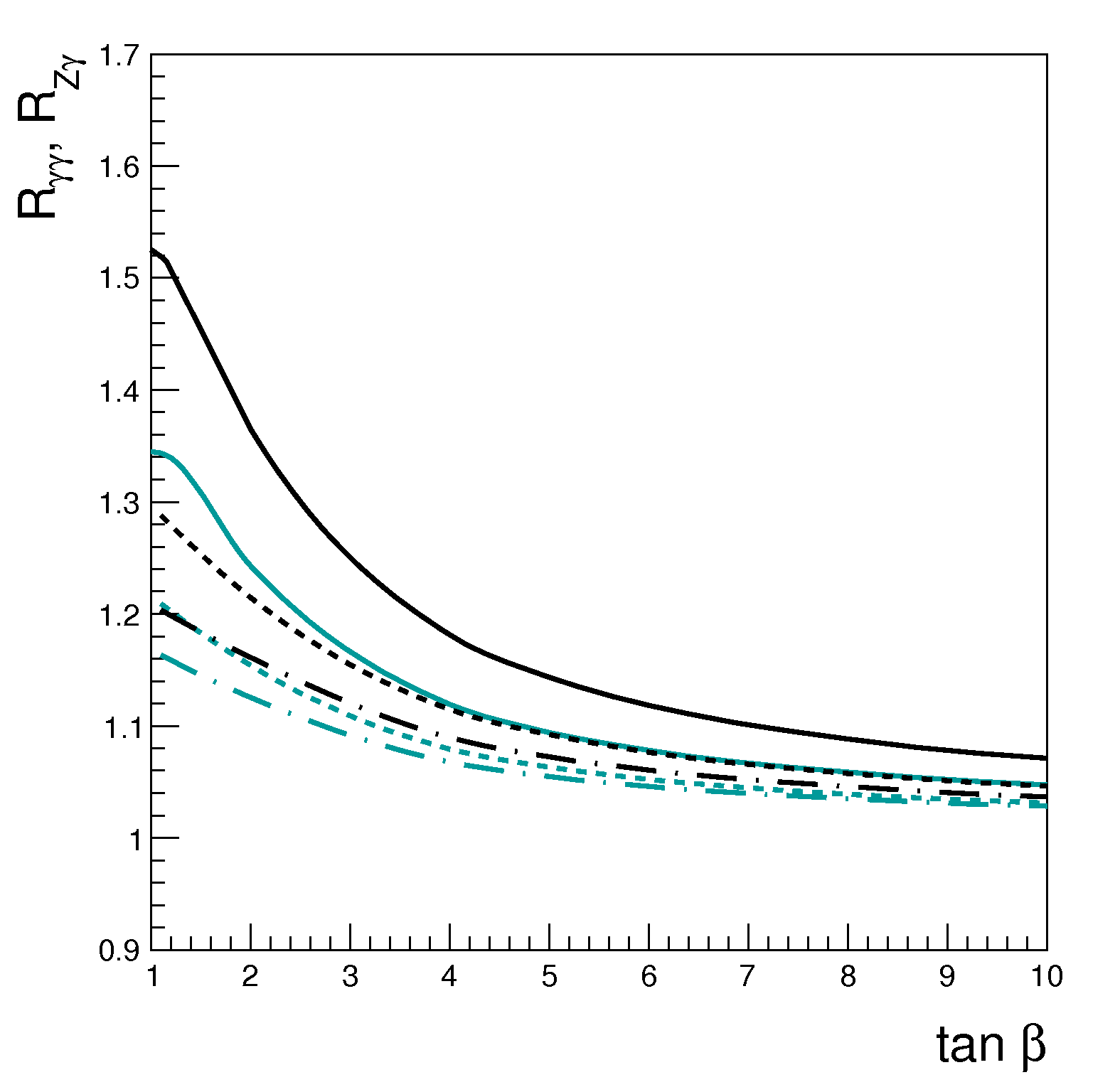}
\end{minipage}
\hspace*{0.2cm}
\begin{minipage}[t]{0.49\textwidth}
\includegraphics[width=1.\columnwidth,trim=0mm 0mm 0mm 0mm, clip]{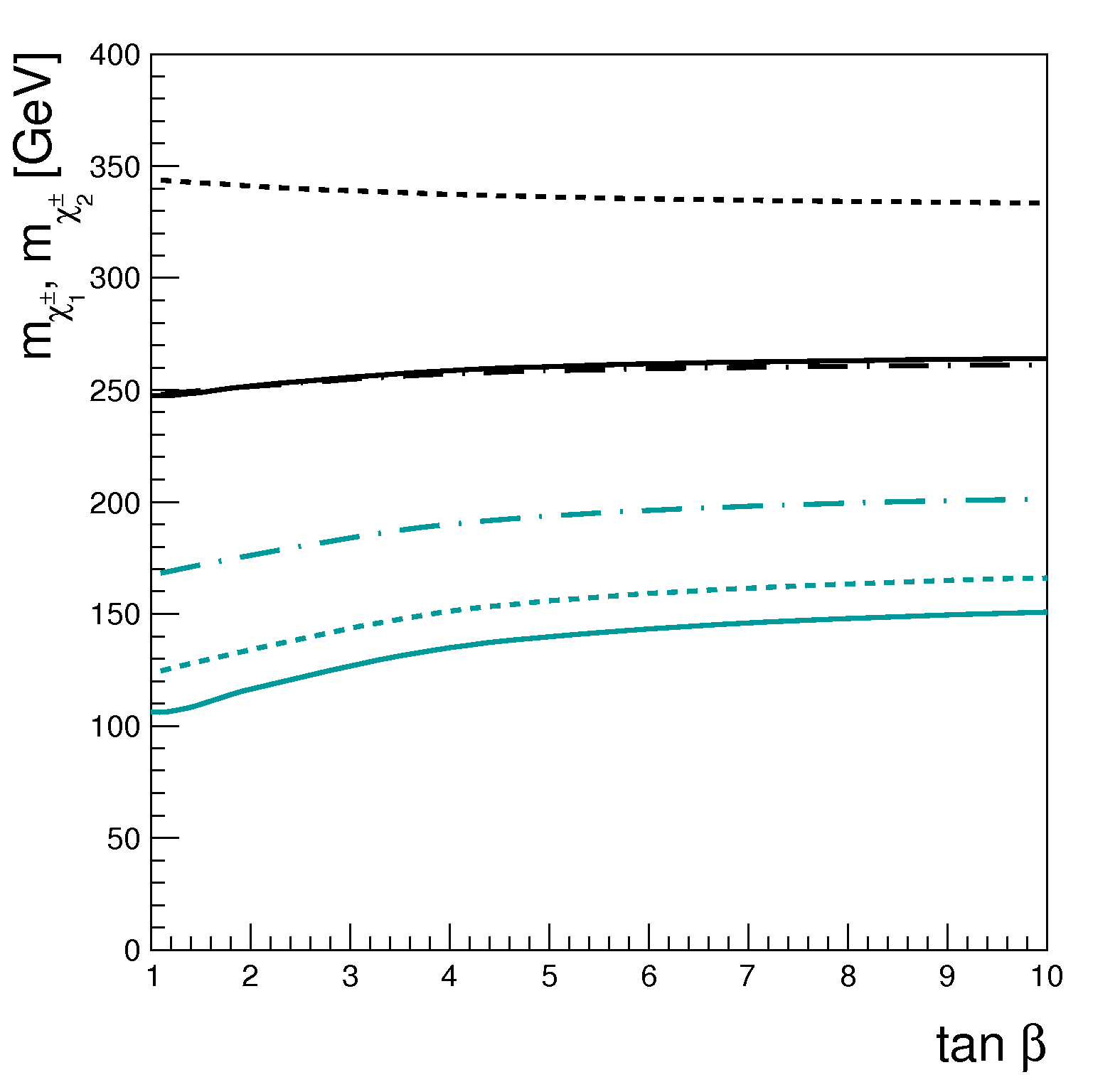}
\end{minipage}
\\
\begin{minipage}[t]{0.49\textwidth}
\centering
\includegraphics[width=1.\columnwidth,trim=0mm 0mm 0mm 0mm, clip]{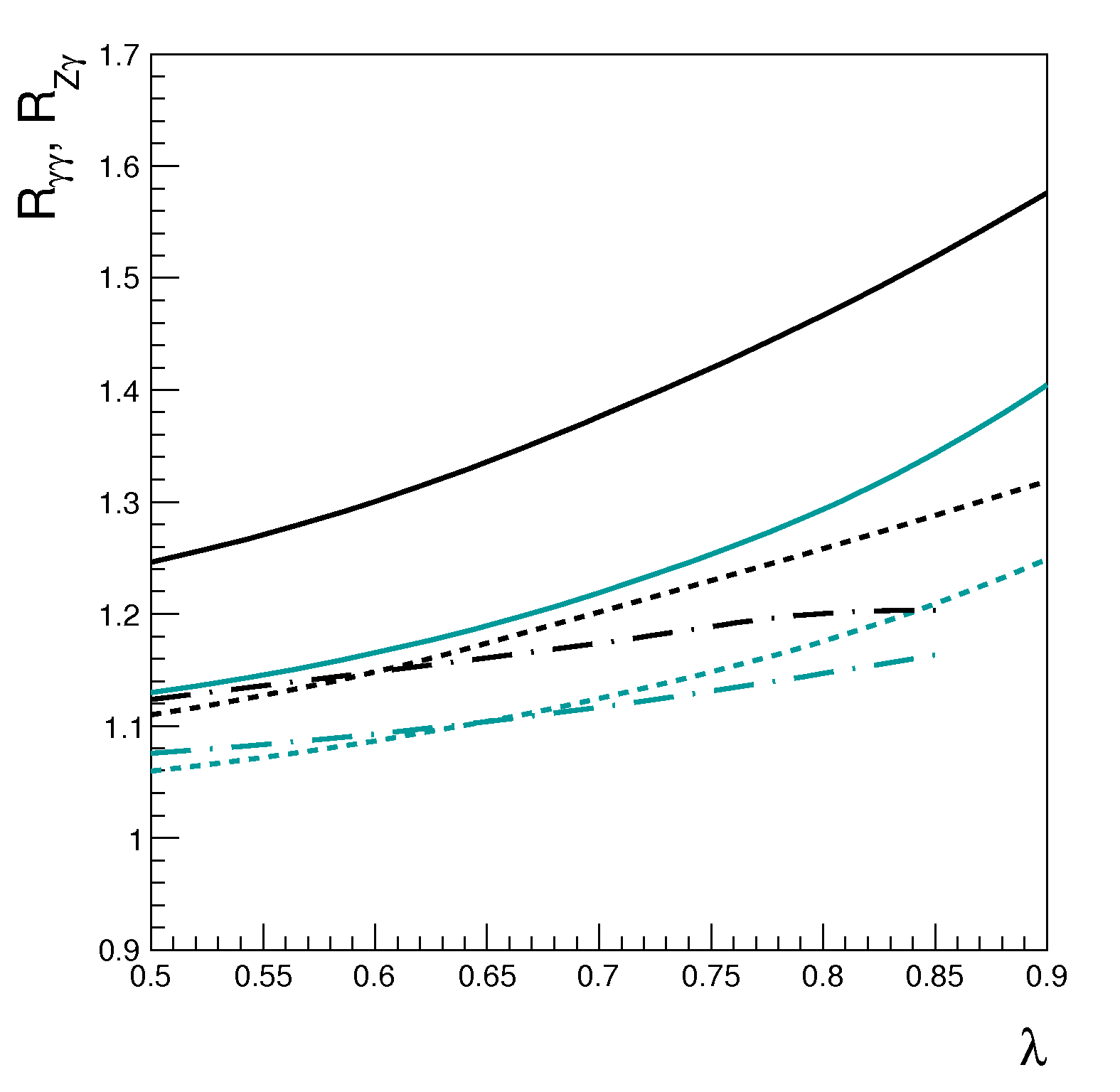}
\end{minipage}
\hspace*{0.2cm}
\begin{minipage}[t]{0.49\textwidth}
\includegraphics[width=1.\columnwidth,trim=0mm 0mm 0mm 0mm, clip]{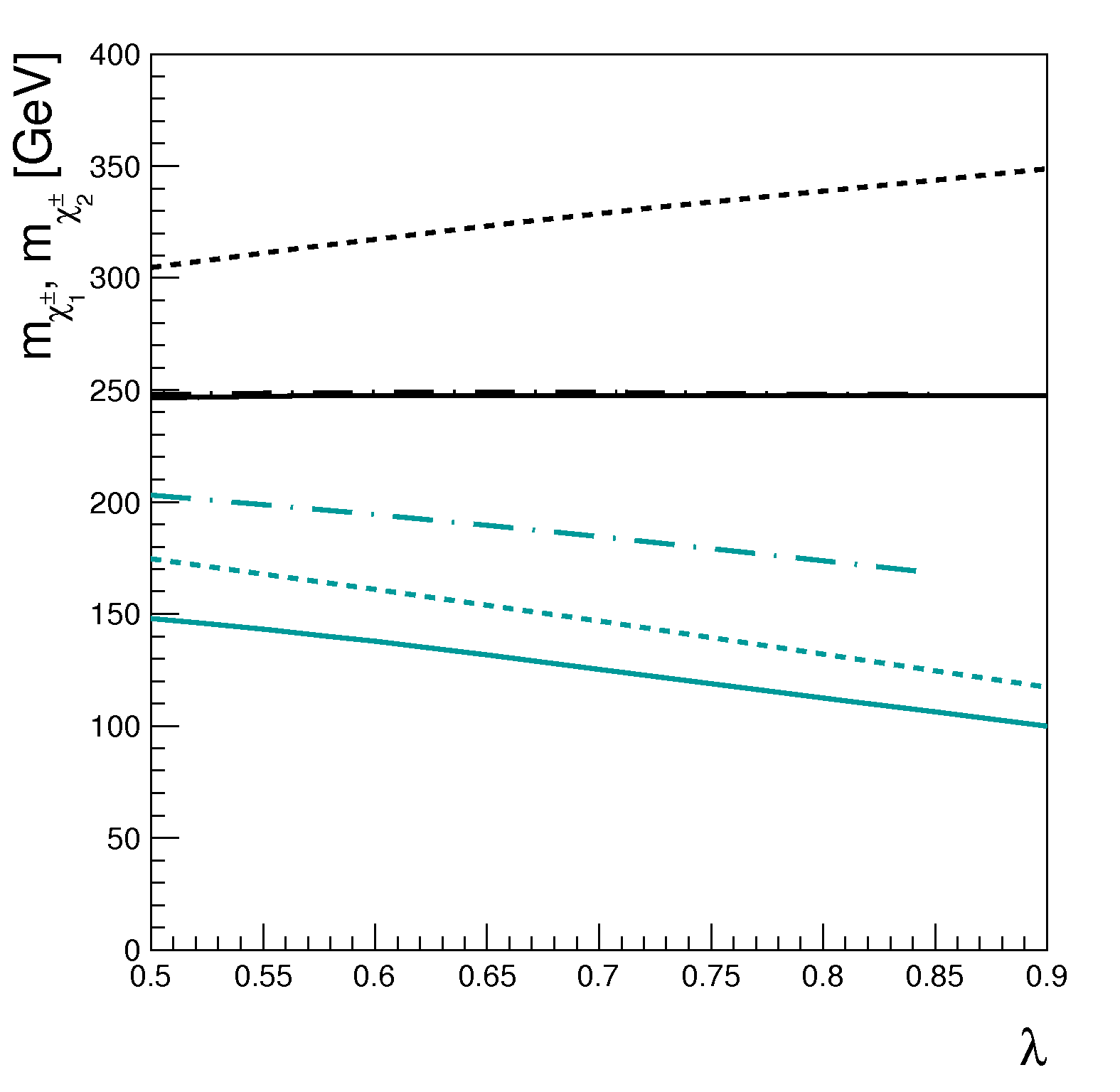}
\end{minipage}
\caption{{\it Top left}: Analytic behavior of $R_{\gamma \gamma}$
  (black) and $R_{Z\gamma}$ (turquoise) as a function of $\tan\beta$ for $\lambda=0.85$. The
  solid lines are for $\mu=\mu_\Sigma=M_2=230\,$GeV (scenario A), the
  dashed lines stand for $\mu=\mu_\Sigma=230\,{\rm GeV},M_2=1\,$TeV
  (scenario B) and the dot-dash lines for $\mu_\Sigma=M_2=230\,{\rm
    GeV}, \mu = 400\,$GeV (scenario C). {\it Top right:} Dependence on
  $\tan\beta$ of the lightest (turquoise) and next to lightest
  (black) chargino masses, which contribute to the $R_{\gamma \gamma}$
  and $R_{Z\gamma}$ shown in the left panel (the solid/dashing code is
  as in the left panel). {\it Bottom:} Same as above as a function of
  $\lambda$ for $\tan\beta=1.1$}.
\label{fig:analytnoDM}
\end{figure}

Within the TMSSM, the possibility of achieving sizeable enhancements
in $R_{\gamma\gamma}$ has been previously highlighted in
refs.~\cite{antonio1, antonio2}. These analyses were performed by
considering tree-level chargino masses and low-energy limit
approximations. They were moreover carried out for some illustrative
parameter regions.  In this section we extend the previous analysis
focused on the $m_A\gg m_h$ regime~\cite{antonio1}. In particular, we
explore a broader parameter space (but still keeping large $m_A$) and
we include the radiative effects discussed in section~\ref{sec:hgg}.
We also present our findings for $R_{Z\gamma}$ in
eq.~\eqref{eq:Rzg}.

Before reporting the result of the full parameter
sampling, it is educative to understand the role of some inputs.  The
essential parameter dependence of $R_{\gamma\gamma}$ and $R_{Z\gamma}$
is shown in the left panels of figure~\ref{fig:analytnoDM}. In the
figure we assume the setting in eq.~\eqref{set1}, as well as
$\lambda=0.85$ in the upper plot and $\tan\beta=1.1$ in the lower
one. At each point the stop parameter $\widetilde m$ is adjusted to
obtain $m_h=126\,$GeV. The signal strengths $R_{\gamma\gamma}$ and
$R_{Z\gamma}$ (black and turquoise lines, respectively) are calculated
for three chargino mass settings: $\mu=\mu_\Sigma=M_2=230\,$GeV (solid
curves; scenario A), $\mu=\mu_\Sigma=230\,{\rm GeV}, M_2=1\,$TeV
(dashed curves; scenario B) and $\mu_\Sigma=M_2=230\,{\rm GeV}, \mu =
400\,$GeV (dotted-dashed curves; scenario C). The corresponding
chargino masses $m_{\widetilde{\chi}^\pm_1}$ and
$m_{\widetilde{\chi}^\pm_2}$ are presented in the right panels by
employing the same mark code of the left plots.

For the parameter choice considered in the figure, the enhancement in
$h\to\gamma\gamma$ is always larger than the one in $h\to
Z\gamma$. Moreover, $R_{\gamma\gamma}$ and $R_{Z\gamma}$ are strongly
correlated and a sizeable enhancement in $R_{\gamma\gamma}$ requires a
departure from the SM also in the $h\to {Z\gamma}$ channel. These
behaviors will be confirmed in the results of the full parameter
sampling.

As figure~\ref{fig:samplesnoDM} shows, in each scenario the largest
$R_{\gamma\gamma}$ and $R_{Z\gamma}$ are achieved by reducing
$\tan\beta$ and increasing $\lambda$, which also corresponds to
requiring less tuning in the electroweak sector
(cf.~eq.~\eqref{tree-level}). The
enhancement is mostly due to the decrease of
$m_{\widetilde{\chi}_{1}^\pm}$ and the consequent smaller suppression
of the loop functions in eqs.~\eqref{Rgg} and \eqref{AZg} (cf.~right
panels of the figure; the masses $m_{\widetilde{\chi}_{2,3}^\pm}$ are
large in the three scenarios and hence provide a subleading
effect). However, also the coupling
$g_{h\widetilde{\chi}_1^\pm\widetilde{\chi}_1^\pm}$ plays an important
role.  This can be deduced by comparing $R_{\gamma\gamma}$ (or
$R_{Z\gamma}$) in different scenarios in correspondence to the same
$m_{\widetilde{\chi}_1^\pm}$ value. For instance, for $\lambda=0.85$
both scenario A with $\tan\beta\simeq 10$ and scenario B with
$\tan\beta\simeq 1.1$ have the same chargino mass
$m_{\widetilde{\chi}_{1}^\pm}\simeq 150\,$GeV but quite different
$R_{\gamma\gamma}$. These observations are in agreement with previous
results obtained for $R_{\gamma\gamma}$~\cite{antonio1,Casas:2013pta,
  Batell:2013bka}.

\begin{figure}[t]
\begin{minipage}[t]{0.49\textwidth}
\centering
\includegraphics[width=1.\columnwidth,trim=0mm 0mm 0mm 0mm, clip]{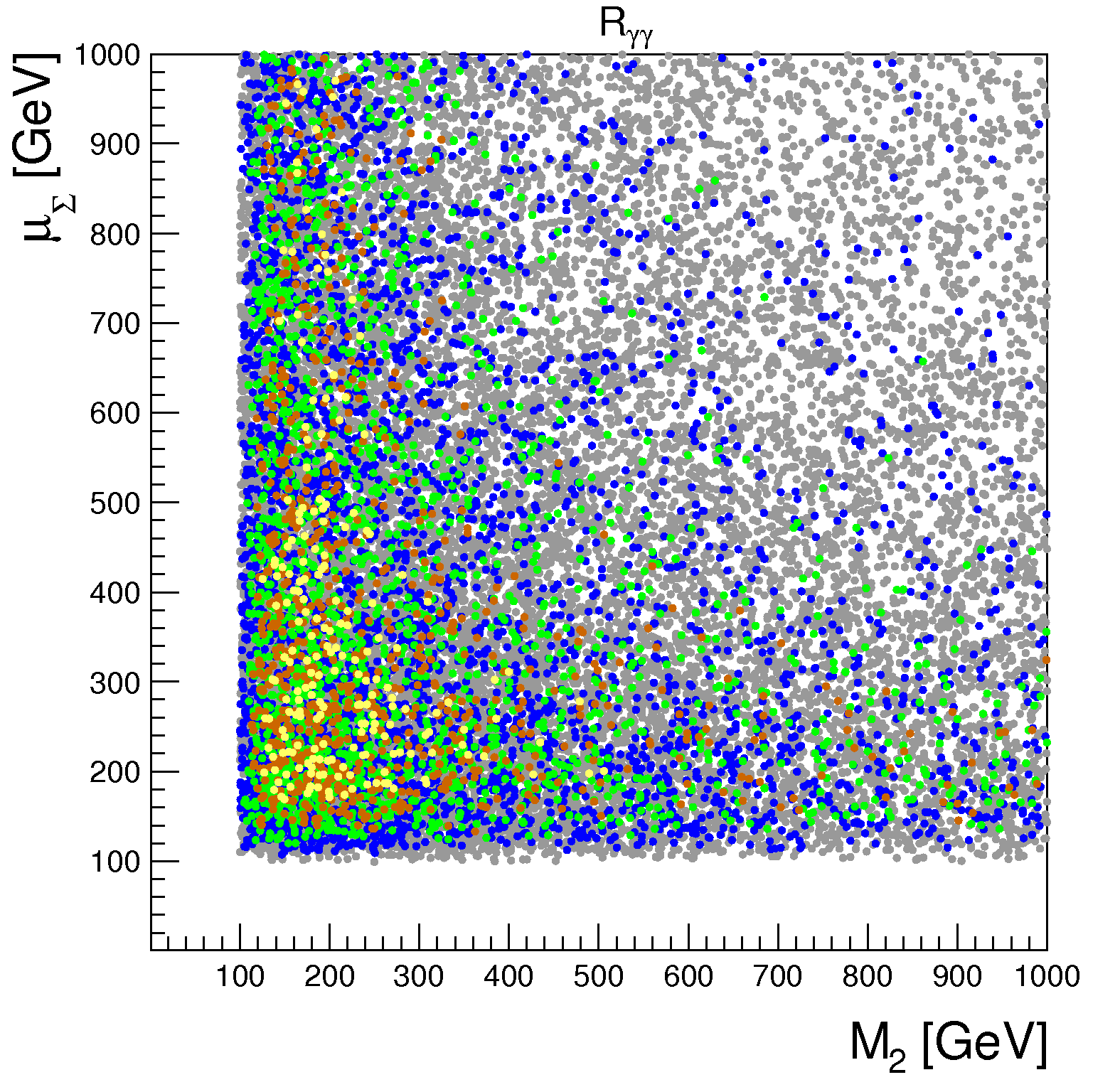}
\end{minipage}
\hspace*{0.2cm}
\begin{minipage}[t]{0.49\textwidth}
\includegraphics[width=1.\columnwidth,trim=0mm 0mm 0mm 0mm, clip]{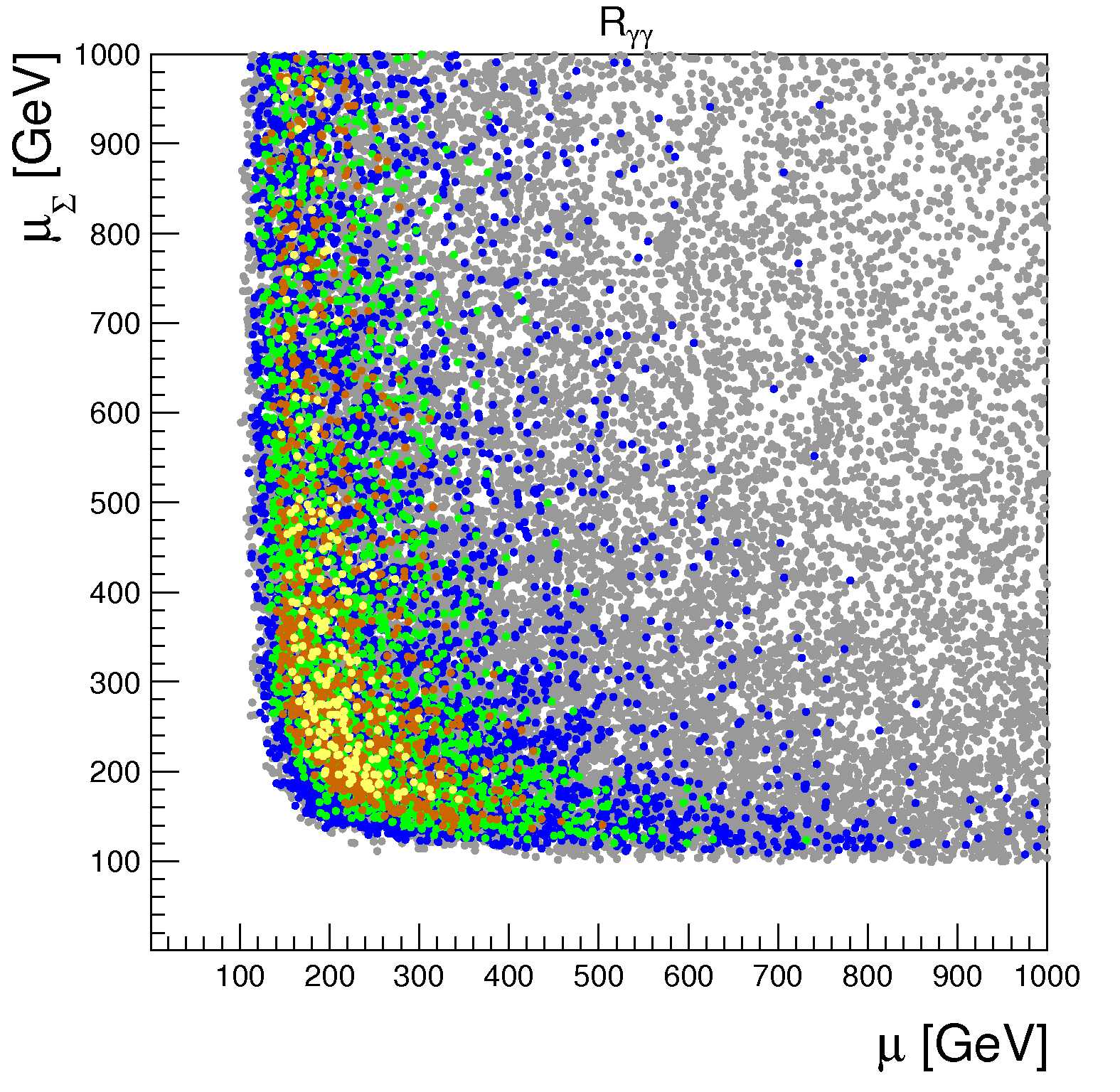}
\end{minipage}
\\
\begin{minipage}[t]{0.49\textwidth}
\centering
\includegraphics[width=1.\columnwidth,trim=0mm 0mm 0mm 0mm, clip]{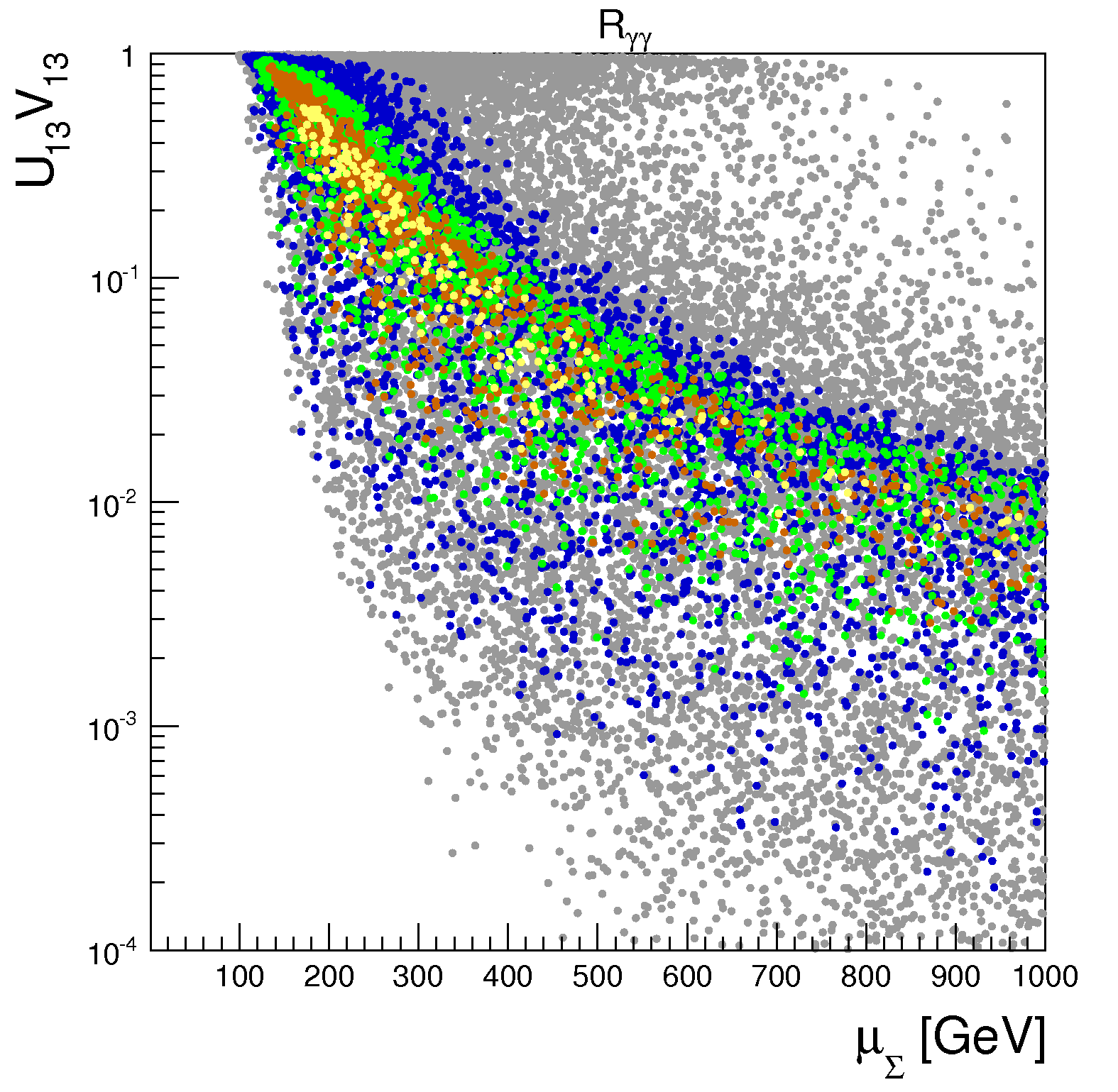}
\end{minipage}
\hspace*{0.2cm}
\begin{minipage}[t]{0.49\textwidth}
\includegraphics[width=1.\columnwidth,trim=0mm 0mm 0mm 0mm, clip]{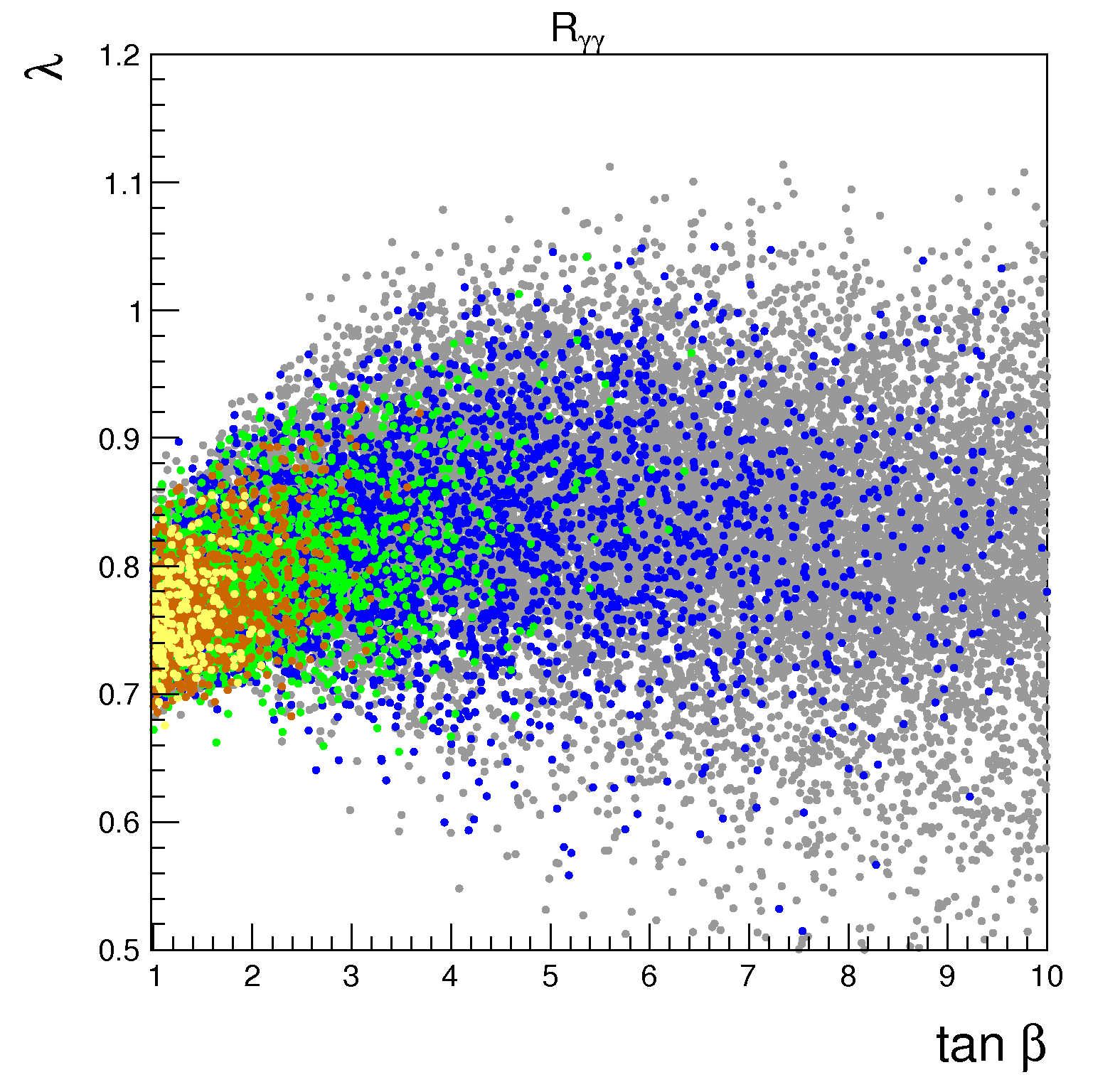}
\end{minipage}
\caption{{\it Top left:} $R_{\gamma \gamma}$ (third direction)
  projected in the $\{\mu_\Sigma - M_2\}$-plane. The values of
  $R_{\gamma \gamma}$ are encoded in the colors:
  $R_{\gamma\gamma}<1.1$ in gray, $1.1\le R_{\gamma\gamma}<1.2$ in blue,
  $1.2\le R_{\gamma\gamma}<1.3$ in green, $1.3\le R_{\gamma\gamma}<1.4$ in
  brown, and $R_{\gamma\gamma}\ge 1.4$ in yellow. {\it Top right:} Same
  as left in the $\{\mu_\Sigma - \mu\}$-plane. {\it Bottom left and
    right:} Same as top left for the Triplino component of the
  lightest chargino as a function of $\mu_\Sigma$ and
  $\{\lambda-\tan\beta\}$-plane respectively.}
\label{fig:samplesnoDM}
\end{figure}

A last remark concerns the parameter range of
figure~\ref{fig:analytnoDM}. We do not enter the regime of
$\tan\beta\simeq 1$ and $\lambda\gtrsim 1$ to achieve larger
enhancements. Besides the reasons previously provided, there is a
further issue that imposes such a restriction: the more the tree-level
Higgs mass is boosted, the smaller the radiative corrections have to
be not to overstep the $m_h\simeq 126$\,GeV constraint.  In
particular, this may require stop masses below the bound of
table~\ref{tab:co}, as it happens in scenario C at $\lambda\gtrsim
0.85$ with $\tan\beta\lesssim 1.1$. Of course, slightly bigger
enhancements would exist by allowing for $m_{\widetilde t}\ll
650\,$GeV and/or $m_\Sigma\ll 5\,$TeV.  However, since prooving the
experimental suitability of such modifications would require specific
collider analyses, and lowering $m_\Sigma$ may also increase the
electroweak fine tuning, we do not further discuss this
possibility.

The two samples obtained with $\mathcal{L}_{\rm Coll}(d|\theta_i)$ are
presented in figure~\ref{fig:samplesnoDM}. The amount of diphoton
enhancement is encoded in the color of the points:
$R_{\gamma\gamma}<1.1$ in gray, $1.1<R_{\gamma\gamma}<1.2$ in blue,
$1.2<R_{\gamma\gamma}<1.3$ in green, $1.3<R_{\gamma\gamma}<1.4$ in
brown, and $R_{\gamma\gamma}> 1.4$ in yellow (this color code will be
maintained in the rest of the paper). The bottom left panel of the
figure proves that the presence of the Triplino is fundamental to
achieve $R_{\gamma\gamma}\gtrsim 1.3$. Indeed, if the Triplino
component of the lightest chargino is negligible, $R_{\gamma\gamma}$
falls to MSSM-like values, i.e.~$R_{\gamma\gamma}\lesssim
1.2$~\cite{Casas:2013pta}~\footnote{This MSSM upper bound on
    $R_{\gamma\gamma}$ is obtained by assuming a mass spectrum similar
    to ours, where only charginos can enhance the diphoton rate. Due to
    different crucial assumptions, we do not compare our results to
    those of refs.~\cite{Carena:2011aa,Batell:2013bka}.}. However, in
  some extreme cases, Triplino effects may be still present even when
  $\mu_\Sigma$ is quite large and the Triplino component of the
  lightest chargino is subdominant (but not negligible), as the case
  $\mu_{\Sigma}\simeq 900\,$GeV and $R_{\gamma\gamma}\gtrsim 1.4$
  shows.

The features of the parameter regions where the yellow points
accumulate can be explained as follows. As observed in
figure~\ref{fig:analytnoDM}, large diphoton enhancements are allowed
for small $\tan\beta$ and rather large $\lambda$ (cf.~bottom right
panel). In such a case, the relation $\lambda>g_2$
arises. Consequently, in the Higgs-chargino-chargino coupling the
contribution proportional to $\lambda$ can push
$g_{h\widetilde{\chi}^\pm_1\widetilde{\chi}^\pm_1}^{L,R}$ above the maximal value obtained in
the MSSM. This of course occurs only if both Triplino and Higgsino
components are unsuppressed.  Large $R_{\gamma\gamma}$ enhancements
then require $\mu$ and $\mu_\Sigma$ at the electroweak scale (cf.~top
right panel). On the other hand, also the MSSM-like contribution of
the Higgs-chargino-chargino coupling can provide an additional boost
to $g_{h\widetilde{\chi}^\pm_1\widetilde{\chi}^\pm_1}^{L,R}$ if the Wino mixing is
sizeable. Therefore, also $M_2$ has to be small to achieve maximal
enhancements (cf.~top left plot). In particular, in order to minimally
suppress the loop function $A_{1/2}(\tau_{\widetilde\chi_i^\pm})$, the
parameters have to be correlated in such a way that
$m_{\widetilde{\chi}_1^\pm}\approx 101\,$GeV.

The effect of the Higgs and stop mass constraints is pointed out in
the bottom right panel of the figure.  The region with small $\lambda$
and small $\tan\beta$ is not populated because the tree-level Higgs
mass~\eqref{tree-level} is very small. In such a case, only large stop
loop corrections to $m_h$ could push it up to 126~GeV, but such
corrections are not allowed due to the $\widetilde m$ range of
table~\ref{tab:priors}. On the contrary, in the upper empty area with
small $\tan\beta$, the tree-level Higgs mass is too large. In this
case stop loop corrections have to be small not to overstep the Higgs
mass constraint, but they are incompatible with the bound
$m_{\widetilde t_1}> 650\,$GeV. Curiously, the mass bound cuts off
most of the parameter space where the TMSSM exhibits a Landau pole at
a scale $Q\lesssim 10^8$\,GeV (see ref.~\cite{antonio1} for estimates
of the Landau pole scale). We stress however that the border of this
empty region could be mildly moved by considering larger values of
$M_2,\mu,$ and $\mu_\Sigma$ than those in
table~\ref{tab:priors}. Heavier charginos would indeed provide bigger
(negative) radiative correction to $m_h$ that should be compensated by
slightly larger stop masses to keep $m_h\approx
126\,$GeV~\cite{Bandyopadhyay:2013lca}.
\begin{figure}
\begin{minipage}[t]{0.49\textwidth}
\centering
\includegraphics[width=1.\columnwidth,trim=0mm 0mm 0mm 0mm, clip]{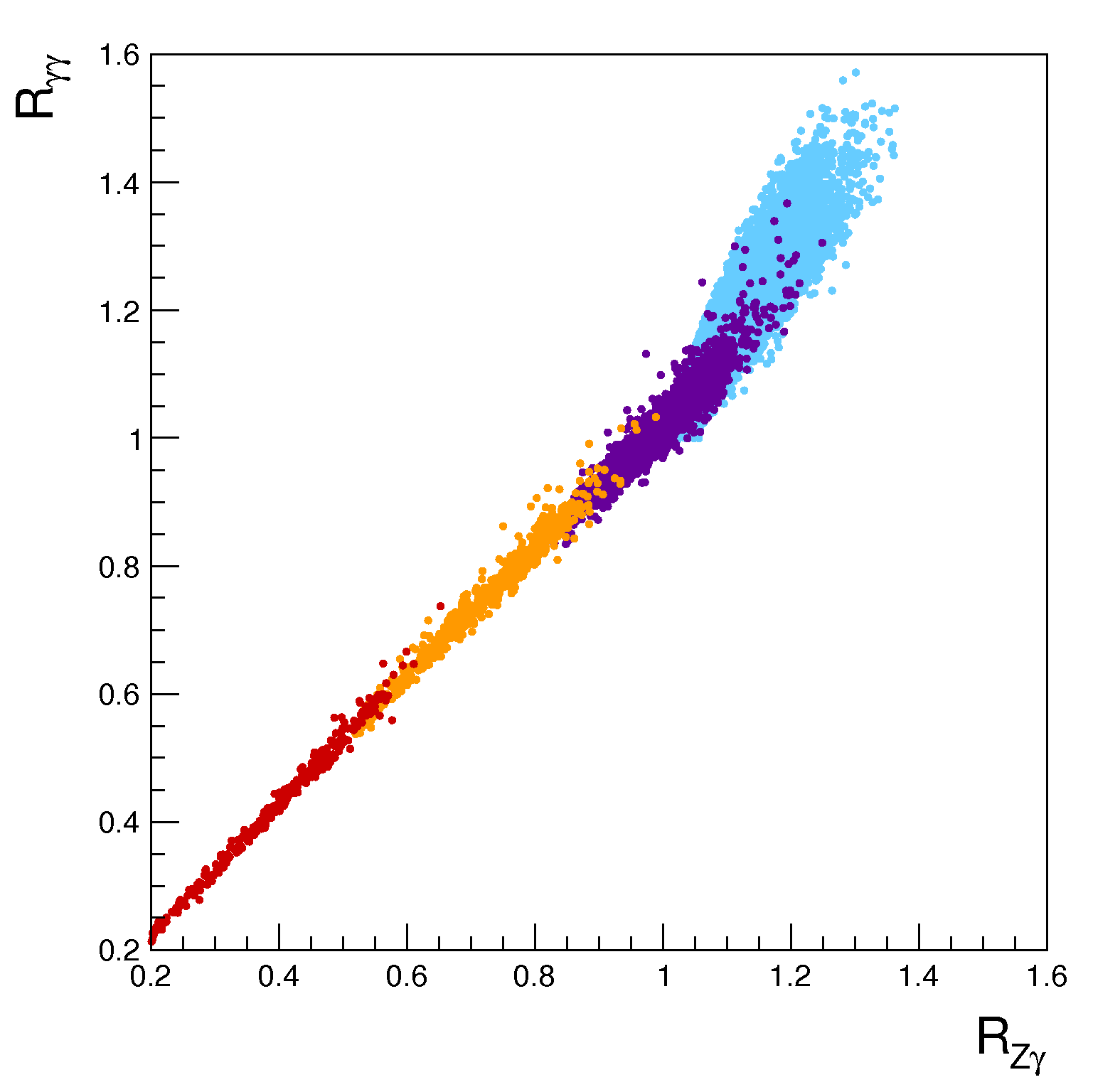}
\end{minipage}
\hspace*{0.2cm}
\begin{minipage}[t]{0.49\textwidth}
\includegraphics[width=1.\columnwidth,trim=0mm 0mm 0mm 0mm, clip]{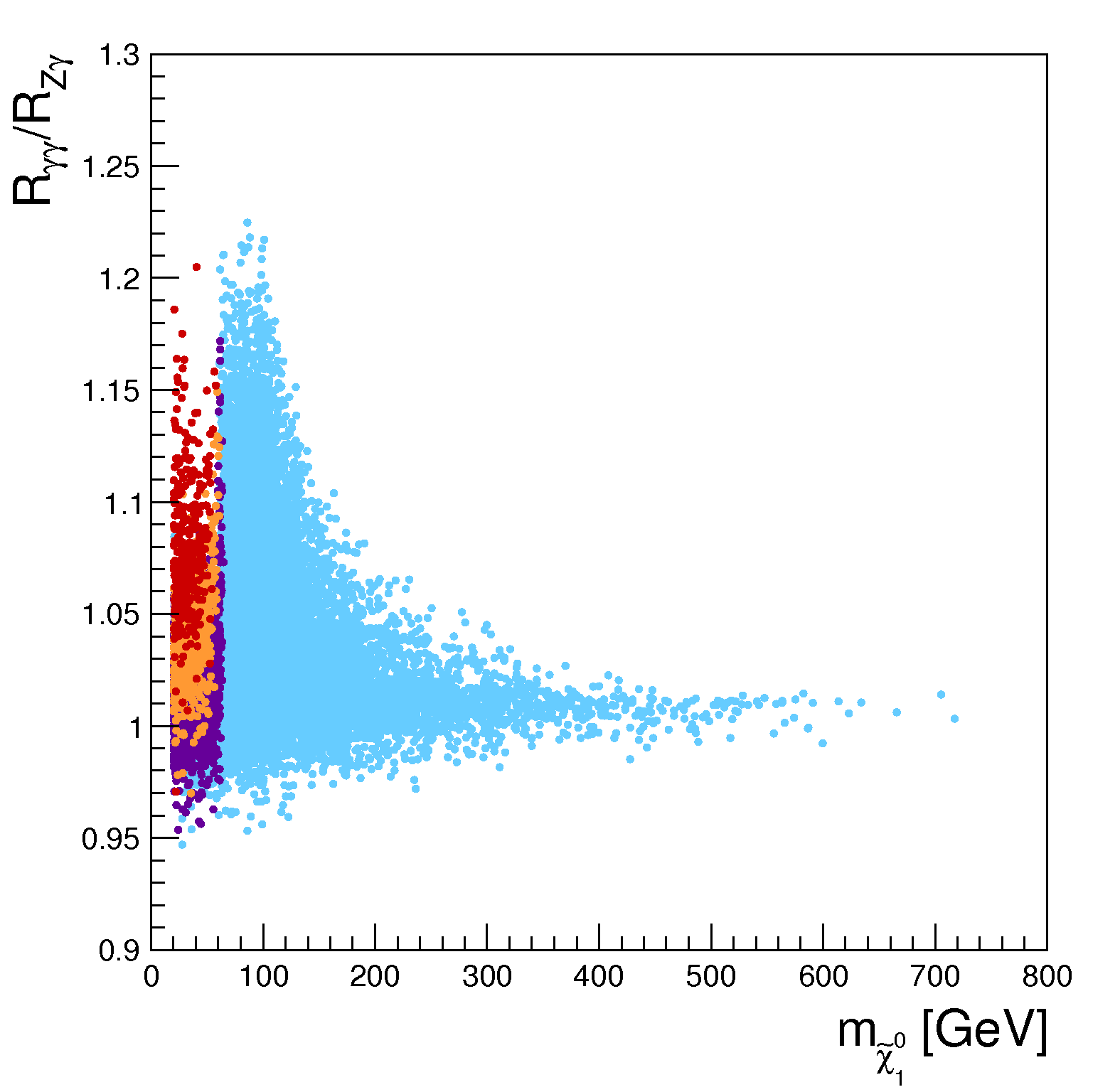}
\end{minipage}
\caption{Correlation between $R_{\gamma\gamma}$ and $R_{Z\gamma}$ for
  the equal weight posterior sample. The color code is: light blue for ${\rm
    BR}(h \to \widetilde{\chi}^0_1\widetilde{\chi}^0_1) < 1\%$, violet
  for $1\%<{\rm BR}(h \to \widetilde{\chi}^0_1\widetilde{\chi}^0_1) <
  20\%$, dark yellow for $20\%<{\rm BR}(h \to
  \widetilde{\chi}^0_1\widetilde{\chi}^0_1) < 50\%$ and red for
  ${\rm BR}(h \to \widetilde{\chi}^0_1\widetilde{\chi}^0_1) > 50\%$.}
\label{fig:gggznoDM}
\end{figure}

Although figure~\ref{fig:samplesnoDM} presents only $R_{\gamma\gamma}$
results, the above considerations apply also to $R_{Z\gamma}$. Indeed,
in the TMSSM $R_{\gamma\gamma}$ and $R_{Z\gamma}$ are tightly
correlated. This is proven in figure~\ref{fig:gggznoDM}, which
displays the values of $R_{\gamma\gamma}$ and $R_{Z\gamma}$ arising in
the samples 1 and 2. Since the sampling also explores the region with
small $M_1$, the $h \to \widetilde{\chi}^0_1\widetilde{\chi}^0_1$
channel can be open. Depending on the value of ${\rm BR}(h \to
\widetilde{\chi}^0_1\widetilde{\chi}^0_1)$, the points in the figure
are colored as follows: ${\rm BR}(h \to
\widetilde{\chi}^0_1\widetilde{\chi}^0_1) < 1\%$ in light blue,
 $1\%\le{\rm BR}(h \to
\widetilde{\chi}^0_1\widetilde{\chi}^0_1) < 20\%$ in violet, $20\%\le{\rm
  BR}(h \to \widetilde{\chi}^0_1\widetilde{\chi}^0_1) < 50\%$ in dark yellow
and ${\rm BR}(h \to \widetilde{\chi}^0_1\widetilde{\chi}^0_1) \ge
50\%$ in red (this color code will be followed in the rest of the
paper).

In the left panel of the figure, the light blue area is not aligned with the
remaining region. We have investigated the origin of this feature and
it seems related to the different kinds of configurations of chargino
parameters in that region. In fact, in the upper-left part
of the light blue area, the typical chargino configuration yields to very
large $g_{h\widetilde{\chi}^0_1\widetilde{\chi}^0_1}^{L,R}$ (for $M_1\lesssim
100$\,GeV). Therefore, only when $m_{\widetilde{\chi}^0_1}$ is tuned
just below the $m_h/2$ threshold, ${\rm BR}(h \to
\widetilde{\chi}^0_1\widetilde{\chi}^0_1)$ is small.  Unless of this
rare accident, ${\rm BR}(h \to \widetilde{\chi}^0_1\widetilde{\chi}^0_1)$ is either
huge or zero. Consequently, for the configurations of chargino
parameters that populate the upper-left part of the light blue area,
$R_{\gamma\gamma}$ and $R_{Z\gamma}$ are typically either larger or
much smaller than one, and an empty region at $R_{\gamma\gamma}\approx
1$ and $R_{Z\gamma}\approx 0.9$ is thus produced.

The opposite effect instead occurs in the lower part of the light blue
region: the typical chargino parameters yield tiny
$g_{h\widetilde{\chi}^0_1\widetilde{\chi}^0_1}^{L,R}$ and hence ${\rm BR}(h \to
\widetilde{\chi}^0_1\widetilde{\chi}^0_1)$ is small for very most of
the values that $M_1$ can assume. For these configurations of chargino
parameters, therefore, $R_{\gamma\gamma}$ and $R_{Z\gamma}$ do not
jump from one value to a very different one at the threshold
$m_{\widetilde{\chi}^0_1}\sim m_h/2$ but slowly change as function of
$M_1$. For this reason the violet region is abundantly populated by such
chargino configurations.

Finally, let us summarize the most striking result of this
section. From our analysis we obtain the following TMSSM bounds (see
figure~\ref{fig:samplesnoDM}):
\be
R_{\gamma\gamma}\lesssim 1.6~,
\qquad R_{Z\gamma}\lesssim 1.4~,
\qquad 0.95\lesssim R_{\gamma\gamma}/R_{Z\gamma}\lesssim 1.2
\hspace{1.5cm} {\rm (no~DM~obs.)}
\ee
and we stress the tight degree of correlation between the two
loop-induced processes.We will comments on the future experimental
  implications of these bounds  in
the conclusions (section~\ref{sec:Concl}).

%%%%%%%%%%%%%%%%%%%%%%%%%%%%%%%%%%%%%%%%%%%
\section{DM phenomenology and constraints on $R_{\gamma\gamma}$ and $R_{Z\gamma}$}
\label{sec:dm}

In this section we present the TMSSM phenomenology in the presence of
a neutralino DM candidate. As previously motivated, we require that no
supersymmetric particle but neutralinos and charginos interferes
during freeze-out, to achieve the correct relic density. To
understand the relevant consequences of introducing the Triplino
component, we first analyze the Wino decoupling limit.
\begin{figure}[t]
\begin{minipage}[t]{0.49\textwidth}
\centering
\includegraphics[width=1.\columnwidth,trim=0mm 0mm 0mm 0mm, clip]{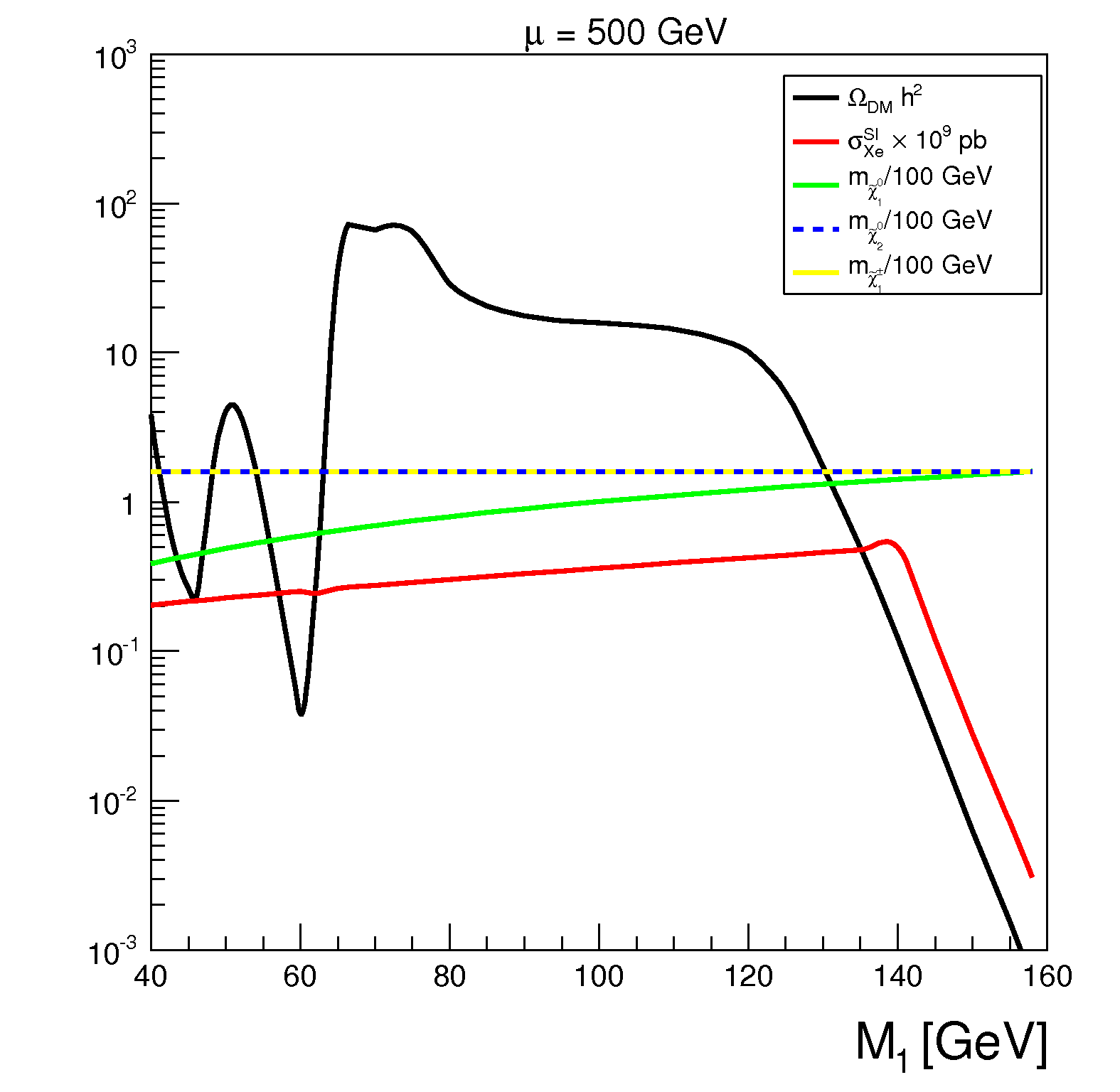}
\end{minipage}
\hspace*{0.2cm}
\begin{minipage}[t]{0.49\textwidth}
\includegraphics[width=1.\columnwidth,trim=0mm 0mm 0mm 0mm, clip]{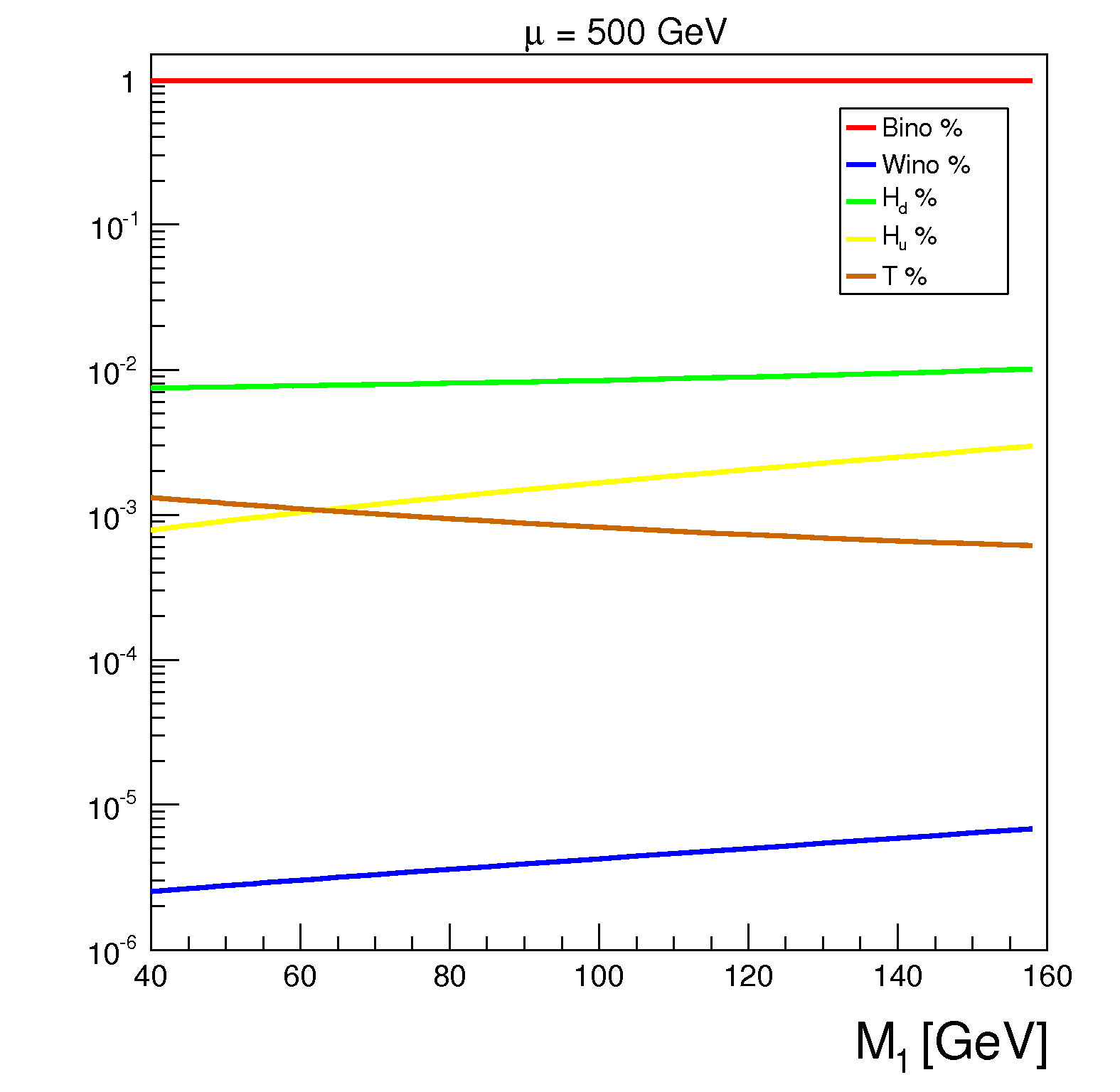}
\end{minipage}
\caption{{\it Left:} Dependence of several physical quantities on
  $M_1$: the black line denotes $\Omega_{\rm DM} h^2$, the red
  $\sigma^{SI}_{Xe} \times 10^{9} \rm pb$, the green the LSP, the blue
  $m_{\widetilde{\chi}^0_2}$, the yellow
  $m_{\widetilde{\chi}^\pm_1}$. The lightest chargino is degenerate
  with the next-to-lightest neutralino. {\it Right:} Component of LSP
  as a function of $M_1$: in red is the Bino fraction, in brown the
  Triplino, in yellow and green the two Higgsino components and in
  blue the Wino fraction (as labelled in the caption). In both panels
  $\mu=500$ GeV, $M_2=1.5$ TeV, $\mu_{\Sigma}=180$ GeV, $\tan \beta =
  2.9$ and $\lambda=0.88$.} \label{fig:proof} 
\end{figure}

In figure~\ref{fig:proof} (left panel) the behaviors of the
lightest-neutralino relic density and SI cross-section (black and red
lines, respectively) for the limit $M_2\gg1\,$TeV are depicted as a
function of $M_1$. The corresponding masses
$m_{\widetilde\chi_1^0},m_{\widetilde\chi_2^0},m_{\widetilde\chi_1^\pm}$
(left panel) and the lightest neutralino compositions (right panel)
are also displayed by the mark code reported in the legends. The
choice $\mu=500\,$GeV, $\mu_{\Sigma}=180\,$GeV, $\tan\beta=2.9$ and
$\lambda=0.88$ is assumed.  The SI cross-section is normalized to
$10^{-9}$ pb, which is close to the maximum of LUX sensitivity given
by $\sigma^{\rm SI}\lesssim 8 \times 10^{-10}$\,pb at
$m_{\widetilde\chi_1^0}\sim 50$ GeV~\cite{Akerib:2013tjd}.

It results that at low $M_1$ the lightest neutralino, which is almost
pure Bino, overcloses the Universe until it reaches the Higgs
resonance. In this region the lightest neutralino can provide the
correct relic density and, moreover, its SI cross-section is below the
LUX upper bound (i.e.~the red curve is below 0.8).  This occurs
because the Higgsino components in the coupling
$g_{h\chi^0_1\chi^0_1}$ is enough suppressed to be compatible with LUX
results. The Higgs pole region is only mildly sensitive to the
presence of the Triplino.  For different parameter configurations, the
correct relic density is achieved also at the $Z$ boson
resonance. We will discuss these two poles more in detail in
section~\ref{sec:hp}.

Above the Higgs resonance, the relic density increases until it
reaches the opening of the $W^+W^-$ annihilation channel and then
decreases. It reaches the experimental value when the coannihilation
with the lightest chargino $\widetilde \chi^\pm_1$ (and marginally
with $\widetilde \chi^0_2$) becomes efficient enough.  Since the field
$\widetilde \chi^\pm_1$ is dominantly Triplino (we are assuming
$\mu\gg\mu_\Sigma$), the coannihilation cross section strictly
depends on the tuning between $\mu_{\Sigma}$ and $M_1$. In particular, the
correct relic density occurs for $M_1<\mu_{\Sigma}$ and the LSP is Bino-like
(cf.~right panel). Since in this region also the LUX constraint is
fulfilled, it results that in the TMSSM a well-tempered Bino-Triplino
neutralino can be a good DM candidate.

The behaviors of the relic density and SI cross section shown in the
figure is then a proof of concept for the DM in the TMSSM.  Indeed we
find two qualitatively-different regions where the LSP satisfies the
DM constraints. In the next two sections we discuss
them in detail, still in the $M_2$ decoupling limit.

\subsection{Well-tempered `Bino-Triplino' neutralino}
\label{sec:wtbt}

As it is well known, in MSSM scenarios with well-tempered neutralinos
the correct relic density is achieved by a tuning of the Bino and Wino
(or Higgsino) mass parameters to get an opportune balance between the
large annihilation cross-sections of the Bino and the small ones of
the Wino (or Higgsino)~\cite{ArkaniHamed:2006mb}. In the TMSSM with
$M_2$ above the TeV scale, the role of the Wino is replaced by the
Triplino, which still has gauge interactions with the $W$ bosons.
%
\begin{comment}
 but suppressed by the chirality factors
%
\bea
&-\frac{i}{2} g_2 \Big(2 U^*_{j 1} N_{{i 2}}  + 2 U^*_{j 3} N_{{i 5}}  + \sqrt{2} U^*_{j 2} N_{{i 3}} \Big)\Big(\gamma_{\mu}\cdot\frac{1-\gamma_5}{2}\Big)\\ 
  & + \,-\frac{i}{2} g_2 \Big(2 N^*_{i 2} V_{{j 1}}  + 2 N^*_{i 5} V_{{j 3}}  - \sqrt{2} N^*_{i 4} V_{{j 2}} \Big)\Big(\gamma_{\mu}\cdot\frac{1+\gamma_5}{2}\Big)
\eea 
%
\end{comment}
%
The channels contributing to the relic density are the chargino
annihilation into $W^+W^-,ZZ$ followed by the coannihilations
$\widetilde{\chi}^0_1 \widetilde{\chi}^\pm_1 \to Z W^\pm,
q\bar{q}'$. The relevance of the former processes with respect to the
latter ones depends on the exact hierarchy between $M_1$ and
$\mu_\Sigma$. The $\mu$ parameter is instead constrained by
LUX. Indeed, due to the LUX bound the Higgsino components of the LSP
have to be small in order to suppress the $g_{h\widetilde\chi^0_1\widetilde\chi^0_1}$
coupling that is the main responsible for the SI cross section via
Higgs exchange. This is illustrated in
figure~\ref{fig:b2}, where the SI cross section is plotted as a
function of $\mu_\Sigma$.

In all panels of the figure we fix $\tan\beta=2.9$, $\lambda=0.88$ and
$M_2=1.5\,$TeV.  Besides the quantities shown in
figure~\ref{fig:proof}, also the values of $R_{\gamma\gamma}$ are
displayed (for the color code of each quantity see the legend). At
each point the parameter $M_1$ is adjusted just below $\mu_\Sigma$ to
reproduce the observed relic density.
\begin{figure}[t]
\begin{minipage}[t]{0.49\textwidth}
\centering
\includegraphics[width=1.\columnwidth,trim=0mm 0mm 0mm 0mm, clip]{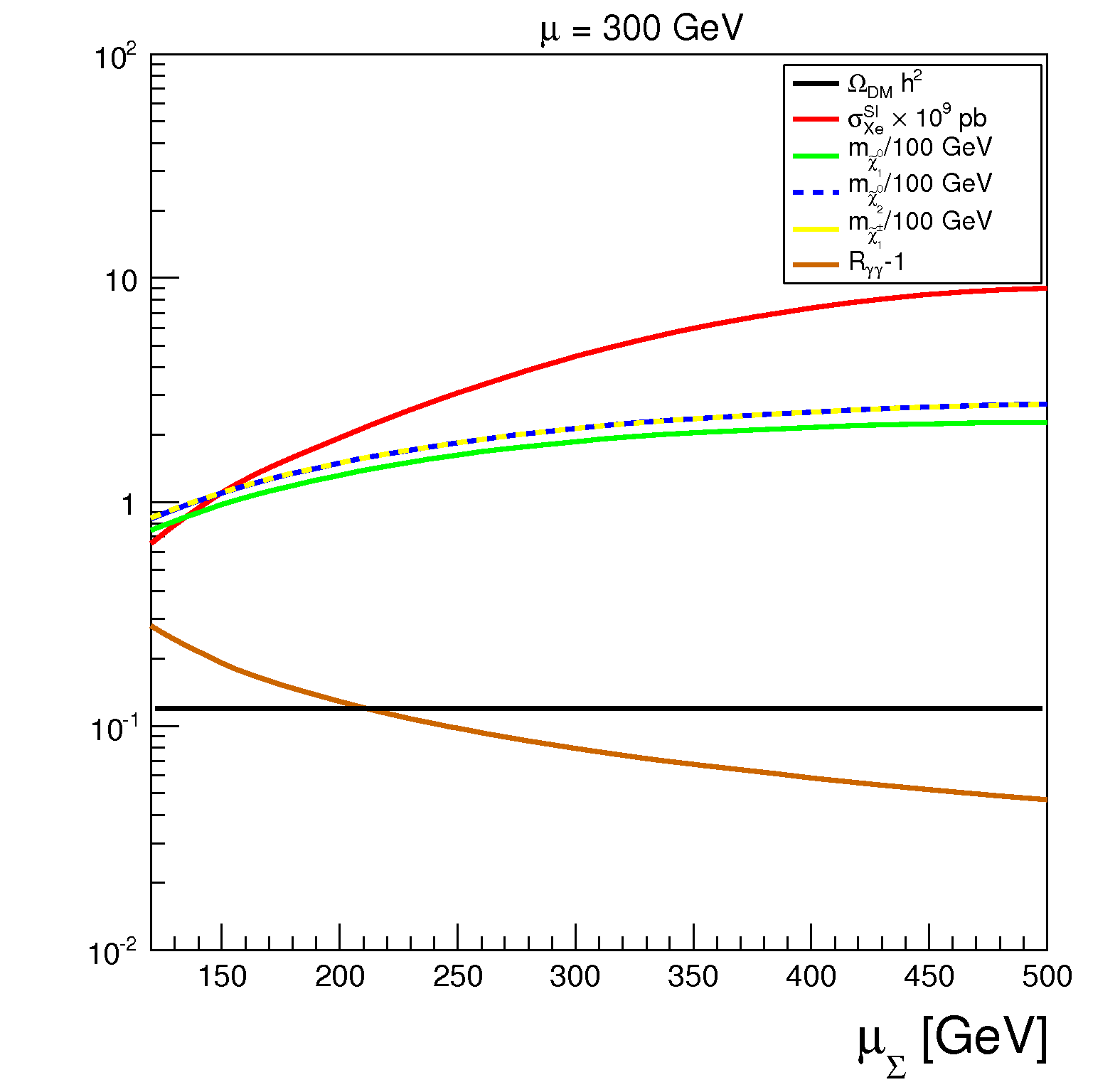}
\end{minipage}
\hspace*{0.2cm}
\begin{minipage}[t]{0.49\textwidth}
\includegraphics[width=1.\columnwidth,trim=0mm 0mm 0mm 0mm, clip]{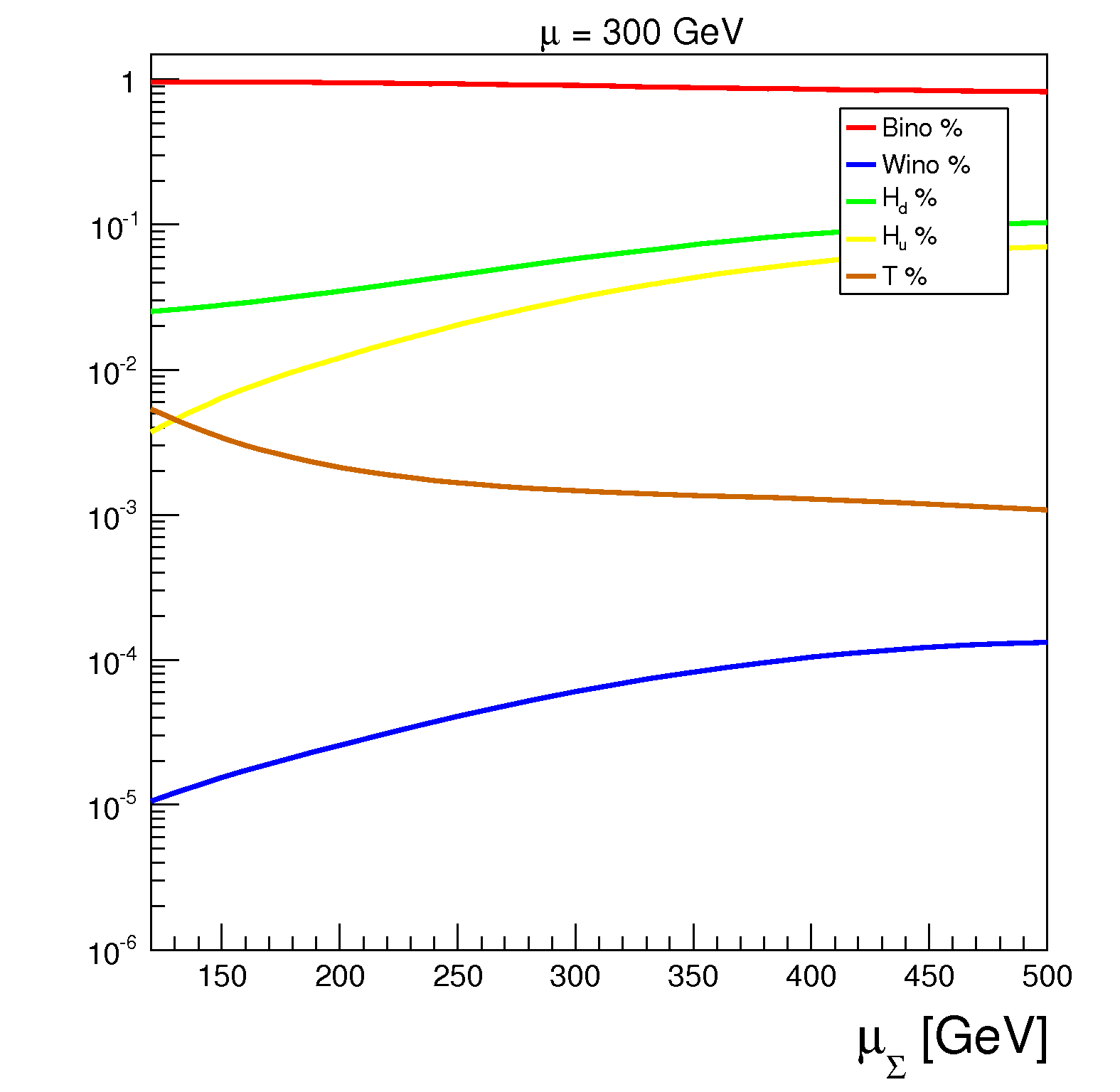}
\end{minipage}
\\
\begin{minipage}[t]{0.49\textwidth}
\centering
\includegraphics[width=1.\columnwidth,trim=0mm 0mm 0mm 0mm, clip]{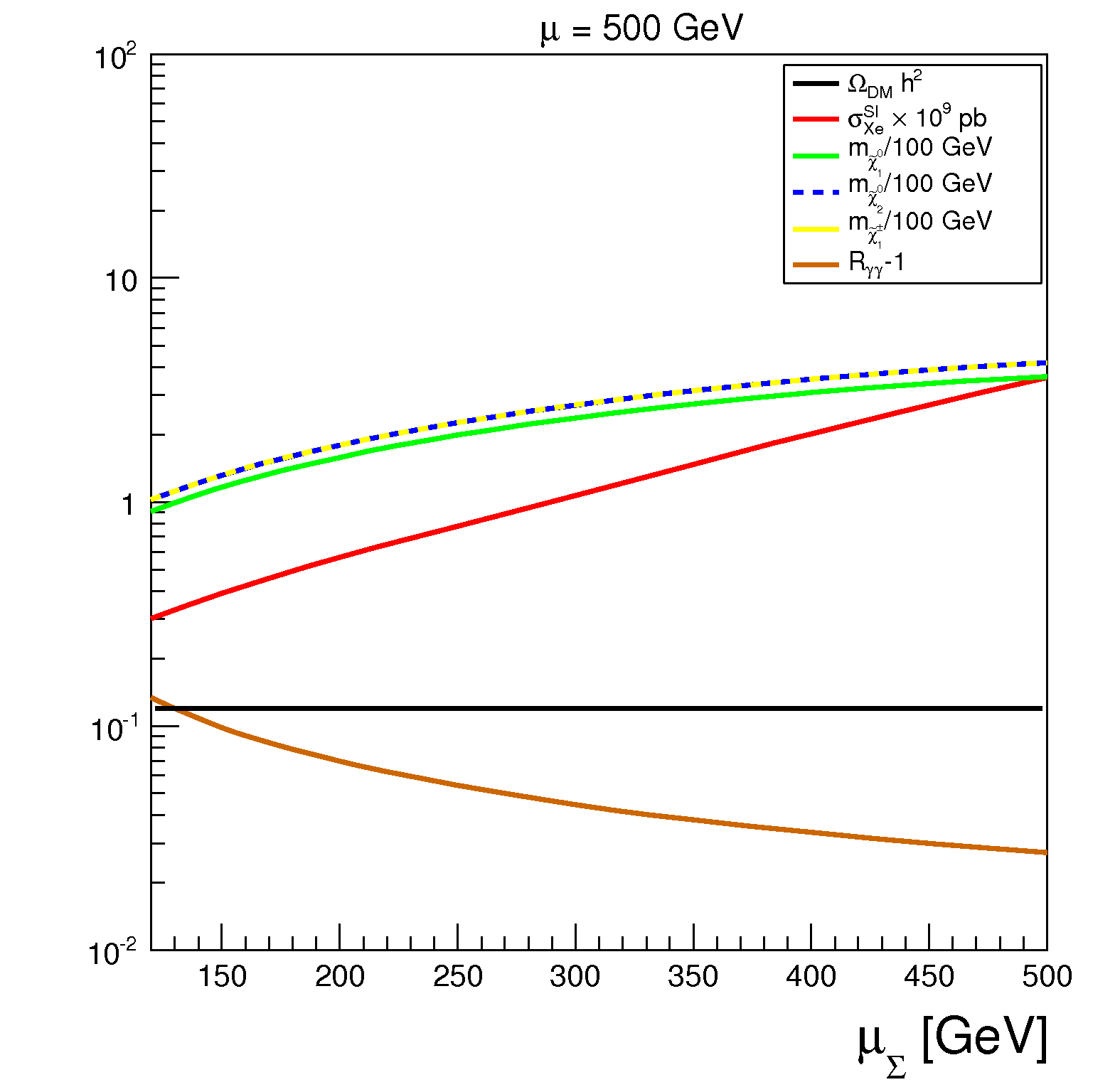}
\end{minipage}
\hspace*{0.2cm}
\begin{minipage}[t]{0.49\textwidth}
\includegraphics[width=1.\columnwidth,trim=0mm 0mm 0mm 0mm, clip]{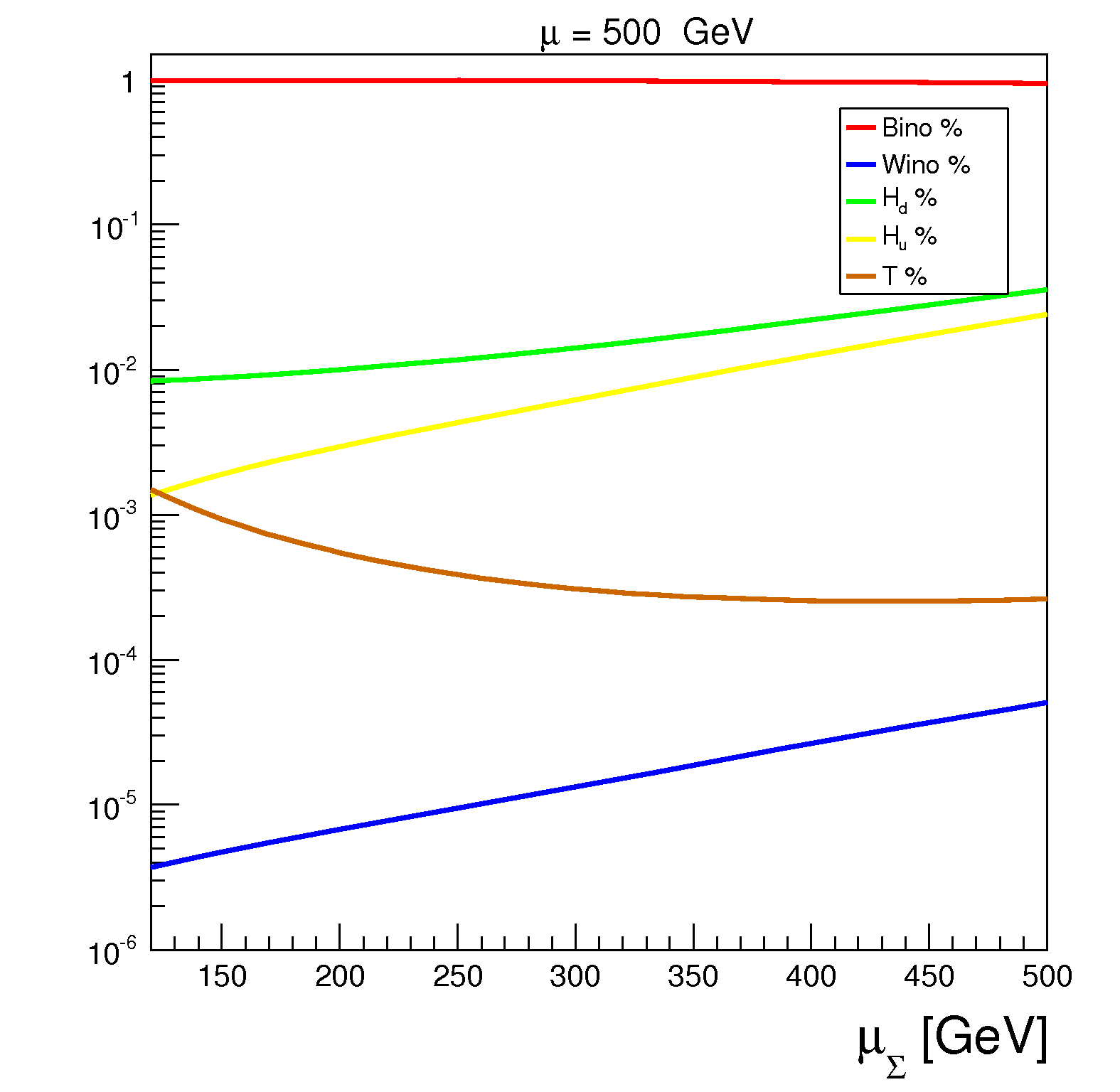}
\end{minipage}
\caption{{\it Top left}: Dependence of $\sigma^{SI}_{Xe}$ (red line),
  $R_{\gamma\gamma}-1$ (brown line), $m_{\widetilde{\chi}^0_1}$ (green
  line), $m_{\widetilde{\chi}^\pm_1}$ (yellow line) and
  $m_{\widetilde{\chi}^0_2}$ (blue line) on $\mu_\Sigma$. $M_1$ is
  adjusted to satisfy $\Omega_{\rm DM} h^2=0.12$ (black solid line)
  for the well-tempered neutralino. The other parameters are
  $\mu=300$\,GeV, $\lambda=0.88$, $\tan\beta= 2.9$ and $M_2=1.5$ TeV.
  {\it Top right}: LSP composition as a function of $\mu_\Sigma$, as
  labelled in the caption. The other parameters are as in the left
  panel. {\it Bottom:} As above but for $\mu=500$ GeV.} \label{fig:b2}
\end{figure}
For $\mu=300$ GeV (top panels), the LSP is mostly Bino but the amount
of its subdominant components vary at different $\mu_\Sigma$. For
light LSPs, the Triplino mixing is comparable to the Higgsino ones and
the SI cross section is below the LUX limit. As soon as both Higgsino
components reach the Triplino one, the SI cross-section is excluded by
the LUX bound. By increasing $\mu$ to 500 GeV (bottom panels), the
Higgsino mixings at a given $\mu_\Sigma$ become smaller than in the
$\mu=300\,$GeV case, and the SI remains below the LUX bound in a wider
range: $\widetilde{\chi}_1^0$ satisfies all DM constraints in the mass
range of 90\textdiv200 GeV. The Higgsino components tend to be always larger than the Triplino one, as the Triplino connects with the Bino only via the Higgsino mixing (see~eq.~\eqref{eq:mnmass}). The figure confirms
as well that the mechanism that provides the relic density is a
balance between annihilation and coannihilation with the lightest
chargino, as both particles are close in mass. The contribution of
$\widetilde{\chi}_2^0$ in coannihilation is marginal and depends
strongly on the exact mixing.

Of course, the minimal $\mu$ value that LUX allows depends on the
parameters that we have kept fixed in the figure. In particular, the
LUX bound on $\mu$ becomes stronger at small $\tan\beta$
(see~eq.\eqref{eq:ghh}). This anti-correlation is discussed more in
detail in section~\ref{sec:DMgggz}. However, we can anticipate that
it affects negatively the enhancements of both $\gamma\gamma$ and
$Z\gamma$ Higgs signals. Indeed, as previously discussed, large
$R_{\gamma\gamma}$ and $R_{Z\gamma}$ require either $\tan\beta$ and
$\mu$ to be small. This is confirmed by the brown line in the left
panel of figure~\ref{fig:b2}: for the considered parameter set, the
maximal $R_{\gamma\gamma}$ drops as $\mu$ goes from 300\,GeV to
500\,GeV. 

\subsection{DM at the Higgs and Z resonances}\label{sec:hp}
\begin{figure}[t]
\begin{minipage}[t]{0.49\textwidth}
\centering
\includegraphics[width=1.\columnwidth,trim=0mm 0mm 0mm 0mm, clip]{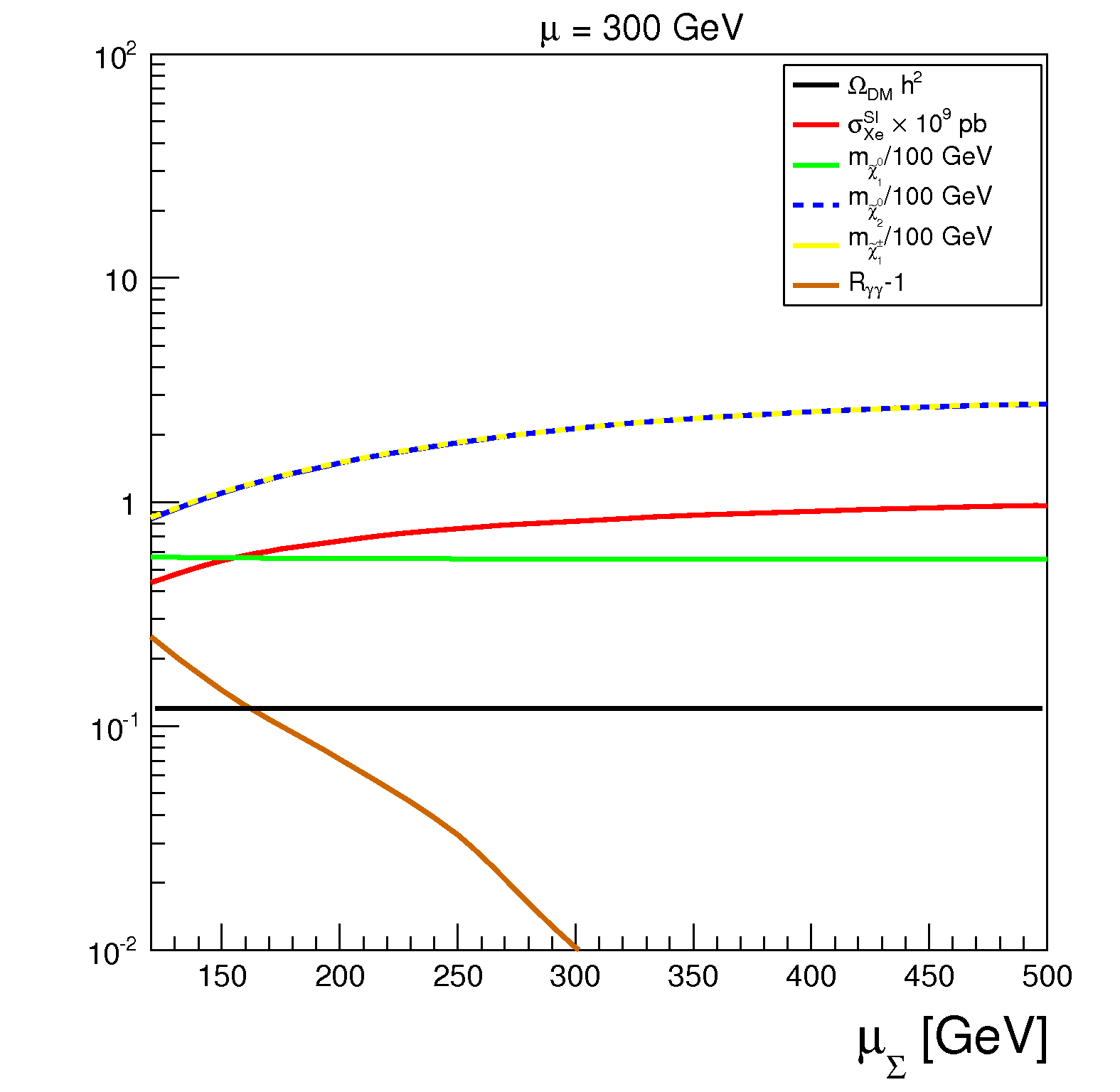}
\end{minipage}
\hspace*{0.2cm}
\begin{minipage}[t]{0.49\textwidth}
\includegraphics[width=1.\columnwidth,trim=0mm 0mm 0mm 0mm, clip]{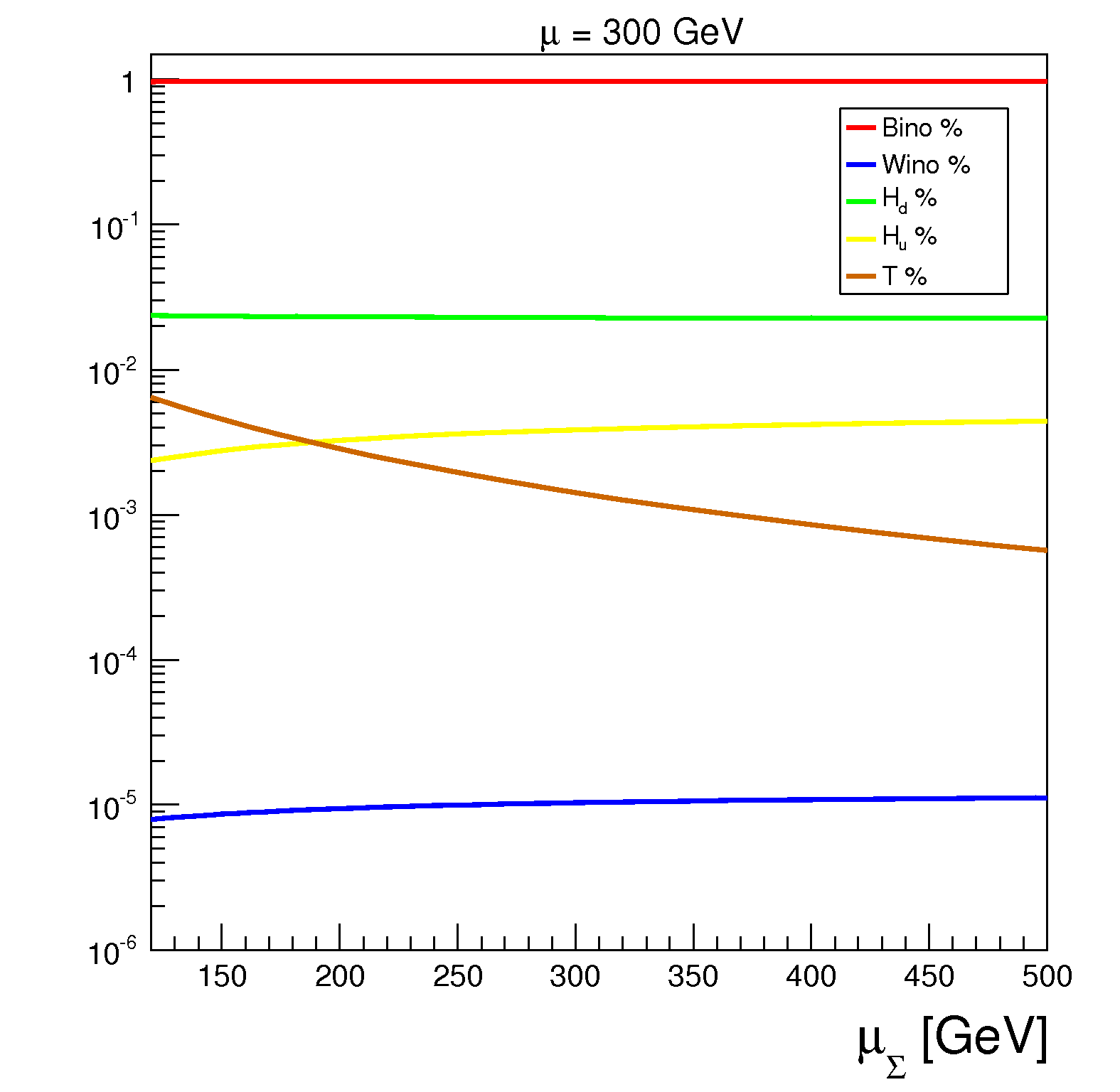}
\end{minipage}
\\
\begin{minipage}[t]{0.49\textwidth}
\centering
\includegraphics[width=1.\columnwidth,trim=0mm 0mm 0mm 0mm, clip]{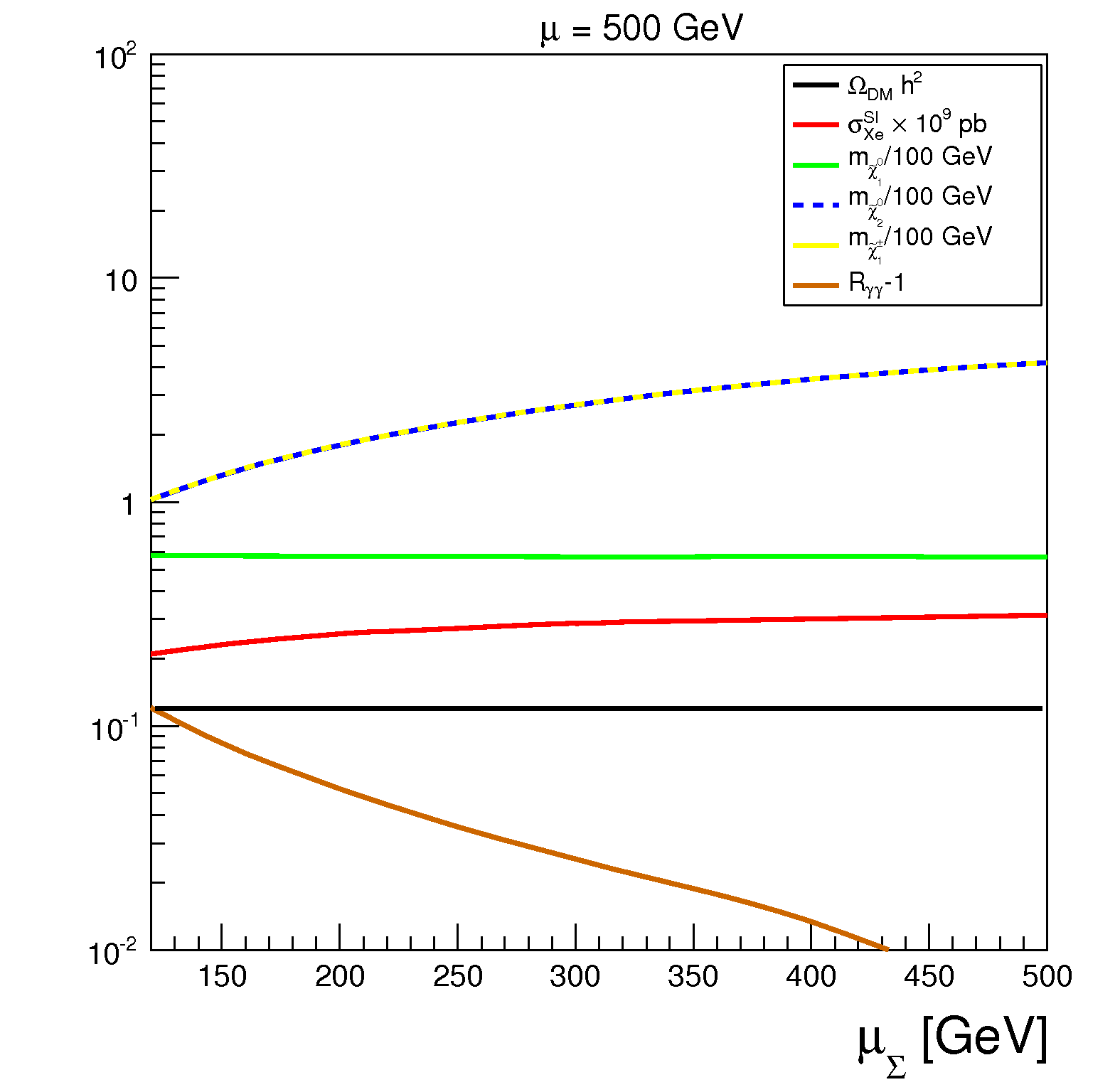}
\end{minipage}
\hspace*{0.2cm}
\begin{minipage}[t]{0.49\textwidth}
\includegraphics[width=1.\columnwidth,trim=0mm 0mm 0mm 0mm, clip]{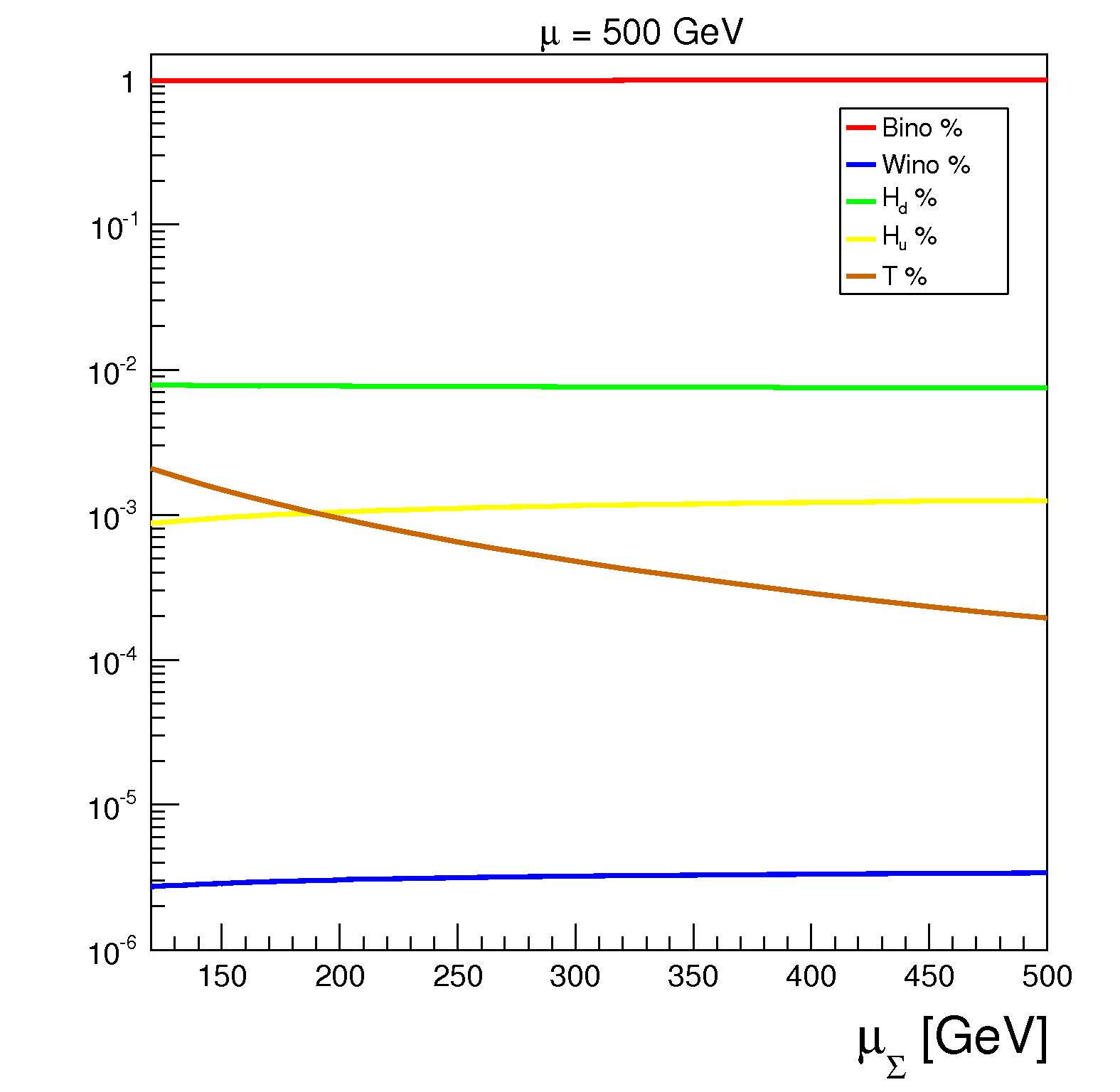}
\end{minipage}
\caption{{\it All panels:} Same as figure~\ref{fig:b2} for the Higgs pole. $M_1$ is chosen to satisfy the relic density and to be $M_1 \simeq m_h/2$.}
\label{fig:b1}
\end{figure}

The Higgs and $Z$ boson resonances are fine-tuned regions as they rely
on the fact that for $M_1 \sim m_h/2$ and $M_1 \sim m_Z/2$ the
annihilation cross-section gets enhanced, hence decreasing the relic
density. We first comment on the Higgs pole.

The case of the Higgs resonance is peculiar because the phenomenology
of the LSP can be reconducted to one coupling 
  only. The vertex Bino-Higgsino-Higgs is responsible for both the
annihilation ($\widetilde{\chi}^0_1 \widetilde{\chi}^0_1 \to h \to q
\bar{q}$) and the SI scattering cross-section since the neutralino is
mostly pure Bino. Hence the key parameters are $M_1$ and $\mu$,
whereas there is a minor dependence on both $\mu_\Sigma$ and $M_2$.
Similarly to the case of the well-tempered neutralino, the $\mu$
parameter is constrained by the LUX bound, as illustrated in
figure~\ref{fig:b1} where at each point $M_1$ is tuned at the Higgs
resonance to achieve the observed relic density. (The plotted
quantities and their color code are as in figure~\ref{fig:b2}).
Indeed for $\mu=300$ GeV (top panels), the SI cross-section is only
marginally compatible with the LUX constraint at large $\mu_\Sigma$
and clearly a small decrease in $\mu$ will exclude these points
(cf.~top and bottom left panels). The behavior of the SI cross-section
is only mildly dependent on $\mu_\Sigma$, as it is almost flat over
all $\mu_\Sigma$ range. This is even more manifest for $\mu=500$ GeV
(lower panels). For such a $\mu$ value the SI cross section is well
below the experimental bound.  From the right panels it is clear that
the LSP is almost pure Bino and that the the dominant annihilation
channels is a Higgs exchange on $s$-channel. Indeed,
  due to the large mass gap between the lightest neutralino and the other
  charginos and neutralinos (left panel), coannihilation is
completely irrelevant.
 
In the Higgs resonance region one might expect to have large
$R_{\gamma \gamma}$ and $R_{Z\gamma}$ signal strengths because the DM
phenomenology is not tightly bounded to the Triplino component. In
other words the $\mu_\Sigma$ parameter is not correlated to
$\sigma_{\rm Xe}^{\rm SI}$ or $\Omega_{\rm DM} h^2$, and therefore can
take low values such that the lightest chargino mass is close to the
LEP bound. However, the anti-correlation between $\tan\beta$ and $\mu$
mentioned in section~\ref{sec:wtbt}, is present in this region as
well. Therefore, the enhancement in the $R_{\gamma \gamma}$ turns out
to be at most $\sim 10\%$ for $\mu=300\,$GeV and negligible for
$\mu=500$ GeV, as indicated by the brown line in the left panel of
figure~\ref{fig:b1}. We will discuss this issue in detail in
section~\ref{sec:DMgggz}.
 
A similar reasoning applies to the $Z$ resonance region, with the
difference that in that region the process that fixes the relic
density, which is proportional to the $Z$-Higgsino coupling of the LSP
(given in eq.~\eqref{eq:Zlr}), is uncorrelated from the SI elastic
cross-section.
\subsection{DM in the TMSSM: global survey}\label{sec:run}
\begin{figure}[t]
\begin{minipage}[t]{0.49\textwidth}
\centering
\includegraphics[width=1.\columnwidth,trim=0mm 0mm 0mm 0mm, clip]{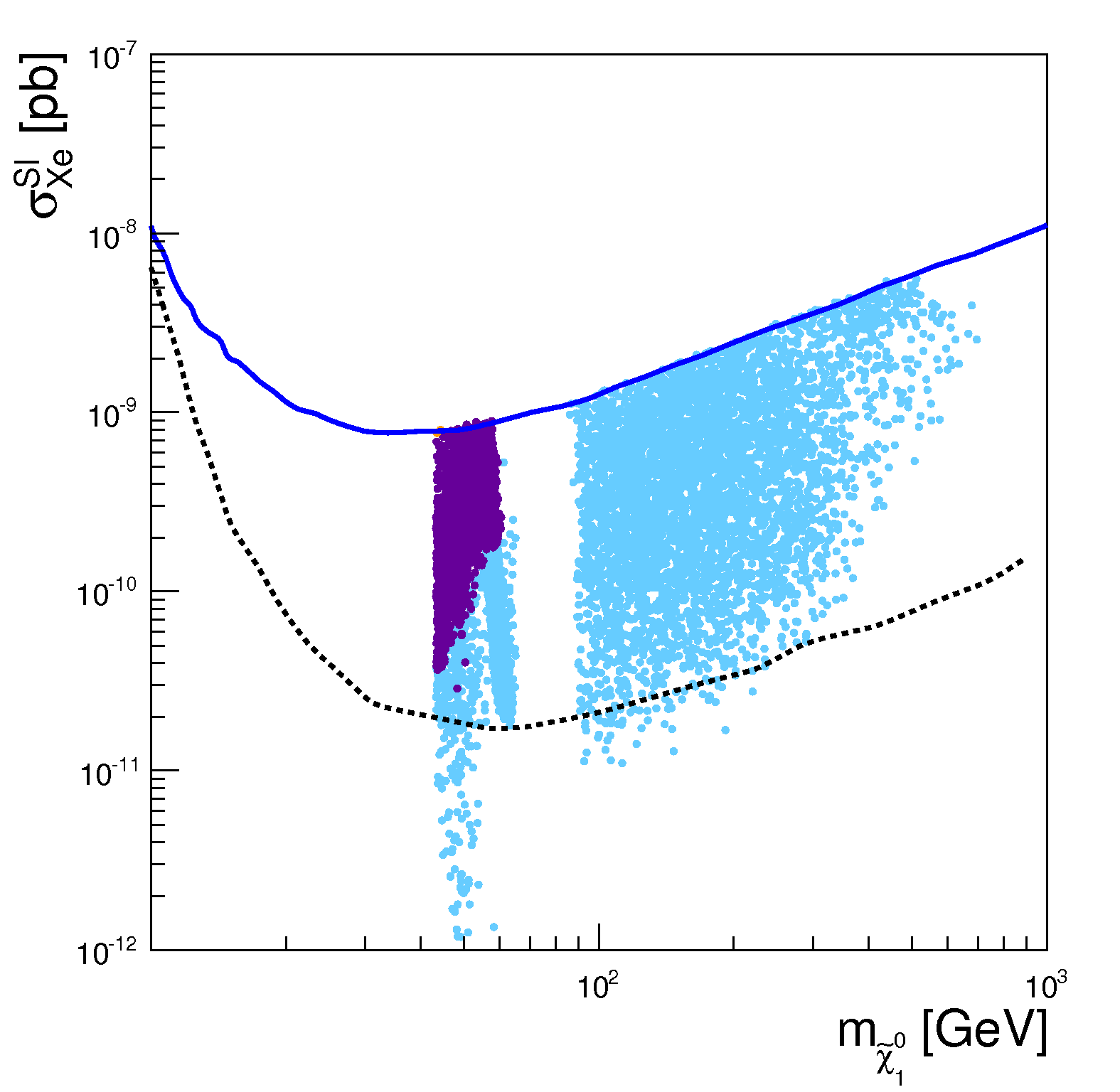}
\end{minipage}
\hspace*{0.2cm}
\begin{minipage}[t]{0.49\textwidth}
\includegraphics[width=1.\columnwidth,trim=0mm 0mm 0mm 0mm, clip]{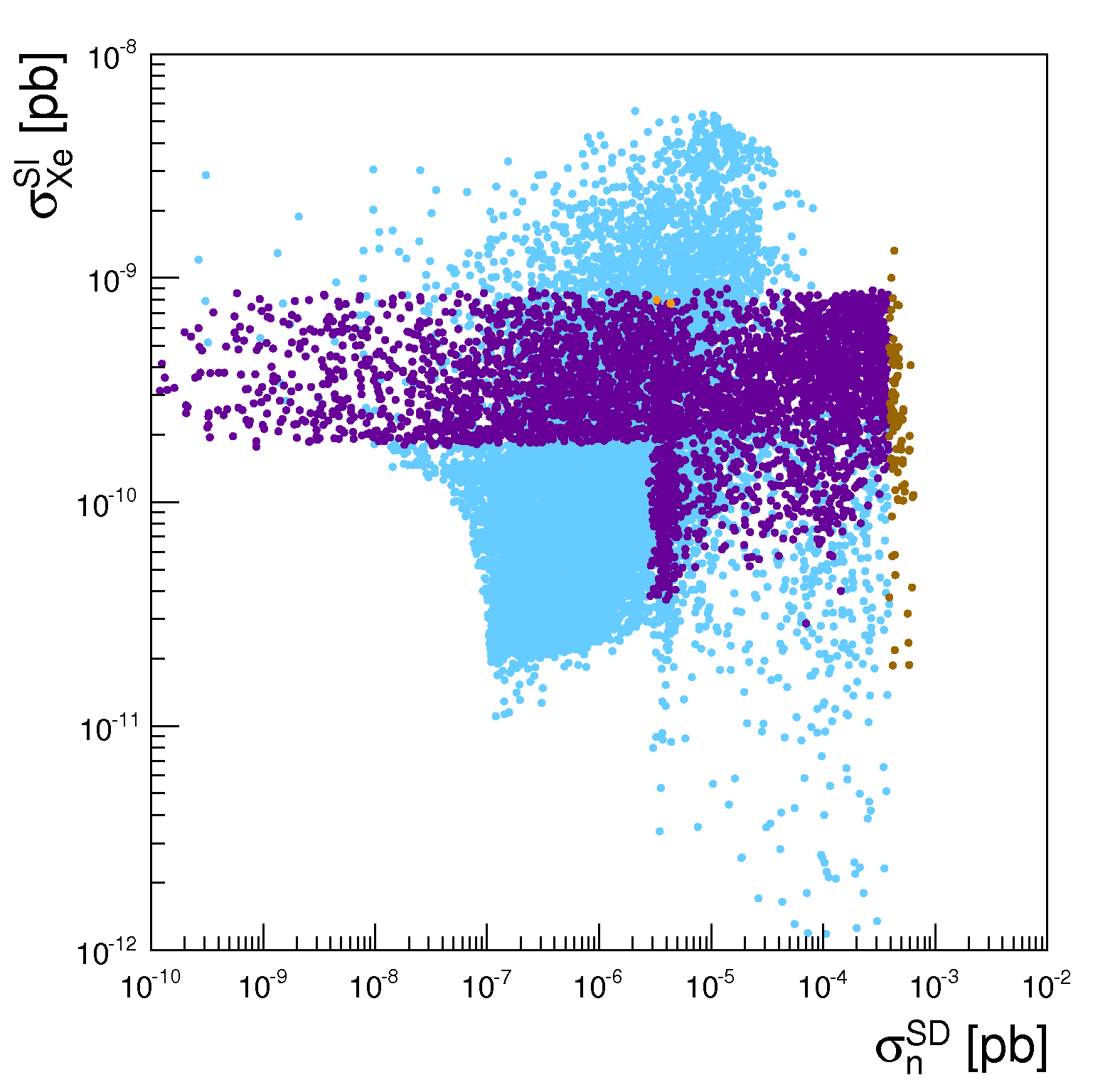}
\end{minipage}
\caption{{\it Left:} Equal weight posterior sample in the
  $\{\sigma^{\rm SI}_{\rm Xe}-m_{\widetilde{\chi}^0_1}\}$-plane. The
  solid blue line stands for the LUX exclusion limit, while the black
  dotted line is the projected sensitivity of XENON1T. The color code
  is as in figure~\ref{fig:gggznoDM} and indicates the Higgs into
  invisible branching ratio percentage when the channel is open. {\it
    Right:} Same as left in the $\{\sigma^{\rm SI}_{\rm
    Xe}-\sigma^{\rm SD}_n\}$-plane. The brown points stand for points
  at odds with the XENON100 exclusion bound for $\sigma_n^{\rm
    SD}$.}\label{fig:mDMSI}
\end{figure}

In this section we present the results of a comprehensive sampling of
the TMSSM parameter space, using the likelihood $\mathcal{L}_{\rm
  DM}(d|\theta_i)$ and the prior ranges described in
section~\ref{sec:num}.

Figure~\ref{fig:mDMSI} (left panel) shows the mass value for which the
LSP is a viable DM candidate compatible with LUX (blue solid). As
discussed above, there are two separate regions: one with the
resonances at $40\,$GeV$\lesssim m_{\widetilde\chi_1^0}\lesssim
70\,$GeV, and one with a well-tempered neutralino at
$m_{\widetilde\chi_1^0}\gtrsim 90\,$GeV. The apparent upper limit at
about 600\,GeV is an artifact of the prior range choice for the mass
parameters. Notice that almost all the parameter space is in the
sensitivity range of XENON1T~\cite{Aprile:2012zx} (black dotted line),
so an effective TMSSM might be probed by DM direct searches in 5-7
years time~\footnote{However, we stress that the accuracy of the
  present analysis does not allow for refined comparison between the
  TMSSM and XENON1T. Indeed, our estimate of the SI cross section does
  not take into account loop-induced corrections of the order of
  $\mathcal O(10^{-11})$ pb~\cite{Cirelli:2005uq}.}. In our sample the
minimal values of $\sigma^{\rm SI}_{\rm Xe}$ correspond to the
contribution to the SI cross-section due to squarks exchange, when the
Higgsino component start to be negligible. The value is similar in all
the sample as the squark sector is kept heavy. More interestingly, the
requirement of having the LSP as good DM candidate sets as well an
upper bound on the Higgs invisible branching ratio ${\rm BR}(h \to
\widetilde{\chi}_1^0 \widetilde{\chi}^0_1) < 20\%$ (violet points),
which is comparable to current LHC
bounds~\cite{Espinosa:2012vu,Belanger:2013kya,Giardino:2013bma,Ellis:2013lra} (indeed there are the very few dark yellow points
with ${\rm BR}(h \to \widetilde{\chi}_1^0 \widetilde{\chi}^0_1) >
20\%$).  This illustrates the complementarity between DM direct
searches and colliders: significant values of the Higgs invisible
  width can be fully probed by XENON1T.

In the right panel of figure~\ref{fig:mDMSI} we show the $\sigma^{\rm
  SI}_{\rm Xe}$ versus the SD cross-section on neutron (which is
equivalent to the one on proton). As in the previous figures, violet
points represent parameter configurations with ${\rm BR}(h \to
\widetilde{\chi}_1^0 \widetilde{\chi}^0_1) >1\%$, whereas brown points
correspond to SD cross-section values at odds with the XENON100
exclusion bound, whose strongest limit $\sigma^{\rm SD}_n \lesssim 3
\times 10^{-4}$\,pb is at $m_{\widetilde{\chi}^0_1}\sim50\,$GeV.
We do not remove the brown points from our samples as the nuclear uncertainties on the structure functions for the nucleons are large and can affect the predicted number of events by a factor of 3-4~\cite{Cerdeno:2012ix,Aprile:2013doa,Arina:2013jya}.
In addition the predictions for the SD on proton give the same value, however these values are below the COUPP sensitivity, which reaches the maximum at  $\sigma^{\rm SD}_p\sim 5 \times
10^{-3}\,$pb for $m_{\widetilde{\chi}^0_1}\sim40\,$GeV. These points
with large $\sigma^{\rm SD}_{p,n}$ are associated to the $t$-channel
$Z$ boson exchange and arise when Higgsino components are
sizable. They correspond to the $Z$ resonance region, where the
coupling $g_{Z\widetilde\chi^0_1 \widetilde\chi^0_1}$ is large. In the
case of SD scattering there is a one-to-one correspondence with the
annihilation cross-section mediated by the $Z$ boson.} The main bulk
of the sample is however below the current SD bounds, because the
coupling to the $Z$ boson tends to be suppressed for most of the LSP
composition, being predominantly Bino-like. The sharp cut on the upper
values of SD and SI when the Higgs channel into invisible is open is
due to the LUX bounds. XENON1T will be less sensitive to SD
interaction (perhaps $\sim 10^{-6}$ pb), hence the model is more
likely to be tested with the SI cross-section and the brown points
will all be probed, independently of the nuclear uncertainties.

\begin{figure}[t]
\begin{minipage}[t]{0.49\textwidth}
\centering
\includegraphics[width=1.\columnwidth,trim=0mm 0mm 0mm 0mm, clip]{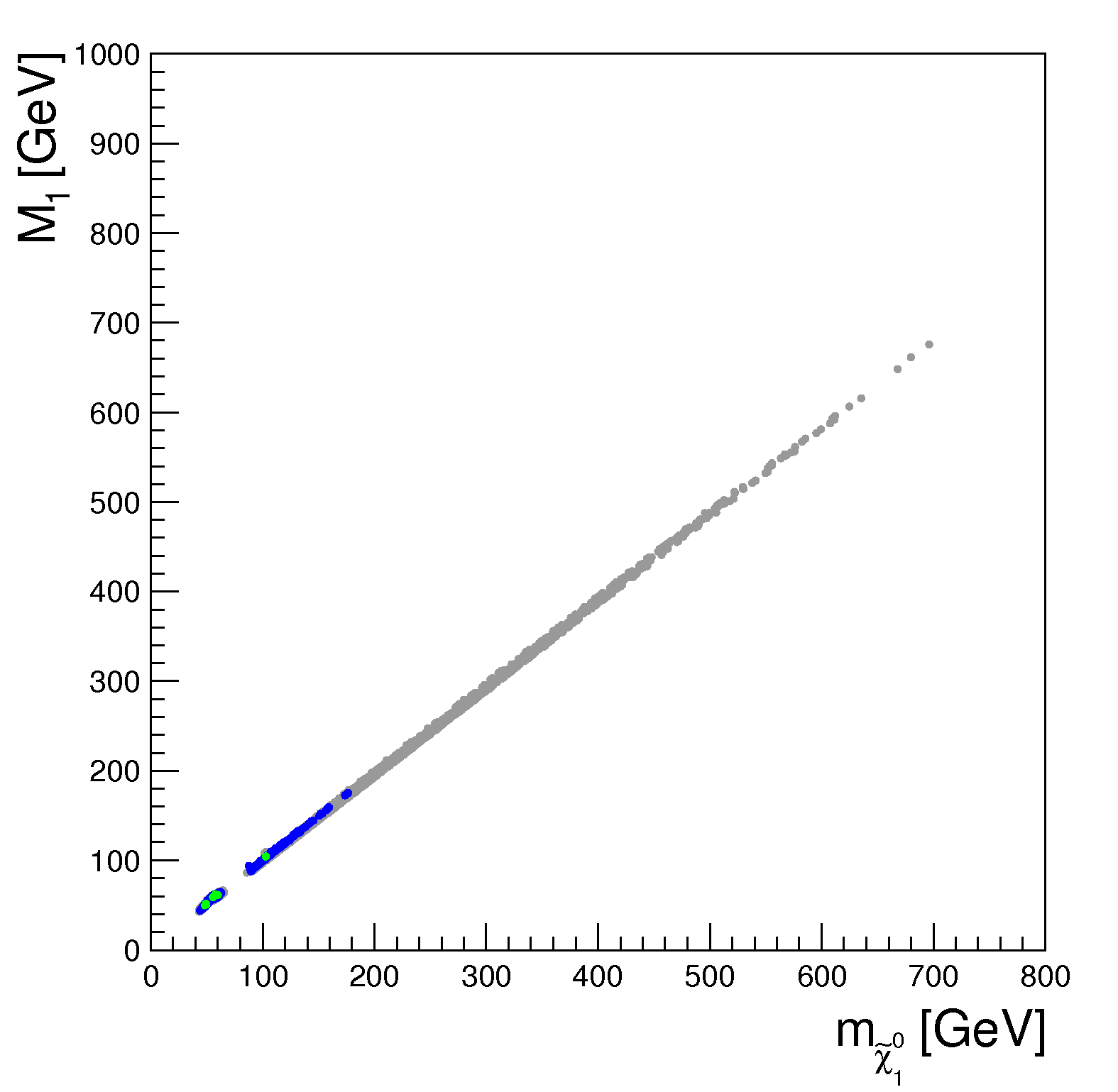}
\end{minipage}
\hspace*{0.2cm}
\begin{minipage}[t]{0.49\textwidth}
\includegraphics[width=1.\columnwidth,trim=0mm 0mm 0mm 0mm, clip]{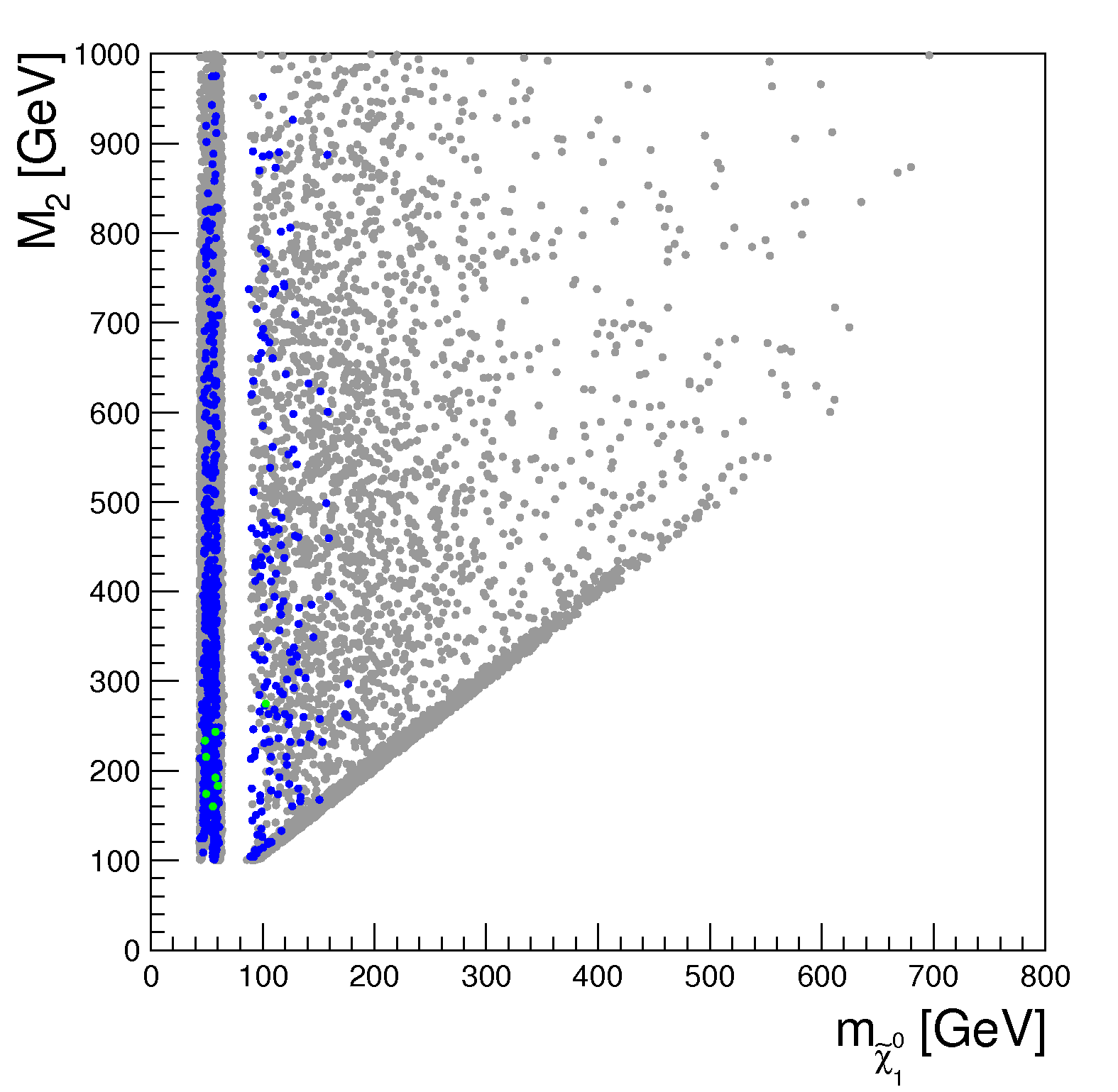}
\end{minipage}
\\
\begin{minipage}[t]{0.49\textwidth}
\centering
\includegraphics[width=1.\columnwidth,trim=0mm 0mm 0mm 0mm, clip]{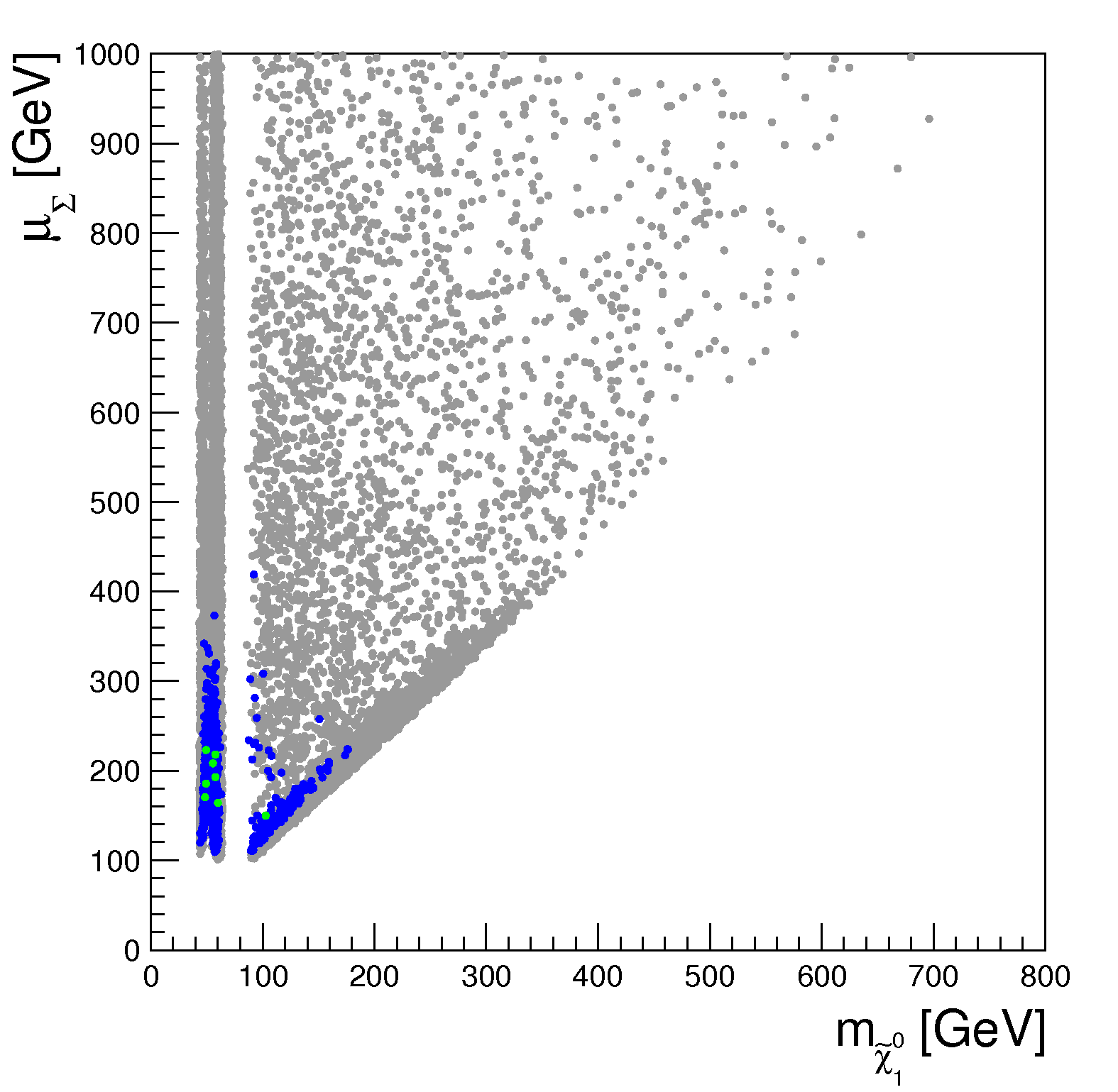}
\end{minipage}
\hspace*{0.2cm}
\begin{minipage}[t]{0.49\textwidth}
\includegraphics[width=1.\columnwidth,trim=0mm 0mm 0mm 0mm, clip]{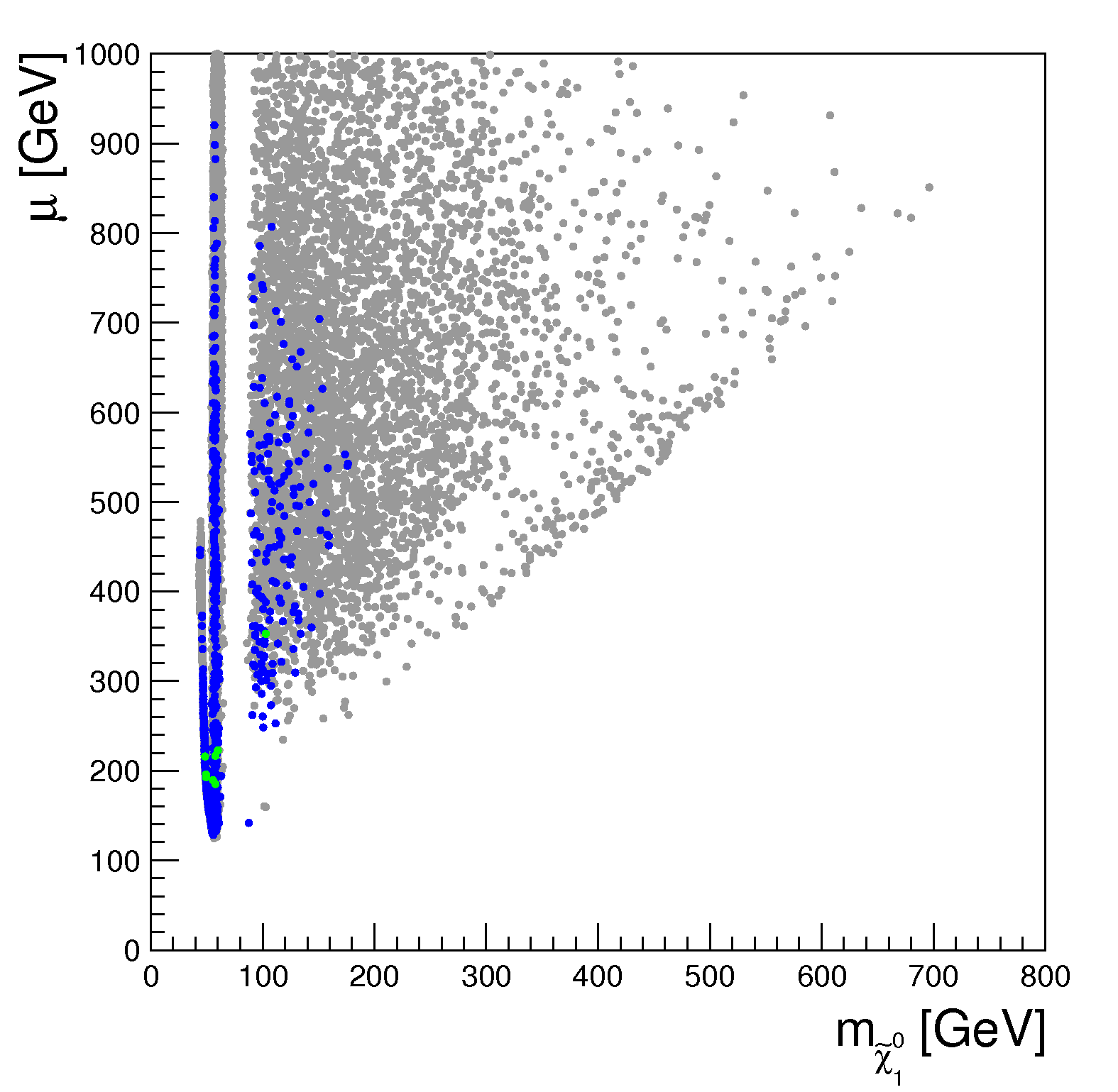}
\end{minipage}
\caption{{\it Top left}: $M_1$ dependence of the mass of the LSP
  $m_{\widetilde{\chi}^0_1}$. {\it Top right:} Same as left for
  $M_2$. {\it Bottom:} Same as top left for $\mu_\Sigma$ and
  $\mu$. The color code is as in
  figure~\ref{fig:samplesnoDM}.}\label{fig:n1comp}
\end{figure}

When $M_2$ is decreased to the same scale as the other chargino
  parameters, the phenomenology of the DM is wider and the tight
bound on the lower limit of $\mu$ described above
(sections~\ref{sec:wtbt} and~\ref{sec:hp}) is relaxed by the
additional admixture with the Wino component. This is illustrated in
figure~\ref{fig:n1comp}, where we display the LSP mass versus the
neutralino and chargino parameters as labelled.
From the top left panel, it is clear that the LSP is mostly Bino, as
$m_{\widetilde{\chi}^0_1}$ and $M_1$ follow each other over all the
allowed range. From the other three panels it is striking that the
Higgs and $Z$ resonances are mostly independent from
$M_2$ and $\mu_\Sigma$ as they can acquire approximately any allowed
value.

The case of the well-tempered neutralino, where all possible
combinations of compositions for the are available, is more
interesting. Indeed the LSP can be mixed Bino-Triplino, as discussed
above, but it can never be Triplino dominated because the LSP would be
the corresponding chargino. Successful DM candidates can also show up
as MSSM-like states, that is Bino-Wino, in which case the relic
density is achieved by neutralino annihilation into $W^+W^-$ and
coannihilation with the lightest chargino producing
$q\bar{q}'$. These MSSM-like scenarios are however less appealing
because the condition $M_1\sim M_2$ is not recovered by the usual
supersymmetry-breaking mechanisms. Of course, a large portion of the
parameter space presents mixed Bino-Triplino-Wino LSP, for which the
dominant annihilation and coannihilation channels are
$\widetilde{\chi}^0_1 \widetilde{\chi}^0_{1,2} \to W^+ W^-$ and
$\widetilde{\chi}^0_1 \widetilde{\chi}^\pm_1 \to q \bar{q}'$. Due to
the LUX bound only $\mu$ larger than about $300$\,GeV is
allowed. Moreover, the relic density is achieved by a mixture of
annihilation and coannihilation. More specifically the main processes
are $\widetilde{\chi}^0_1 \widetilde{\chi}^0_1 \to h Z, q \bar{q}$ and
$\widetilde{\chi}^0_1 \widetilde{\chi}^\pm_1 \to Z W^\pm$. Instead,
when all components (Bino-Triplino-Wino-Higgsino) in the LSP are
sizable, the main annihilation channels are the following:
$\widetilde{\chi}^\pm_1 \widetilde{\chi}^\pm_1 \to W^\pm W^\pm$,
$\widetilde{\chi}^0_1 \widetilde{\chi}^\pm_1 \to h W^\pm$ and
$\widetilde{\chi}^0_1 \widetilde{\chi}^0_1 \to h Z$.

\subsection{DM implications on $R_{\gamma \gamma}$ and $R_{Z\gamma}$}\label{sec:DMgggz}
\begin{figure}[t]
\begin{minipage}[t]{0.49\textwidth}
\centering
\includegraphics[width=1.\columnwidth,trim=0mm 0mm 0mm 0mm, clip]{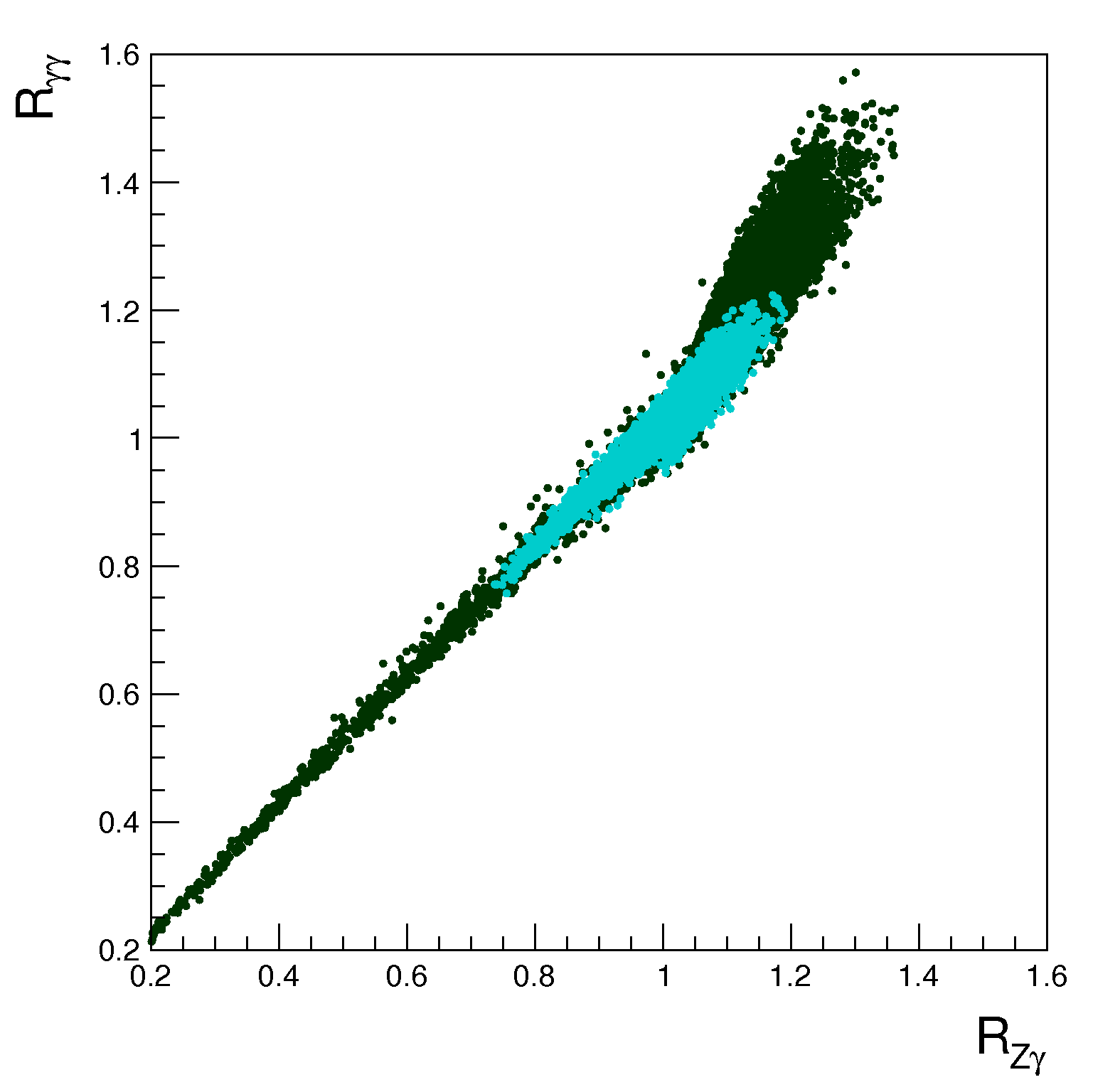}
\end{minipage}
\hspace*{0.2cm}
\begin{minipage}[t]{0.49\textwidth}
\includegraphics[width=1.\columnwidth,trim=0mm 0mm 0mm 0mm, clip]{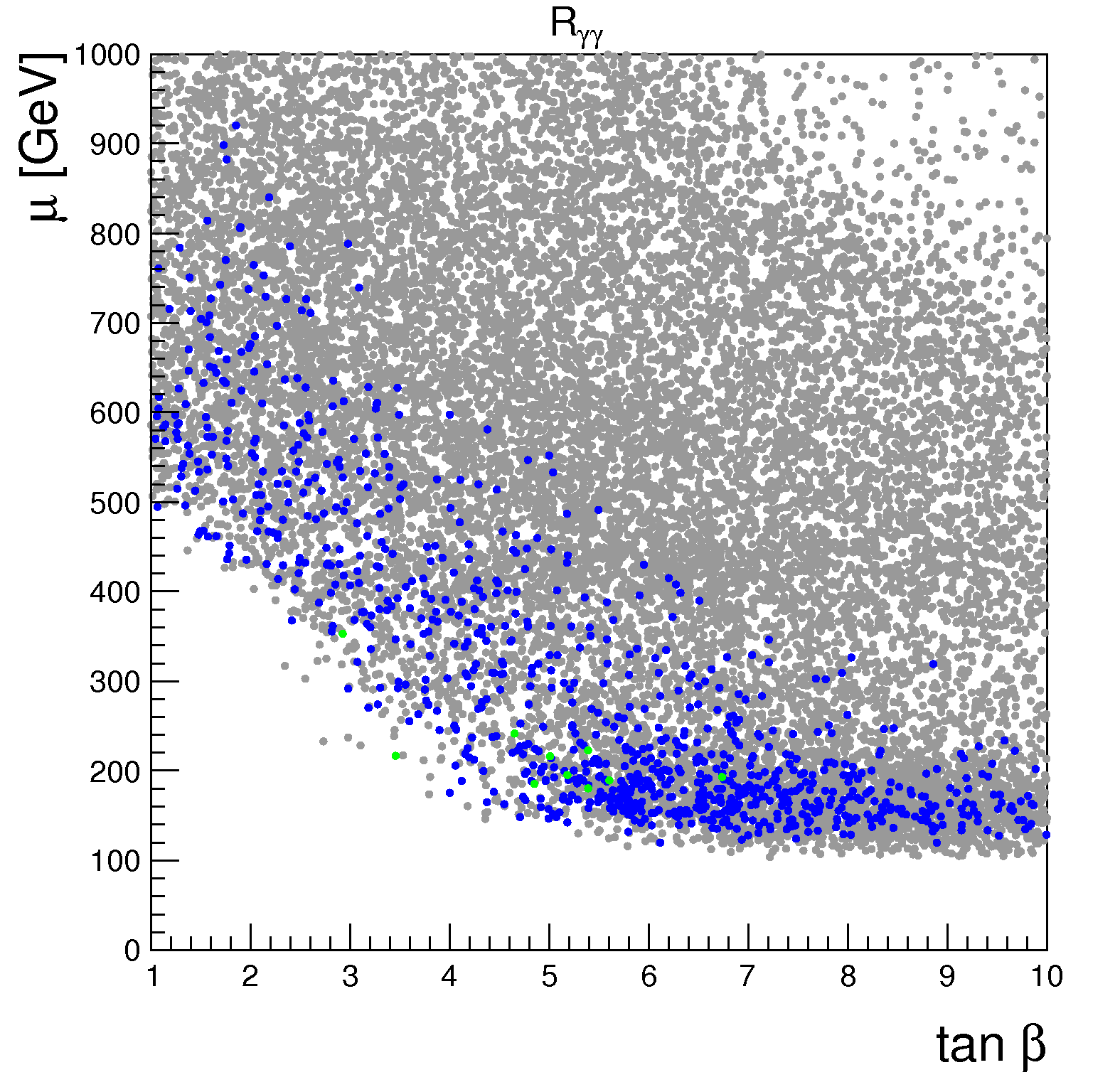}
\end{minipage}
\caption{{\it Left}: Correlation between $R_{\gamma\gamma}$ and $R_{Z\gamma}$ for the equal weight posterior sample for the no DM case (green points) and for the DM case (cyan points). {\it Right:} Correlation between $\tan\beta$ and $\mu$ for the equal weight posterior sample for the DM case. Same color code as in figure~\ref{fig:samplesnoDM}.}\label{fig:gggzDM1}
\end{figure}

Figure~\ref{fig:gggzDM1} shows the $\gamma\gamma$ signal strength
versus the $Z\gamma$ one, similarly to figure~\ref{fig:gggznoDM}. The
possible $R_{\gamma\gamma}$ and $R_{Z\gamma}$ that can be achieved in
the TMSSM where the DM constraints are satisfied, are displayed in
cyan. They are superposed to the points of figure~\ref{fig:gggznoDM}
where no DM observable is imposed (here displayed in
  green).  The DM constraints alleviate the change in slope that
arises in the correlation plot without DM constraints (cf.~left panel
of figure~\ref{fig:gggznoDM}): the cyan points follow a smooth pattern
with respect to the dark green ones. As previously discussed, the
missing dark green zone in the upper part of that plot is due to a
Higgs branching ratio varying very fast as soon as it is kinematically
open. There, to get small reduction in the signal strength, $M_1$
should lie exactly on the threshold value, which is a very infrequent
situation. On the contrary, with the LSP being DM, the relic density
constraint requires $M_1 \sim m_{h,Z}/2$ (or the well-tempered
neutralino conditions) and therefore a large portion of the sampled
parameter space is concentrated in the pole regions.

\begin{figure}[t]
\begin{minipage}[t]{0.32\textwidth}
\includegraphics[width=0.99\columnwidth,trim=0mm 0mm 0mm 0mm, clip]{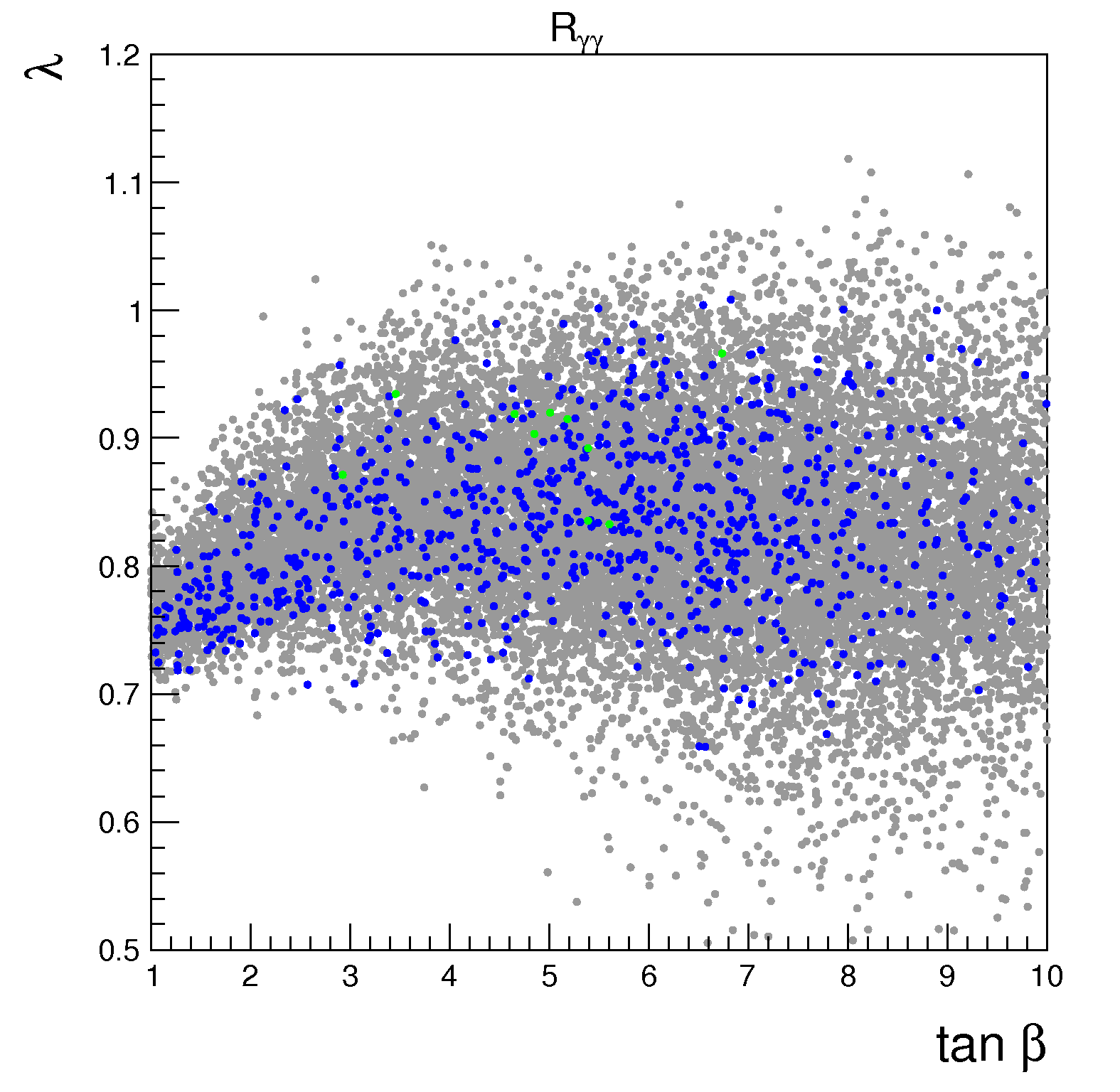}
\end{minipage}
\begin{minipage}[t]{0.32\textwidth}
\centering
\includegraphics[width=0.99\columnwidth,trim=0mm 0mm 0mm 0mm, clip]{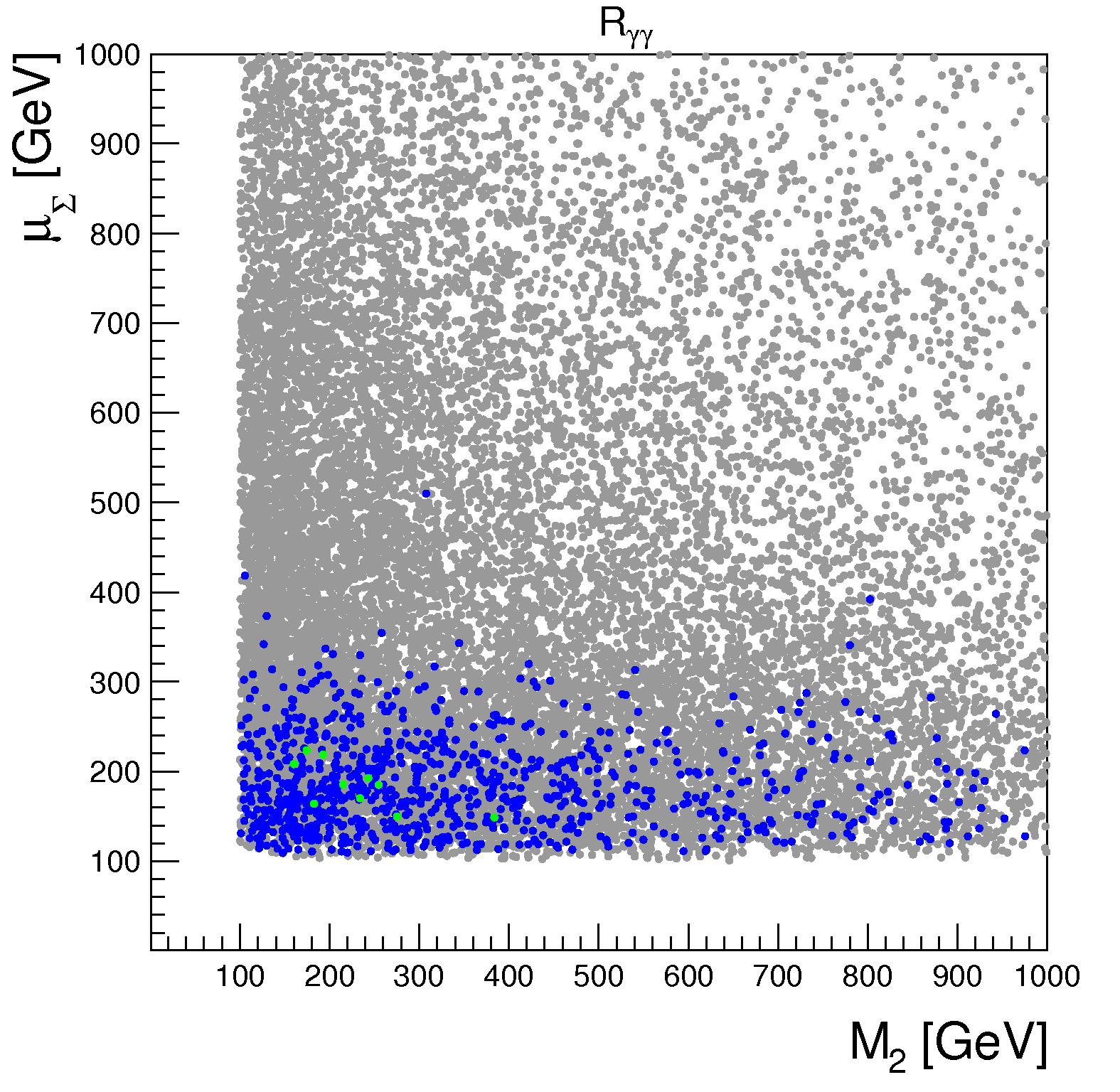}
\end{minipage}
\begin{minipage}[t]{0.32\textwidth}
\includegraphics[width=0.99\columnwidth,trim=0mm 0mm 0mm 0mm, clip]{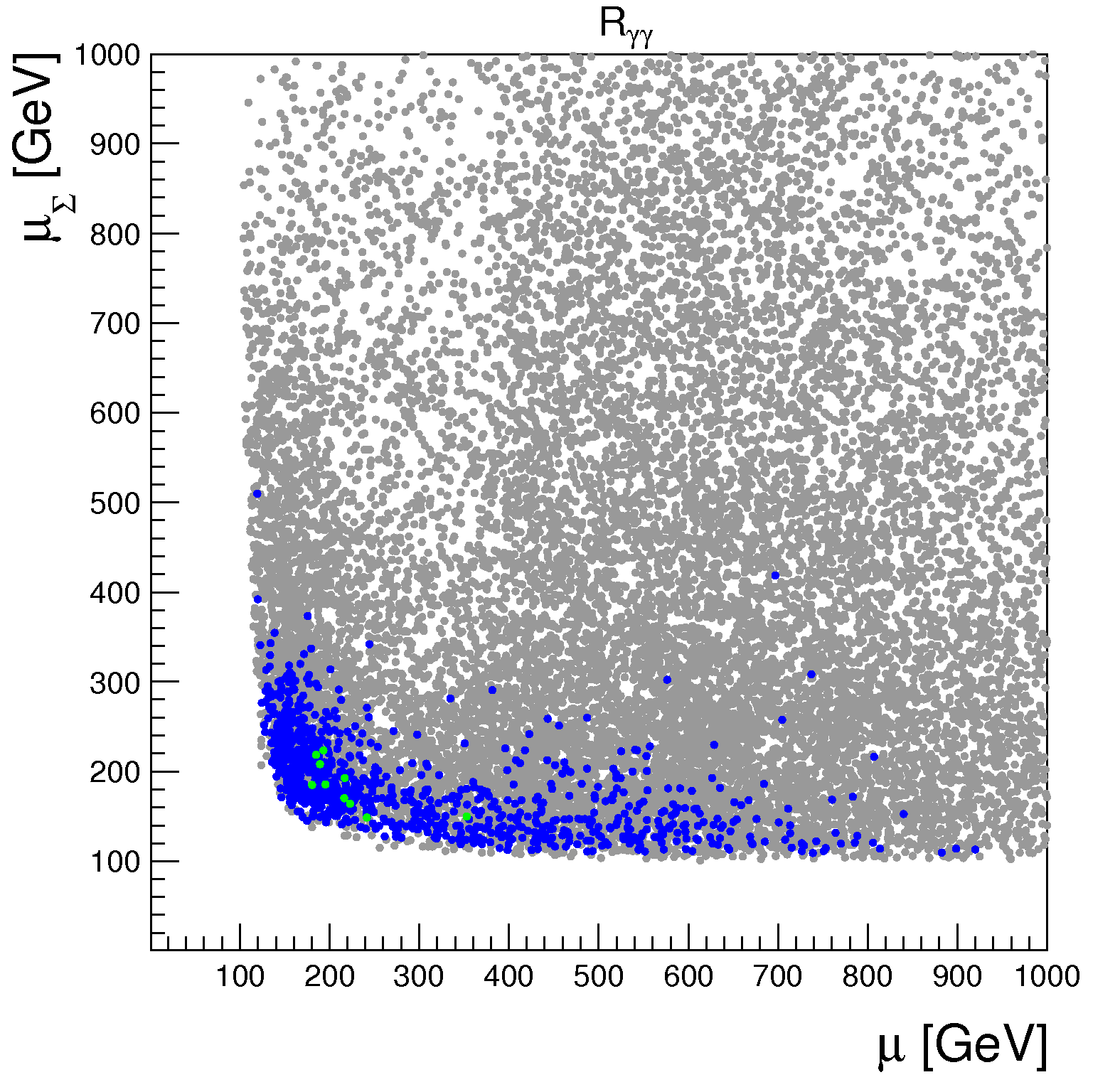}
\end{minipage}
\caption{{\it Left:} $R_{\gamma \gamma}$ (third direction) projected in the $\{\lambda-\tan\beta\}$-plane. Same color code as in figure~\ref{fig:samplesnoDM}. {\it Central and right:} Same as left in the $\{\mu_\Sigma - M_2\}$ and $\{\mu_\Sigma - \mu\}$ planes.}\label{fig:gggzDM2}
\end{figure}

On the other hand, on general basis, the DM constraints are not
encouraging about the collider Higgs phenomenology. Indeed, our analysis
leads to
\be
R_{\gamma\gamma}\lesssim 1.25~,
\qquad R_{Z\gamma}\lesssim 1.2
\hspace{1.5cm} {\rm (with~DM~obs.)~.}
\label{maxDM}
\ee
The most stringent constraint on the parameter space where large
$R_{\gamma\gamma}$ and $R_{\gamma\gamma}$ are achieved, is the LUX
bound on SI cross-section, which rules out the configurations where
either $\mu$ and $\tan\beta$ are simultaneously small: the
anti-correlation between these two variables is striking from the
right panel of figure~\ref{fig:gggzDM1}~\footnote{This correlation is
  proper of the MSSM and has been noticed for instance in
  ref.~\cite{Farina:2011bh}.}. In particular, for $\tan\beta\simeq 1$
the LSP is a viable DM candidate only for $\mu > 500$ GeV, i.e. it is
incompatible with the ballpark that provides the largest possible
enhancements (see figure~\ref{fig:gggznoDM}).  The possibility of
achieving sizeable loop-induced decays of the Higgs and at the same
time a successful neutralino DM particle, starts arising at $\tan\beta
\gtrsim 3$ and $\mu\gtrsim 300$ GeV: this is exactly the region that
saturates the bounds in eq.~\eqref{maxDM}, as shown by the green
points in the right panel (these points have $R_{\gamma\gamma}>
20\%$). The mild enhancement of 10\% is viable in all $\tan\beta$
range, as it is due to small values of $\mu_\Sigma$. This is
illustrated in figure~\ref{fig:gggzDM2}. In the left panel we show the
projection of the $R_{\gamma \gamma}$ values as a function of
$\lambda$ and $\tan\beta$: the range of the Triplino coupling
$\lambda$ that provides the enhancements is limited with respect to
figure~\ref{fig:gggznoDM}, being scattered at around
0.8\textdiv0.9. Signal strengths larger than one can only be achieved
for small values of $\mu_\Sigma$ (central panel); they are
however only marginally sensitive to $M_2$. The same points are
instead concentrated to both small values of $\mu$ and $\mu_\Sigma$
(right panel) and they mostly correspond to the Higgs and $Z$
resonance regions. This is confirmed by looking at the same sparse
green/blue points in figure~\ref{fig:n1comp}, which are mostly
concentrated at values of $m_{\widetilde{\chi}^0_1}$ around
  40\textdiv70\,GeV.
%

%%%%%%%%%%%%%%%%%%%%%%%%%%%%%%%%%%%%%%%%%%%
\section{Conclusions}
\label{sec:Concl}

We are entering the era of precision Higgs physics: the measurements
of the Higgs mass, couplings and decay modes have already started to be
highly sensitive to new physics beyond the SM. In this paper we have
considered the Higgs phenomenology of the $Y=0$ triplet extension of
the MSSM, dubbed TMSSM, in which the new coupling between the triplet
and the MSSM Higgses can alleviate the little hierarchy problem and
modify the chargino and neutralino sector.

We have first accurately determined the couplings and pole masses
  of the stops, charginos, neutralinos and lightest CP-even Higgs
  $h$. Then, we have tackled the subtle effects of the
Triplino in the $h \to \gamma\gamma$ and $h \to Z\gamma$ loop-induced
processes. We have shown that the additional Triplino component in the chargino
sector provides a maximal enhancement of 60\% in the $R_{ \gamma
  \gamma}$ signal strength, which is slightly larger than previously
estimated (i.e.~$R_{\gamma\gamma}\lesssim 1.45$)~\cite{antonio1}. An
enhancement up to 40\% can be achieved in the $R_{ Z\gamma}$ signal
strength, which we find to be highly correlated with the diphoton
channel, even though it is always smaller than
$R_{\gamma\gamma}$. The parameter region leading to the largest $R_{ \gamma
  \gamma}$ and $R_{ Z\gamma}$ is characterized by $\tan\beta\lesssim
2$ and $\mu\sim\mu_\Sigma\sim M_2\sim 250$\,GeV, and in particular by
light charginos close to the LEP bound. The enhancement in the TMSSM
is significantly larger than the one achievable in the MSSM ($\sim
20\%$ for $R_{\gamma\gamma}$) for the same chargino lower mass
bound~\cite{Casas:2013pta}. The measurements of these processes are
likely to improve in the next years. LHC is indeed expected to probe
the SM prediction of $\Gamma(h\to Z\gamma)$ once $\mathcal
O(100~$fb$^{-1})$ data is collected~\cite{Campbell:2013hz}, and to
measure the $g_{h\gamma\gamma}$ effective coupling within a 10\%
accuracy after a high luminosity 3000~fb$^{-1}$
run~\cite{Peskin:2012we}. With these further data the Higgs diphoton
signal strength will plausibly converge to the SM value. In such a
case, sizeable deviations in $h\to Z\gamma$ would not be compatible
with the TMSSM. On the contrary, if data will still exhibit a positive
deviation from the SM, there would be a clear indication of physics
beyond the SM. The above predictions and the tight correlation between
$R_{\gamma\gamma}$ and $R_{Z\gamma}$ could be thus crucial to rule out
or provide hints for the scenario considered here.

Besides the Higgs decays, we have investigated the DM phenomenology in
the TMSSM, focusing on the interplay of the neutralino and chargino
sectors enlarged by the triplet components. Similarly to the MSSM, the
LSP is a viable DM candidate in the Higgs or $Z$ pole region, and in
the so-called well-tempered regime. The Higgs and Z pole regions are
characterized by a Bino DM and are poorly sensitive to the Triplino,
as the Higgs-Higgsino-Bino is the only relevant coupling. However, the
well-tempered neutralino, where the LSP achieves the correct relic
density via coannihilation with the lightest chargino, presents a new
feature. Indeed the Triplino component of the LSP can substitute the
Wino in the well-tempered neutralino and can solve the problem of
having $M_1\sim M_2$ from grand unified model perspective. Indeed the
requirement of DM comes at the expenses of satisfying the LUX
exclusion limit for SI elastic cross-section on nuclei. The dominant
contribution is due to Higgs exchange, which imposes a lower bound on
$\mu$. Interestingly we found that this has an impact for the
$R_{\gamma\gamma}$ and $R_{Z\gamma}$ enhancements: the Higgs-chargino
coupling is reduced as well suppressing the signal strengths to at
most $20\%$. Notice that these values are once again larger than the
ones provided by the MSSM with DM constraints~\cite{Casas:2013pta},
when the Higgs production is SM-like.

The scenario considered here nicely illustrates the complementarity of
DM direct searches with LHC. For instance the next generation of
direct detection experiments, such as XENON1T, will probe a consistent
portion of the neutralino TMSSM parameter space. Moreover it will be
capable of constraining the Higgs invisible decay branching ratio up
to 1\%, in a time scale comparable to the LHC one. In general the
TMSSM is less constrained by current LHC bounds on simplified models
or supersymmetric searches. Indeed the presence of the Triplino can
modify the couplings and the decay modes. This has been already
  observed for stops in the TMSSM~\cite{deBlas:2013epa} even though a
  precise estimate of their current mass bound is still missing. On
  the other hand no study exists for the chargino and neutralino mass
  bounds. Although we have checked that the present
  constraints~\cite{Aad:2014nua,Khachatryan:2014qwa} do not apply to
  our analysis ballpark, a dedicated investigation would be required
  in order to accurately determine the allowed parameter
  region. Present data should primarily affect the chargino parameter
  region with light lightest-neutralino and with small $h\to Z\gamma$
  and $h\to \gamma\gamma$ enhancements. With more LHC data strongest
  bounds are expected, in particular for the DM mass close to the $Z$
  or $h$ resonance. On the other hand, in order to probe the
  coannihilation region (where the spectrum is compressed), ILC data
  and analyses similar to that proposed in ref.~\cite{Porto:2014fca}
  would be crucial.

%%%%%%%%%%%%%%%%%%%%%%%%%%%%%%%%%%%%%%%%%%%

\acknowledgments We are very grateful to Florian Staub for useful
explanations about \texttt{SPheno} and \texttt{SARAH} programs. We
also thank the authors of \texttt{CPsuperH}, in particular Jae Sik
Lee, for their help in numerical issues. GN thanks A.~Delgado and
M.~Quiros for useful discussions. CA acknowledges the support of the
ERC project 267117 (DARK) hosted by Universit\'e Pierre et Marie Curie
- Paris 6, PI J. Silk. GN was supported by the German Science
Foundation (DFG) within the Collaborative Research Center 676
``Particles, Strings and Early Universe''. VML acknowledges the
support of the Consolider-Ingenio 2010 programme under grant MULTIDARK
CSD2009-00064, the Spanish MICINN under Grant No. FPA2012-34694, the
Spanish MINECO ÒCentro de excelencia Severo Ochoa ProgramÓ under Grant
No. SEV-2012-0249, and the European Union under the ERC Advanced Grant
SPLE under contract ERC-2012-ADG-20120216-320421.

\bibliography{biblio}
\bibliographystyle{JHEP}

\end{document}